\documentclass[aps,preprint,showpacs,floatfix]{revtex4-1}
\usepackage{graphicx,bm,amsmath,dcolumn,subfigure,multirow,array}
\usepackage[english]{babel}
\usepackage{graphicx}
\usepackage{float}
\usepackage[font=small,labelfont=bf]{caption}
\usepackage{hyperref}
\begin{document}
\newcommand{\be}{\begin{equation}}
\newcommand{\ee}{\end{equation}}
\newcommand{\ea}{{\em et al.}}
\newcommand{\half}{\frac{1}{2}}
\newcommand{\ith}{^{(i)}}
\newcommand{\im}{^{(i-1)}}
\newcommand{\gae}
{\,\hbox{\lower0.5ex\hbox{$\sim$}\llap{\raise0.5ex\hbox{$>$}}}\,}
\newcommand{\lae}
{\,\hbox{\lower0.5ex\hbox{$\sim$}\llap{\raise0.5ex\hbox{$<$}}}\,}
\newcommand{\mat}[1]{{\bf #1}}
\renewcommand*{\thesubfigure}{}

\title{Completely packed O($n$) loop models and their relation with
exactly solved coloring models}
\author{
Yougang Wang$^{1}$, Wenan Guo$^{2,3}$, and Henk W.~J. Bl\"ote$^{1}$}
\affiliation{$^{1}$Lorentz Institute, Leiden University,
P.~O. Box 9506, 2300 RA Leiden, The Netherlands}
\affiliation{$^{2}$Physics Department, Beijing Normal University,
Beijing 100875, P.~R. China}
\affiliation{$^{3}${State Key Laboratory of Theoretical Physics,
Institute of Theoretical Physics, Chinese Academy of Sciences,
Beijing 100190, P.~R. China}} 
\date{\today}

\begin{abstract}
We explore the physical properties of the completely packed O($n$) loop
model on the square lattice, and its generalization to an Eulerian
graph model, which follows by including cubic vertices which connect 
the four incoming loop segments. This model includes crossing bonds as
well. Our study of the properties of this model involve transfer-matrix
calculations and finite-size scaling.  The  numerical results are compared
to existing exact solutions, including solutions of special cases of a
so-called coloring model, which are shown to be equivalent with our
generalized loop model. The latter  exact solutions correspond with seven
one-dimensional branches in the parameter space of our generalized loop model.
One of these branches, describing the case of nonintersecting loops, is
already known to correspond with the ordering transition of the Potts model.
We find that another exactly solved branch, which describes a model with
nonintersecting loops and cubic vertices, corresponds with a first-order
Ising-like phase transition for $n>2$. For $1<n<2$, this branch can be
interpreted in terms of a low-temperature O($n$) phase with corner-cubic
anisotropy. For $n>2$ this branch is the locus of a first-order phase
boundary between a phase with a hard-square lattice-gas like ordering,
and a phase dominated by cubic vertices. The first-order character of this
transition is  in agreement with a mean-field argument.
\end{abstract}
\pacs{05.50.+q, 64.60.Cn, 64.60.Fr, 75.10.Hk}
\maketitle

\section{Introduction}
\label{intro}
Several types of nonintersecting O($n$) loop models can be obtained as a
result of an exact transformation of certain O($n$)-symmetric spin
models \cite{Stanley,Domea,N,BN,N1,KNB}. Most of these models are
two-dimensional, but the transformation is also applicable in
three dimensions \cite{H2O2}. It provides a generalization of the O($n$)
model to non-integer, and even
negative values of $n$. Whereas most existing work is restricted to
nonintersecting loop models, the models can readily be generalized to
include cubic vertices \cite{BNcub} and crossing bonds \cite{MNR}.
These cubic vertices connect to four incoming loop segments, 
and arise naturally when the O($n$) symmetry of the original spin model is
broken by interactions of a cubic symmetry \cite{CG}. The crossing-bond
vertices occur in the loop representation of non-planar O($n$)-symmetric
spin models. 

\begin{figure}
\begin{center}
\subfigure[{\Large {\it z}}]{
\includegraphics[scale=0.12]{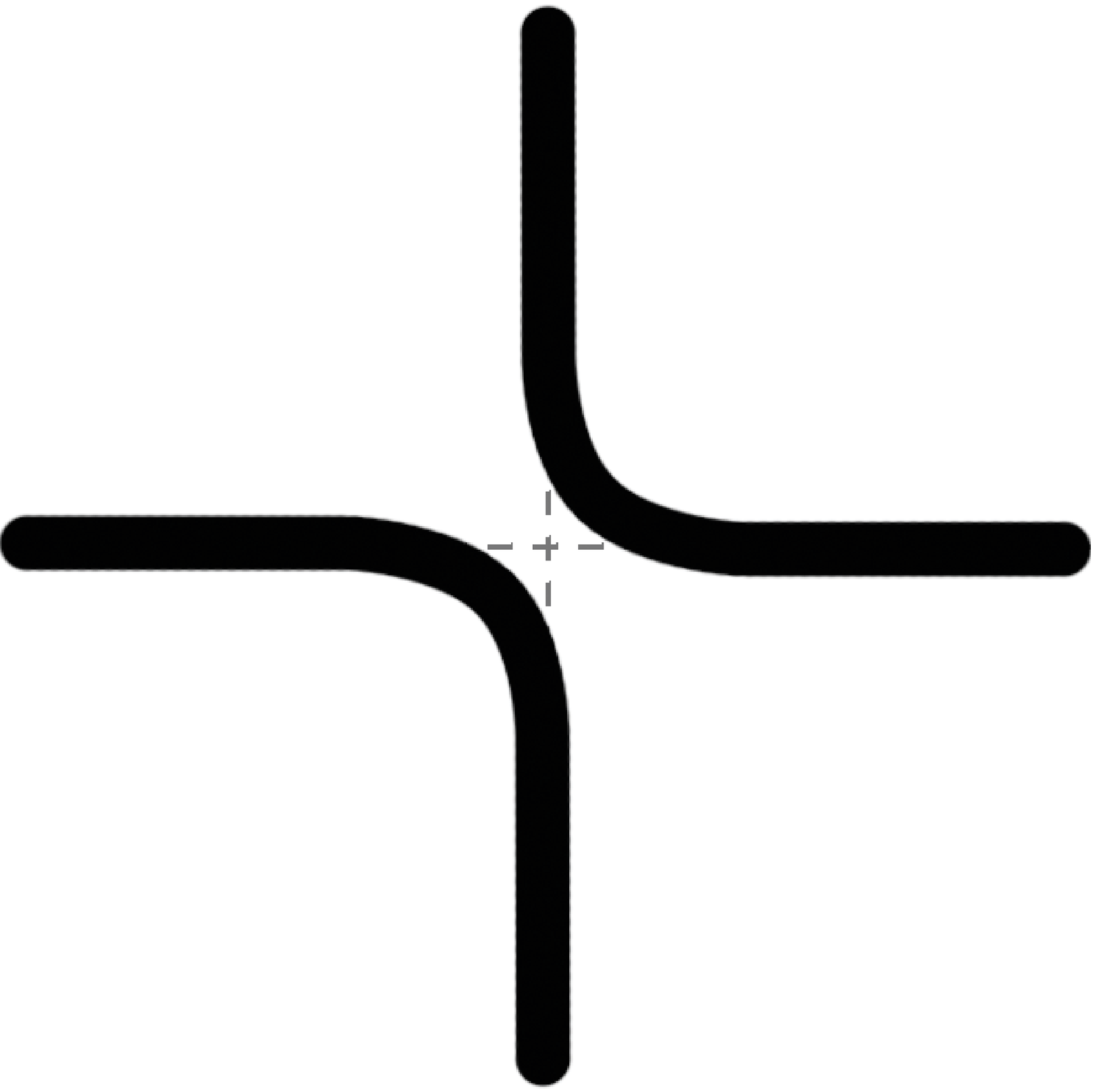}
}
\subfigure[{\Large {\it x}}]{

\includegraphics[scale=0.12]{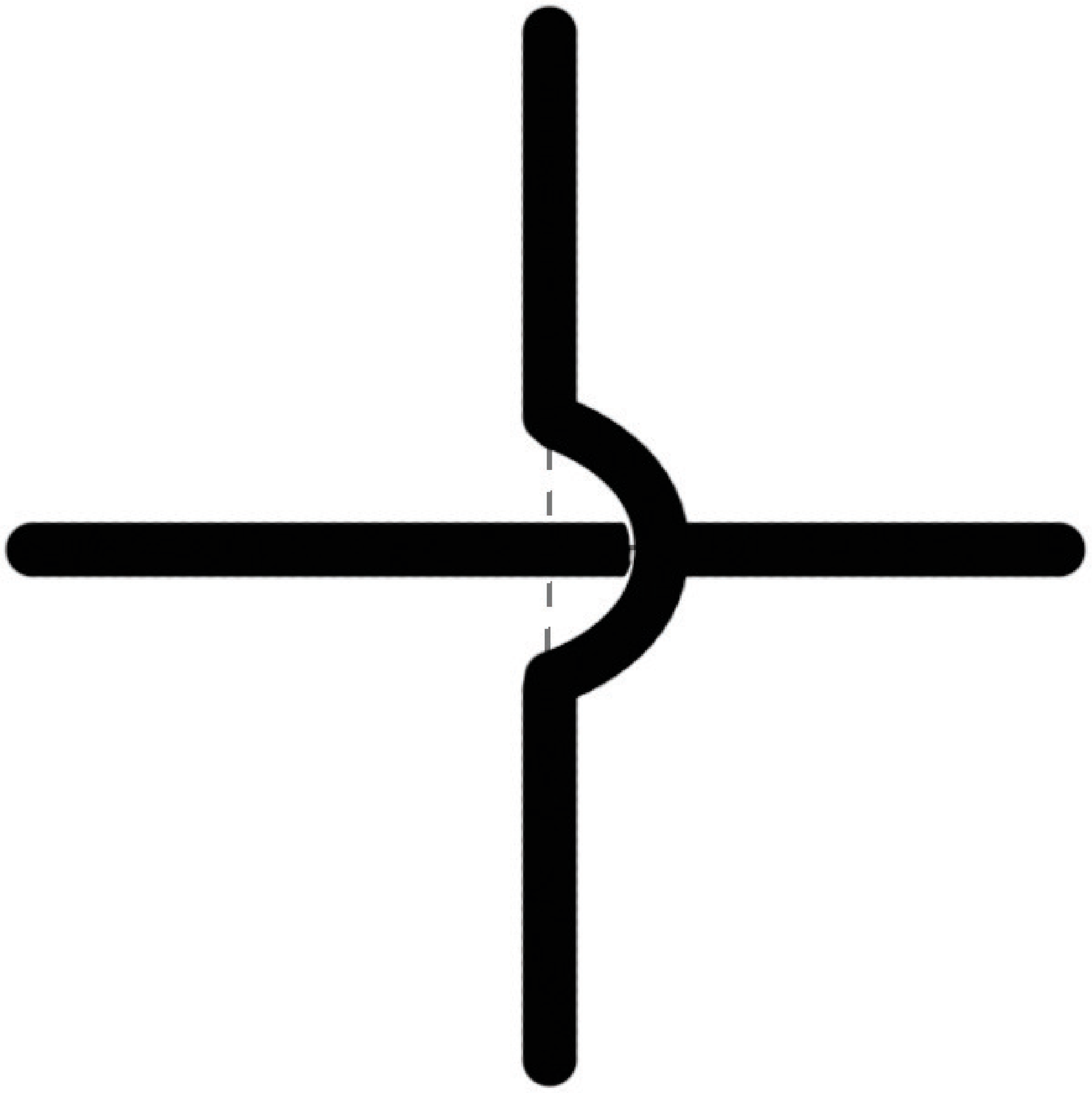}
}
\subfigure[{\Large {\it c}}]{

\includegraphics[scale=0.12 ]{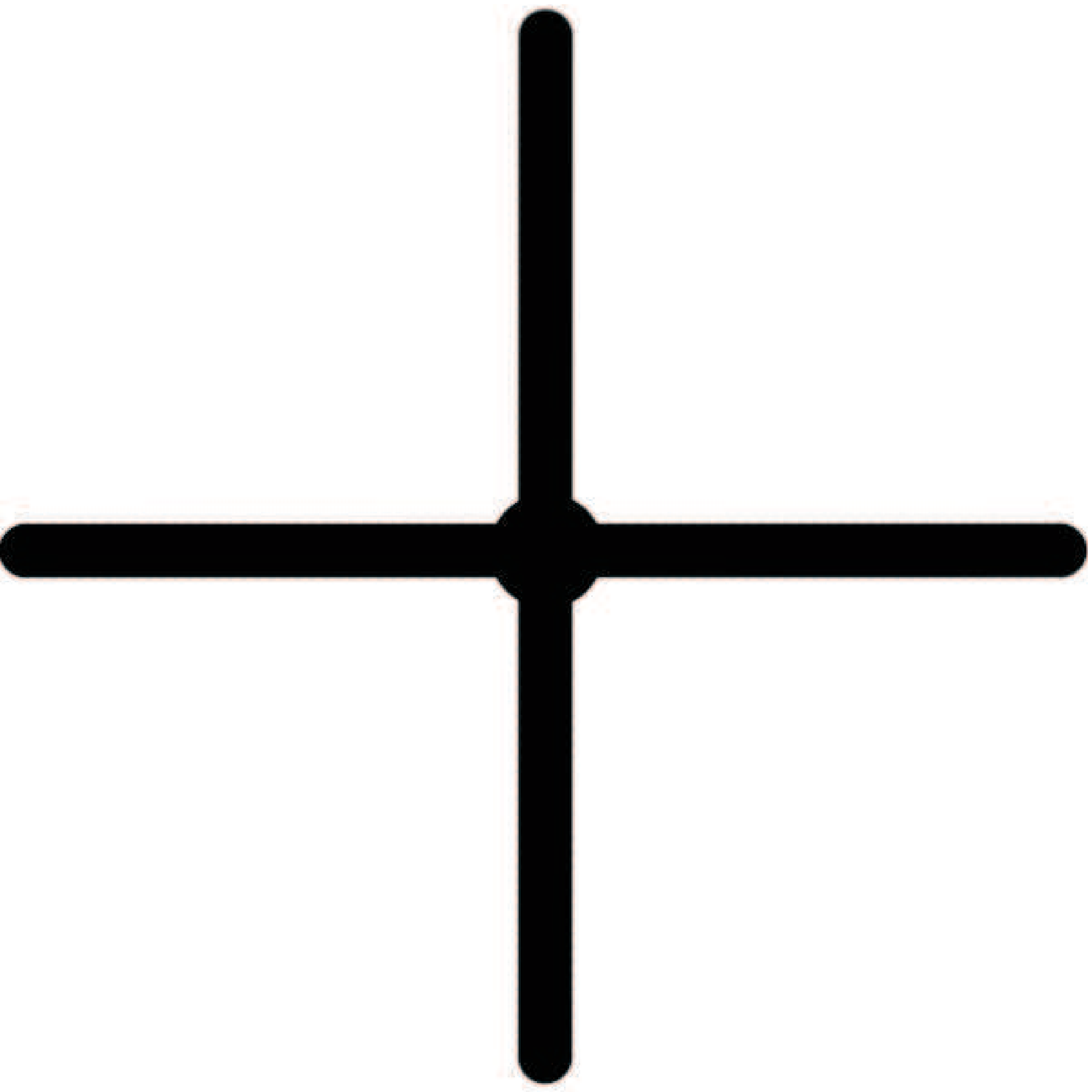}
}
\end{center}
\centering
\centering
\caption{The vertices of the completely packed O($n$) loop model with 
crossing bonds and cubic vertices on the square lattice, together with
their weights. Fourfold rotational symmetry of the model requires that
the weights for the two possible orientations of the $z$-type vertex
are the same.}
\label{vert}
\end{figure}

The presently investigated model is defined in terms of these three types
of vertices on the square lattice. The three types, which are shown in
Fig.~\ref{vert} together with their vertex weights, specify a complete
covering of the lattice edges. In comparison with a recent
investigation \cite{onc} of crossover phenomena in a densely packed
phase of the O($n$) loop model, the present set of vertices
is obtained by excluding those that do not cover all lattice edges.
Due to the absence of empty edges, the physical interpretation in terms
of an O($n$) spin model is  more remote. A formal mapping  of the loop 
model on the spin model leads to a spin-spin interaction energy that can
assume complex values when the relative weight of empty edges becomes
sufficiently small. However, the mapping of  the completely packed
O($n$) loop model on a dilute O($n-1$) loop model model (which was, as
far as we know, first formulated by Nienhuis; see Ref.~\onlinecite{BN})
brings it again closer to the realm of the spin models.

A configuration of the vertices of Fig.~\ref{vert} forms a so-called
Eulerian graph, which is a graph in which only even numbers of loop 
segments can be connected at each vertex.  We denote such graphs as
${\mathcal G}$. The present model may thus be called a completely packed
Eulerian graph model.
The partition sum of this model is defined by 
\begin{equation}
Z_{\rm EG}=\sum_{{\mathcal G}} z^{N_z} c^{N_{c}} x^{N_{x}} n^{N_{n}} \, .
\label{Zsql}
\end{equation}
where the sum on ${\mathcal G}$ is on all possible combinations of vertices.
The exponents $N_z$, $N_c$ and $N_x$ denote the numbers of vertices of types
$z$, $c$ and $x$ respectively, and $N_{n}$ denotes the number of components
of the graph ${\mathcal G}$. A component is a subset of edges connected by
a percolating path of bonds formed by the vertices of ${\mathcal G}$.
Since $Z_{\rm EG}$ is a homogeneous function of the vertex weights, one
may, without loss of generality, scale out one of the weights. We thus
normalize the weight $z$ of the O($n$) vertex describing colliding loop
segments to 1.

At this point it is appropriate to comment on our nomenclature. By
``nonintersecting loops'' we mean configurations consisting of the
type-$z$ vertices in Fig.~\ref{vert}. Since the word ``intersecting''
could be associated with the type-$x$ vertices as
well as the type-$c$ vertices in Fig.~\ref{vert}, we refer to type-$x$
vertices as crossing bonds, and to type-$c$ vertices as cubic vertices.
Thus we may, alternatively, call this model a completely packed loop
model with crossing bonds and cubic vertices, or just a generalized 
loop model. Furthermore we note that the name ``fully packed'' is used
for models in which all vertices are visited, but not all edges are
covered by loop segments \cite{BNFPL,Batch}.

The present work was inspired by a number of existing exact solutions,
in particular of a ``coloring model'' by Schultz \cite{S}, later studied 
in more detail by Perk and Schultz \cite{PS} and others \cite{BVV,VL}, 
see also Fateev \cite{VF}. In the Perk-Schultz model, the edges of
the square lattice receive one out of several colors, in such a way that,
for any given color, an even number of edges connects to each vertex.
Exact solution were found for several different branches of critical
lines that are parametrized by the number of colors.
Our purpose is to explore the physical context and the universal
properties of these exact solutions, and to determine their embedding in
or their intersection with the phase diagram spanned by the parameters
in Eq.~(\ref{Zsql}).

The outline of this paper is as follows. In Sec.~\ref{theory} we
reformulate the Eulerian graph model in terms of the number of loops,
and describe the transformation connecting it to the coloring model.
We review the exact results for the free energy, which apply to several
one-dimensional ``branches'' parametrized by $n$ in the parameter space
of Eq.~(\ref{Zsql}), and which will be relevant in the numerical analysis
of the conformal anomaly along these branches. 
This analysis is based on transfer-matrix calculations, for which 
some technical details are provided in Sec.~\ref{transfmat} and
Appendix~\ref{tmtechnique}. Results
for the free energy of the exactly solved branches are presented in
Sec.~\ref{numeranalf}, and for some scaling dimensions in
Sec.~\ref{numericanx}.  While the exploration of the complete (in fact 
three-dimensional) phase diagram of Eq.~(\ref{Zsql}) is beyond the
scope of the present paper, the embedding of some of the exactly
solved branches in this diagram is investigated in Sec.~\ref{embed}.
Our conclusions are presented and discussed in Sec.~\ref{discus}.

\section{Mappings and existing theory}
\label{theory}
\subsection{Euler's theorem}
Euler's theorem specifies that the number of components satisfies
$N_{n}=N_s-N_b+N_l$ where $N_s$ is the number of sites of the lattice, 
$N_b$ the number of bonds covered by ${\mathcal G}$, and $N_l$ is the
number of loops in ${\mathcal G}$. It simply means that every new
bond decreases $N_{n}$ by one, unless the end points of that bond were
already connected. Application of this theorem to the present model
requires some care because it merges the degrees of freedom of the cubic
model with those of the O($n$) model. Whereas the spins of the cubic
model \cite{BNcub,GQBW} are defined on the vertices of the square lattice,
the spins of the square-lattice O($n$) model \cite{N1,BN} are placed in the 
middle of the edges. Here we will adhere to the description for the 
square-lattice O($n$) loop model, which means that the number $N_s$ of
sites in Euler's relation 
is to be taken as twice the number $N_v$ of vertices. Furthermore,
in this formulation, a cubic vertex consists of {\em three} bonds:
it connects one pair of sites along the $x$ direction, one pair of
sites along the $y$ direction, and it also makes a connection between
both pairs. Thus, for the present model, the number of bonds as required 
in Euler's formula is  $N_b=2N_z+2N_x+3N_c$, and Euler's theorem takes
the form
\begin{equation}
N_{n}=N_s-2N_z-2N_x-3N_c+N_l=N_l-N_c \, ,
\label{Eul}
\end{equation}
where the last step uses $N_s=2 N_v$ and $N_v=N_z+N_x+N_c$ in the
completely packed model.
After substitution of Euler's theorem, the partition sum Eq.~(\ref{Zsql}) 
is thus reformulated as
\begin{equation}
Z_{\rm EG}= Z_{\rm loop}= 
\sum_{{\mathcal G}} z^{N_z} (c/n)^{N_{c}} x^{N_{x}} n^{N_{l}}\,.
\label{Zsqll}
\end{equation}
The Boltzmann weights now only depend on the numbers of vertices
of each type, and on the number of loops.  This formula exposes the 
nature of the partition sum as that of a generalized loop model.
As a consequence of the elimination of the number $N_n$ of components
of the Eulerian graphs, the weight of a cubic vertex now appears as
$c_n\equiv c/n$ instead of $c$.  In this context it is noteworthy that the
cubic weight $c$ used in Ref.~\onlinecite{onc} is equal to $c_n=c/n$ when
expressed in the parameters of the present work.

\subsection{Relation with the coloring model}
The Perk-Schultz coloring model is defined in Refs.~\onlinecite{S,PS} in
terms of bond variables that can assume $n$ different colors. The colors
of the bonds connected to a given vertex are not independent. The number of 
bonds of a given color connected to a vertex is restricted to be even.
Following Ref.~\onlinecite{S}, the vertex weights are denoted
$R^{\lambda \mu}(\alpha \beta)$ where
$\lambda,\mu$ denote the colors of the bonds in the $-x,+x$ directions,
and $\alpha,\beta$ apply to the $-y,+y$ directions respectively.
The color restrictions and symmetries are expressed by
\begin{equation}
R^{\lambda \mu}(\alpha \beta)=
W^{\rm d}_{\alpha \lambda} \delta_{\alpha \beta} \delta_{\lambda \mu}+
W^{\rm r}_{\alpha \beta} \delta_{\alpha \lambda} \delta_{\beta   \mu}+
W^{\rm l}_{\alpha \beta} \delta_{\alpha \mu  } \delta_{\beta \lambda}
\label{cr}
\end{equation}
with 
\begin{equation}
W^{\rm r}_{\alpha \beta}=W^{\rm r} (1-\delta_{\alpha \beta})  \, , ~~~~~
W^{\rm l}_{\alpha \beta}=W^{\rm l} (1-\delta_{\alpha \beta})  \, ,
\label{wr}
\end{equation}
and
\begin{equation}
W^{\rm d}_{\alpha \beta}=W^{\rm d} \delta_{\alpha \beta}+
W^{0} (1-\delta_{\alpha \beta}) \, .
\label{wd}
\end{equation}
Here, the weights are restricted such as to satisfy the permutation
symmetry of all colors, so that all colors are equivalent.
In this work we furthermore impose the additional symmetry condition
\begin{equation}
W^{\rm l} =W^{\rm r} \, ,
\label{lrsym}
\end{equation}
which leads to a set of vertex weights that
is invariant under rotations by $\pi/2$, thereby allowing conformal
symmetry  of the coloring model in the scaling limit. 

The model still contains, besides the number $n$
of colors, three variable parameters $W^{0}$, $W^{\rm d}$ and $W^{\rm r}$.
The  partition sum of the coloring model is defined by
\begin{equation}
Z_{\rm cm}=\sum_{{\mathcal C}}\prod_vR^{\lambda_v\mu_v}(\alpha_v\beta_v)\,,
\label{Zcm}
\end{equation}
where the sum on ${\mathcal C}$ is over all colors of all bonds, and the
product is over all vertices $v$. Each bond variable occurs twice in the
product, once as a superscript and once as an argument of $R$.

\begin{figure}
\begin{center}
\subfigure[{\Large {\it Z}$_1$}]{
\includegraphics[scale=0.08]{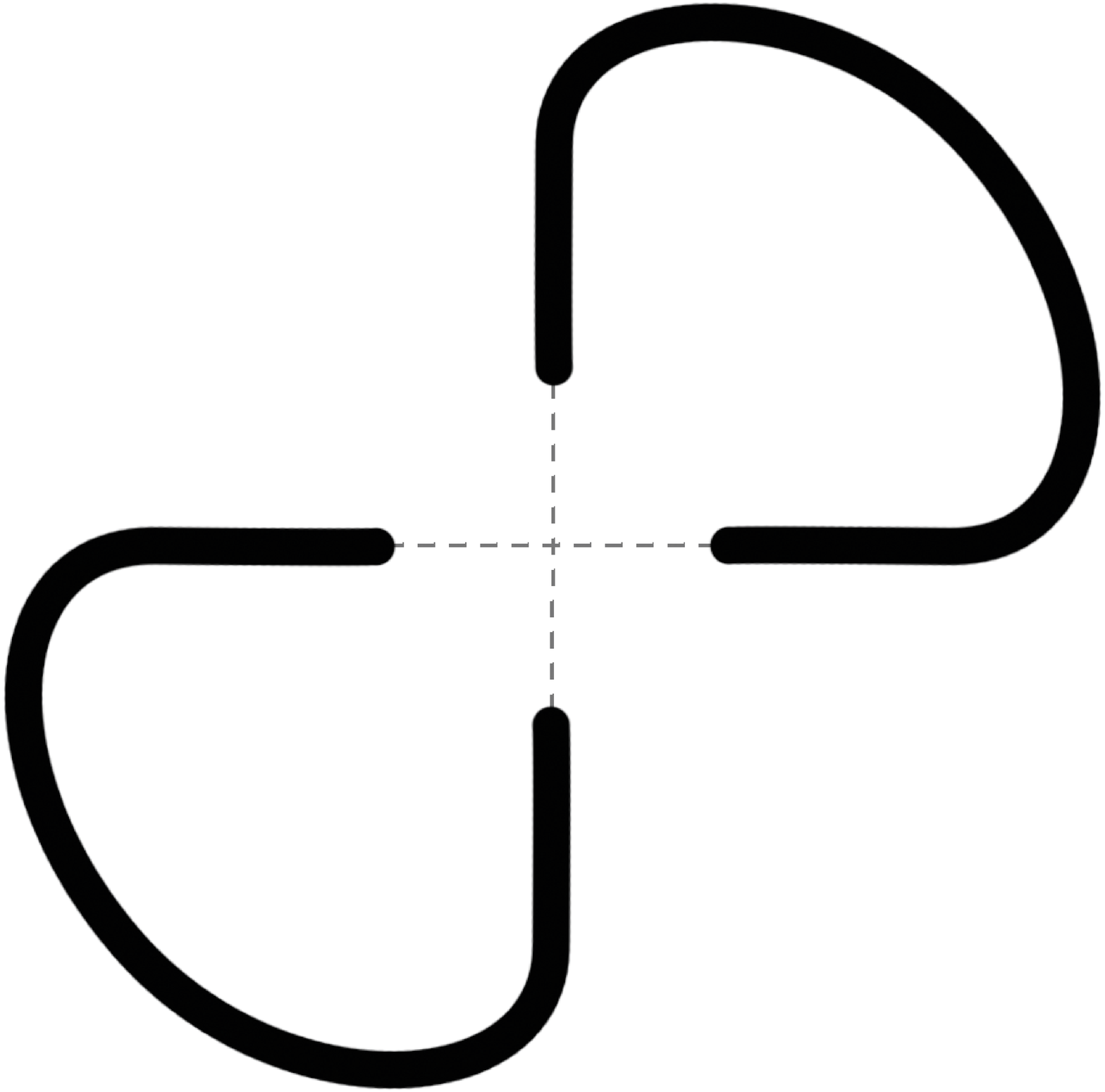}

}
\subfigure[{\Large {\it Z}$_2$}]{
\includegraphics[scale=0.08]{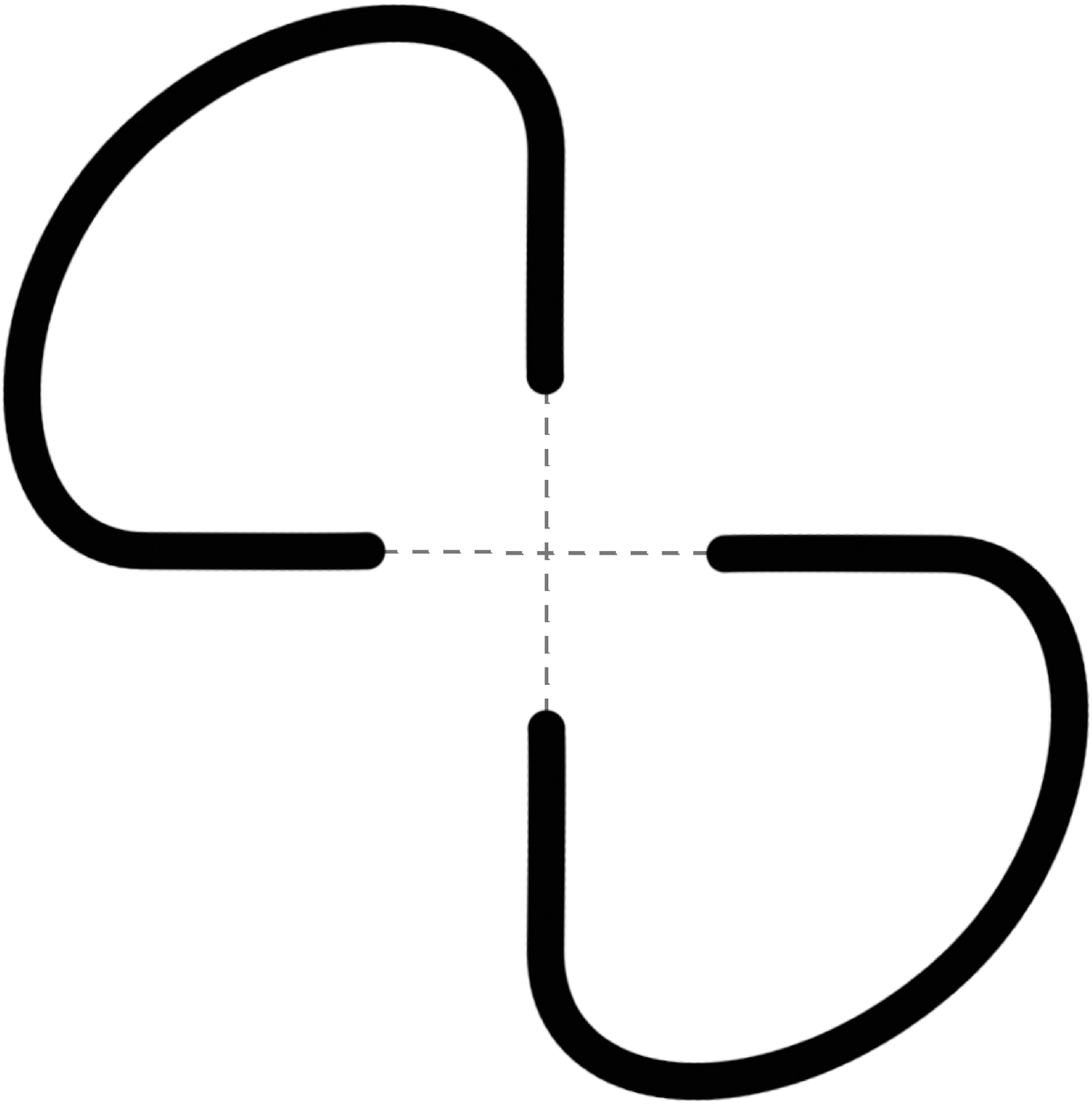}
}
\subfigure[{\Large {\it Z}$_3$}]{
\includegraphics[scale=0.08]{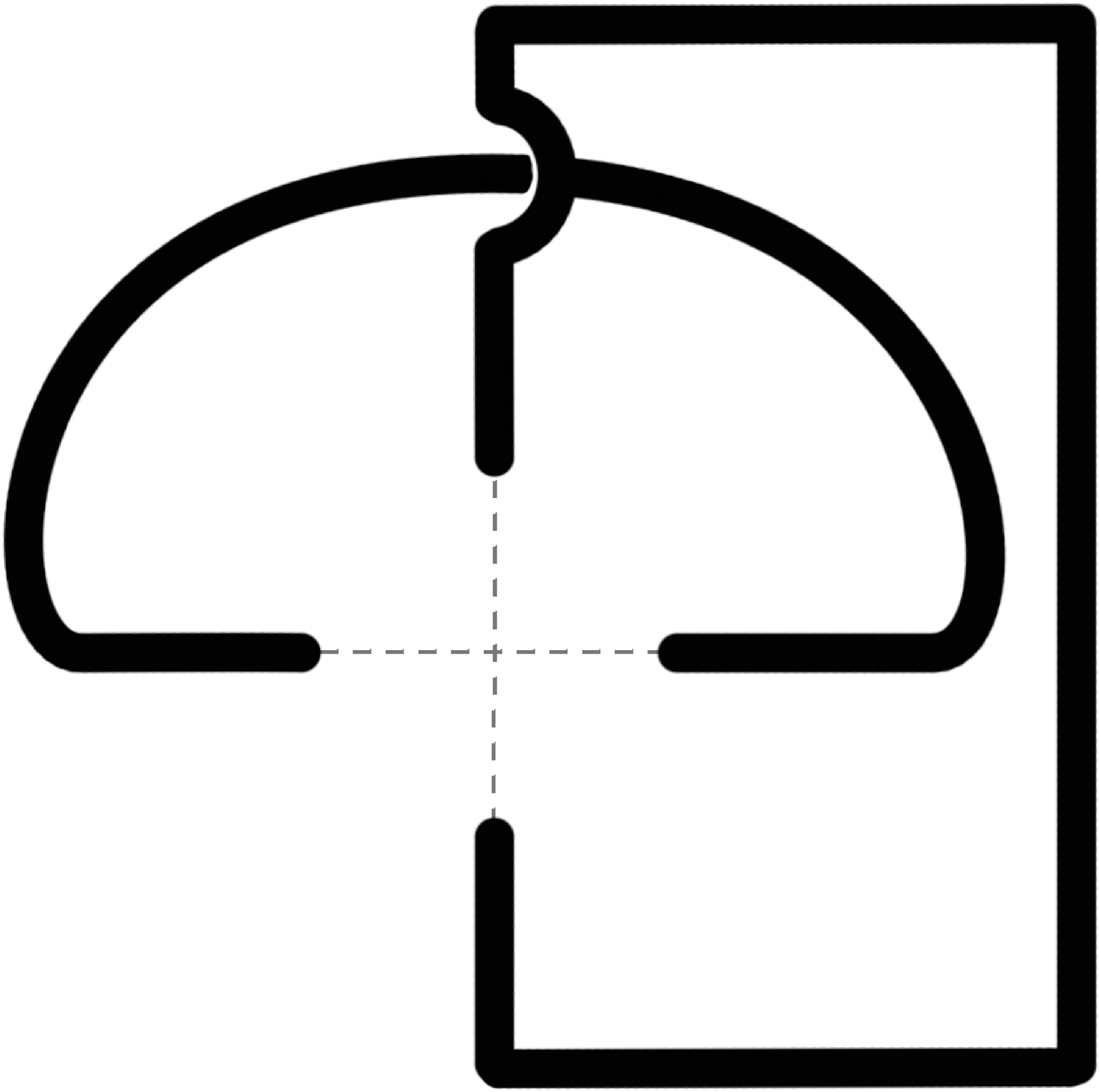}
}
\subfigure[{\Large {\it Z}$_4$}]{
\includegraphics[scale=0.08]{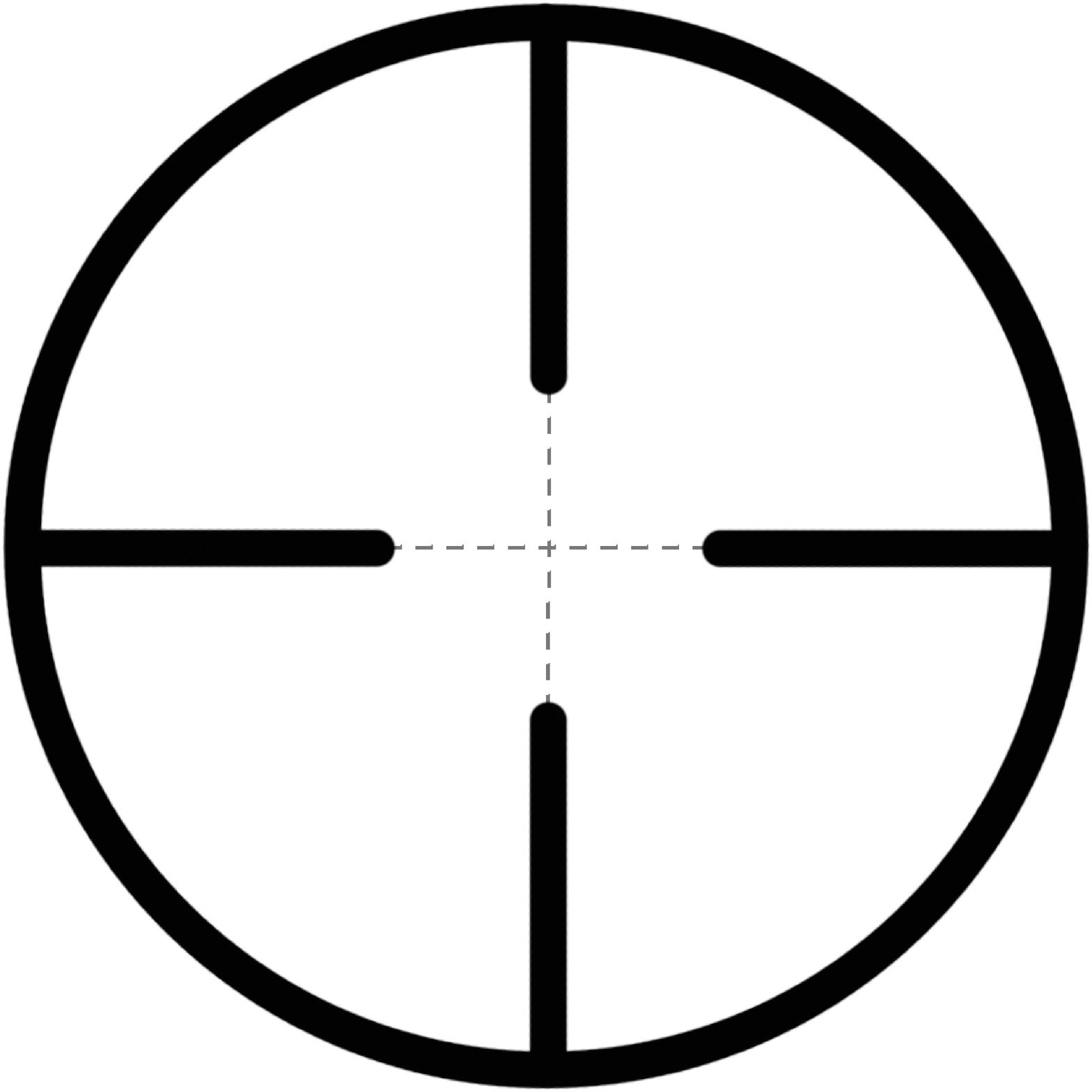}
}
\end{center}
\centering
\centering
\caption{The four different ways in which the incoming bonds at a vertex
can be connected by the remaining configuration of the generalized loop
model. The corresponding restricted partition sums are indicated under
the figures.}
\label{conn}
\end{figure}

In the absence of intersections between different colors, i.e., 
$W^{\rm d}=0$, the coloring model is known to be equivalent with a
Potts model and its Eulerian graph representation \cite{PW}.
Here we provide the exact correspondence between the $W^{\rm d} \ne 0$
coloring model \cite{referee} and the model of Eq.~(\ref{Zsqll}).
This follows simply by the interpretation of the weight $n$ of each
component in Eq.~(\ref{Zsql}) in terms of a summation on $n$ different
colors. Then, the set of configurations of the loop model precisely
matches that of the coloring model with the weights restricted according
to Eqs.~(\ref{cr})-(\ref{wd}). The relation of the parameters $W^{\rm d}$,
$W^{0}$ and $W^{\rm r}$ with $z$, $x$ and $c$ can be obtained from a
comparison between the expressions for the partition sum in terms of the
two types of vertex weights. Consider a loop model configuration and
remove one vertex. The connectivity of the incoming bonds, as determined 
by the surrounding loop model configuration, is denoted by an integer 1-4,
as specified in Fig.~\ref{conn}.
The corresponding restricted partition sums of the generalized loop model
are denoted as $Z_1$ to $Z_4$. They do not yet include the degeneracy
factor $n$ of the incomplete loops connected to the incoming bonds.
In terms of these restricted sums and the local vertex weights of the
coloring model, the partition sum is obtained by summation on the color 
combinations allowed by the diagrams in Fig.~\ref{conn} as
\begin{equation}
Z_{\rm cm}=[(n^2-n)W^{\rm r}+nW^{\rm d}](Z_1+Z_2)+[(n^2-n)W^0+nW^{\rm d}]Z_3
+nW^{\rm d}Z_4 \, .
\label{Zcm1}
\end{equation}
Using instead the local vertex weights of the generalized loop model,
the partition sum follows, taking into account the weight $n$ per component
specified by Eq.~(\ref{Zsql}), as
\begin{equation}
Z_{\rm loop}=[(n^2+n)z+nx+nc](Z_1+Z_2)+n(2z+nx+c)Z_3+n(2z+x+c)Z_4  \, .
\label{Zsqll1}
\end{equation}
The equivalence of both models requires that the prefactors of
$Z_1+Z_2$, $Z_3$ and $Z_4$ are the same in both forms of the partition sum.
These conditions lead to three equations, and subsequent solution shows that
the models are equivalent if the parameters simultaneously satisfy
\begin{equation}
\left. 
\begin{array}{lll}
W^0      &=&x       \\
W^{\rm d}&=&2z+x+c  \\
W^{\rm r}&=&z       
\end{array}
\right\} \, .
\label{map}
\end{equation}
In the representation of Eq.~(\ref{Zsql}), the parameter $n$ describing
the number of colors is no longer restricted to positive integers.

\subsection{The branches resulting from the solution of the coloring model}
\label{Schultz_Solutions}
Several cases of the coloring model were studied analytically by
Schultz \cite{S}. That work provided analytic expressions
for the partition sum per site. Included are results
for a number of index-independent models, i.e., models satisfying
Eq.~(\ref{cr})-(\ref{wd}), so that all colors are equivalent.
As noted above, the present work also
restricts the vertex weight to be invariant under rotations by $\pi/2$,
as required by asymptotic conformal invariance \cite{CI}. This enables
the numerical estimation of some universal quantities as outlined in
Sec.~\ref{transfmat}.

After application of these restrictions, the cases studied by Schultz 
reduce to seven one-dimensional subspaces in the parameter space of
the loop model.
These correspond, after the mapping according to Eq.~(\ref{map}), 
with exactly solved ``branches'' of the generalized loop model of
Eqs.~(\ref{Zsql}) and ~(\ref{Zsqll}).  The vertex weights are shown in
Table \ref{branches} as functions of $n$ for these seven branches. 
These weights are normalized such that $z=1$, except for branches 6 and
7, where $z$ vanishes, and where we use the normalization $x=1$ instead.
Table \ref{branches}  also includes, under ``case'', the notation used in
Ref.~\onlinecite{S} referring to each branch.

\begin{table}
\renewcommand{\arraystretch}{1.5}
\caption {Intersection between the exactly solved subspaces of the
coloring model and the parameter space of the generalized loop model. 
These intersections form seven branches, defined in the first column,
for which we also include the vertex weights. The entries under ``case''
show the labeling used by Schultz \cite{S}, with the characters ``a''
and ``b'' appended, in order to separate the Schultz cases
into branches with single-valued vertex weights.}
\label {branches}
\begin{tabular}{ |c | c || >{\centering\arraybackslash} m{2cm}
| >{\centering\arraybackslash}m{2cm} | >{\centering\arraybackslash}m{2cm} |}
\hline
\multirow{2}{*}{branch} & \multirow{2}{*}{case} &
\multicolumn{3}{c}{vertex weights} \vline \\ \cline{3-5}
      &       &$z$& $x$             & $c$              \\ \hline
    1 & IIA1  & 1 & 0               & 0                \\ \hline
    2 & IIA2a & 1 & 0               & $-1+\sqrt{n-1}$  \\ \hline
    3 & IIA2b & 1 & 0               & $-1-\sqrt{n-1}$  \\ \hline
    4 & IIB1a & 1 &$$\frac{2-n}{4}$$& 0                \\ \hline
    5 & IIB1b & 1 &$$\frac{n-2}{4}$$& $$\frac{2-n}{2}$$\\ \hline
    6 & IIB2a & 0 & 1               & 0                \\ \hline
    7 & IIB2b & 0 & 1               & $-2$             \\ \hline
    
\end{tabular}
\end{table}

\subsubsection{Branch 1}
The exact solution of branch 1 by Schultz \cite{S} is presented in terms
of a quantity denoted there as $f$, which appears to be the per-site
partition function, with the normalization $W^{\rm d}=1$ \cite{PS}.
Branch 1 has nonzero weights only for colliding vertices of the $z$-type,
as shown in Fig.~\ref{vert}.  It thus applies to a completely packed,
nonintersecting loop model.  For $n \geq 0$, this branch appears to be
exactly equivalent with the 6-vertex model, and with the $q=n^2$-state
Potts model at its transition point\cite{BKW}.  Due to these equivalences
much is already known for branch 1. We recall some of these results for
reasons of completeness as well as relevance for the interpretation of
the phase diagram of Eq.~(\ref{Zsql}).

Exact solutions of the aforementioned equivalent  models were already 
given by Lieb \cite{EL} and Baxter \cite{BaxP} respectively.
After taking into account the different normalizations of the vertices,
and the fact that the number of Potts sites
is one half of the number of vertices, the Schultz result for the free
energy per vertex in the range $n>2$ have been shown \cite{BWG} to agree
with the results of Lieb and Baxter in the corresponding parameter range. 
The Schultz result does not apply for $n\leq 2$, but there various other
results for the free energy \cite{EL,BaxP,LW,Baxb,BWG} are available.
In the thermodynamic limit, the following results for the free energy per
vertex apply.
\begin{equation}
f(n)=\half \theta + \sum_{k=1}^{\infty}
\frac{\exp(-k \theta)\,\tanh(k\theta)}{k}\, ,\mbox{\hspace{10mm}} (n>2)\, ,
\label{fgt2}
\end{equation}
with $\theta$ defined by $\cosh \theta = n/2$;
\begin{equation}
f(2)=2 \ln \frac{ \Gamma(1/4)}{2 \, \Gamma(3/4)} \, ;
\label{feq2}
\end{equation}
\begin{equation}
f(n)=\half ~\int_{-\infty}^{\infty}dx \,
\frac{\tanh \mu x \, \sinh(\pi-\mu)x}{x \sinh \pi x}
\, ,\mbox{\hspace{10mm}} (-2<n<2)\, ,
\label{flt2}
\end{equation}
where the parameter $\mu$ is defined by $\mu \equiv {\rm arccos}(n/2)$;
\begin{equation}
f(-2)=0 \, ;
\label{feqm2}
\end{equation}
\begin{equation}
f(n)=\half \tilde{\theta}+ \sum_{k=1}^{\infty}
\frac{\left[-\exp(-\tilde{\theta})\right]^k\, \tanh(k \tilde{\theta})} {k} \,
\, ,\mbox{\hspace{10mm}} (n<-2)\, ,
\label{fltm2}
\end{equation}
where $ \cosh \tilde{\theta}=-n/2$. The expression for $n<-2$ applies \cite{BWG}
to the thermodynamic limit of a system with a number of vertices equal to a
multiple of 4.

The correlation functions are known to follow a power law as a function of
distance  in the critical range $-2 \leq n \leq 2$, and to decay exponentially
for $|n| > 2$. The off-critical phase for large $n$ is known \cite{BWG} to
display the same type of order as the square lattice gas with nearest-neighbor
exclusion.

\subsubsection{Branch 2 and 3}
These branches contain both colliding ($z$-type) and cubic vertices
($c$-type), as shown in Fig.~\ref{vert}, with a weight that depends on $n$.
Their nature differs from branch 1
in the fact that different loops may now have common edges and vertices,
and thus be forced into the same component, according to Eq.~(\ref{Zsql}).

In order to avoid confusion with our notation, we denote the Schultz result
for the per-site partition function as $z_{\rm S}$, instead of $f$ as used
there, which we reserve for the free-energy density.
After substitution of the parameters as determined by Eq.~(\ref{map}) and
Table \ref{branches} into the result \cite{S}  for $z_{\rm S}$ of branches
2 and 3, and some simplification, the free-energy per vertex follows as
\begin{equation}
f=\ln(W^d\, z_{\rm S})\\ =\ln\left\{\,\frac{n-1}{|-1 \pm \sqrt{n-1}|}\,
\prod_{k=1}^{\infty}\left[
\frac{1\mp (n-1)^{-2k-1/2}}{1\mp (n-1)^{-2k-3/2}}\right]^2 \right\}\,,
\label{fbr23}
\end{equation}
where the upper signs in $\pm$ and $\mp$ apply to branch 2, and the lower
signs to branch 3. This result applies to the thermodynamic limit of systems
with an even number of vertices. Its validity cannot extend into the
range $n<2$, since the infinite product vanishes there. For $n \to 2$,
branch 2, the infinite product compensates the divergence of the prefactor.
Since branch 2
intersects with branch 1 at $n=2$, its free energy at $n=2$ is given by
Eq.~(\ref{feq2}). For branch 3, the infinite product assumes the value 2
in the limit $n \to 2$, so that the free energy vanishes in this limit.

\subsubsection{Branch 4 and 5}
\label{RvsS}  
For branch 4, the system contains, in addition to the $z$-type colliding
vertices, also $x$-type crossing-bond vertices (see Fig.~\ref{vert}),
but no cubic vertices.

A problem arises with the free-energy density implied by the result for
subcase IIB1 given by Schultz in Ref.~\onlinecite{S}, since it displays
many divergences as a function of $n$. Furthermore, footnote [64] of
Ref.~\onlinecite{PS}, which applies to this result, allows for the possibility
that it has to be modified. The result given for subcase IIB1 in \cite{S}, 
in terms of the per-site partition sum $z_{\rm S}$, reduces in the present
parameter subspace to
\begin{equation}
z_{\rm S}(n)=\frac{2-n}{4} \,
\frac{\Gamma^2\left(\frac{1}{4}\right)}
       {\Gamma^2\left(\frac{3}{4}\right)}\,
\frac{\Gamma(\half+\alpha) \Gamma(1-\alpha)}
     {\Gamma(\half-\alpha) \Gamma(\alpha)}\, , \mbox{\hspace{10mm}}
     \alpha \equiv \frac{6-n}{4(2-n)} \, .
\label{alphd}
\end{equation}
After two applications of Euler's reflection formula, one finds
\begin{equation}
z_{\rm S}(n)=\frac{2-n}{4} \,
\frac{\Gamma^2\left(\frac{1}{4}\right)}
       {\Gamma^2\left(\frac{3}{4}\right)}\,
\frac{\Gamma^2(\alpha+\half)}{\Gamma^2(\alpha)} \, {\rm ctg}(\alpha\pi) \, .
\label{Sb4}
\end{equation}

An independent calculation of the exact free-energy density of branch 4
is due to Rietman \cite{RPhD}. That result was derived for the intersecting
loop representation in Eq.~(\ref{Zsqll}), whose relation with the coloring
model was not immediately obvious. The Rietman expression for the free
energy is free of divergences for $n<2$.  Numerical evaluation shows
that the Schultz and Rietman results are different, except for a number,
mostly fractional, values of $n$ where they happen to coincide.
We thus attempt to
cast the Rietman result in a similar form as Eq.~(\ref{Sb4}). We denote the 
Rietman result for the per-site partition function as $z_{\rm R}$. It is
equal to \cite{RPhD}, 
\begin{equation}
z_{\rm R}(n)=\frac{4\kappa}{1+\kappa u(1-u)}\frac{\Gamma(1+\frac{u}{2})
\Gamma(\frac{3}{2}-\frac{u}{2})\Gamma(\frac{1}{2}+\frac{1}{2\kappa}+
\frac{u}{2})\Gamma(1+\frac{1}{2\kappa}-\frac{u}{2})}{\Gamma(\frac{1}{2}+
\frac{u}{2})\Gamma(1-\frac{u}{2})\Gamma(\frac{1}{2\kappa}+\frac{u}{2})
\Gamma(\frac{1}{2}+\frac{1}{2\kappa}-\frac{u}{2})}\,.
\label{Rm}
\end{equation}
where $\kappa \equiv 1-\frac{n}{2}$. The variable $u$ parametrizes a class
of commuting transfer matrices and describes the anisotropy of the model
when the two $z$-type colliding vertices are given different weights,
say $z_1$ and $z_2$. In that notation we have $u=z_1/(z_1+z_2)$. For the
present work we thus have $u=\half$.
Substitution of $\kappa$ and $u$ in Eq.~(\ref{Rm}) leads to 
\begin{equation}
z_{\rm R}(n)=16\,\frac{2-n}{10-n}\,
     \frac{\Gamma^2\left(\frac{5}{4}\right)}
{\Gamma^2\left(\frac{3}{4}\right)}\,
\frac{\Gamma^2\left(\frac{3n-10}{4(n-2)}\right)}
{\Gamma^2\left(\frac{n-6}{4(n-2)}\right)} 
=\,\frac{2-n}{10-n}\,
\frac{\Gamma^2\left(\frac{1}{4}\right)}
{\Gamma^2\left(\frac{3}{4}\right)}\,
\frac{\Gamma^2(\alpha+\half)}{\Gamma^2(\alpha)} \,.
\label{zbr4}
\end{equation}
where the last equality uses the definition (\ref{alphd}) of $\alpha$.
A comparison of Eqs.~(\ref{Sb4}) and (\ref{zbr4}) shows that
\begin{equation}
z_{\rm S}(n)=\,\frac{10-n}{2-n}\, {\rm ctg}(\alpha\pi)\, z_{\rm R}(n)\,.
\label{S_Rm}
\end{equation}
The factor $(10-n)/(2-n)$ is equal to the weight ratio $W^{\rm d}/W^0$
in the coloring model. The normalization used by Rietman, namely
$2z+x= W^{\rm d}=1$, thus indicates that the normalization $W^0=1$ was
used for the branch-4 result for $z_{\rm S}$ given in Ref.~\onlinecite{S}.
Furthermore, since we do not expect divergences in the free energy as
associated with the factor ${\rm ctg}(\alpha\pi)$ in $z_{\rm S}$,
we attribute that factor to the ambiguity of the periodic factor
mentioned in footnote [64] of Ref.~\onlinecite{PS}, and thus ignore it.
With these provisions, the results for the per-site partition function
according to Refs.~\onlinecite{S} and \cite{RPhD} become identical.
The free energy density of branch 4 follows as 
\begin{equation}
f(n)=\ln \left[\frac{10-n}{4}z_{\rm R}\right] 
=\ln \left[\frac{2-n}{4} \, 
\frac{\Gamma^2\left(\frac{1}{4}\right)}
{\Gamma^2\left(\frac{3}{4}\right)}\,
\frac{\Gamma^2(\alpha+\half)}{\Gamma^2(\alpha)} \right] \,.
\label{fbr4R}
\end{equation}
This expression is well behaved for $n<2$, but in the range $2<n<6$ it
does not exhibit the expected type of behavior, because the arguments
of the gamma functions can diverge and become negative.
The fact that branches 1 and 4 intersect at $n=2$ allows a consistency
check by taking the limit $n\to 2$ in Eq.~(\ref{fbr4R}). Since $\alpha$
diverges, we may safely apply Stirling's formula. It then appears that the
ratio of the divergent gamma functions just cancels the prefactor $2-n$,
so that we indeed reproduce Eq.~(\ref{feq2}).

The vertex weights for branch 5 differ from branch 4 in the additional
presence of $c$-type cubic vertices. The Schultz result \cite{S} for
branch 5 specifies the same  expression for the partition function as 
for branch 4.
Later we shall compare our numerical results for branches 4 and 5 with
Eq.~(\ref{fbr4R}) for several values of $n$.

\subsubsection{Branch 6 and 7}
The vertex weights of branches 6 and 7, given in Table \ref{branches},
do not depend on $n$, but the
partition sum still contains the loop weight explicitly, and indeed it 
appears in the exact per-site partition sum $z_{\rm S}$ as given by
Schultz \cite{S}.
This result leads to the following free-energy density
\begin{equation}
f(n)=\ln[z_{\rm S}]=\ln\left[\frac{\Gamma\left(\frac{1}{nb}\right)
\Gamma\left(\frac{1+(n-2)b}{nb}\right)}
                      {\Gamma\left(\frac{1-b}{nb}\right)
\Gamma\left(\frac{1+(n-1)b}{nb}\right)}\right] \, .
\label{S6_7}
\end{equation}
where
\begin{equation}
b=\frac{W^{\rm r}}{W^{\rm d}}=\frac{z}{2z+x+c}=0 \, .
\nonumber
\end{equation}\
Since we impose rotational symmetry over $\pi/2$ on the vertex weights
by Eq.~(\ref{lrsym}), and moreover $W^{\rm l} W^{\rm r}=0$ for branches
6 and 7,  we arrive at the special point  $b=W^{\rm r}=z=0$. We thus take 
the limit $b \to 0$ in Eq.~(\ref{S6_7}):
\begin{equation}
f(n)=\lim_{b\to 0}\left[ \ln\Gamma\left(\frac{1}{nb}\right)+
\ln\Gamma\left(\frac{1+(n-2)b}{nb}\right)-
    \ln\Gamma\left(\frac{1-b}{nb}\right)-
\ln\Gamma\left(\frac{1+(n-1)b}{nb}\right)\right].
\label{fr}
\end{equation}\
Each of the arguments of the gamma functions diverges the limit of $b=0$.
We apply Stirling's formula and neglect terms that vanish for $z \to \infty$: 
\begin{eqnarray}
f(n)=\lim_{b\to 0}\mbox{\large\{}&\mbox{\large (}&\frac{2-nb}{2nb}\,
[-\ln(nb)]-\frac{1}{nb}+\frac{1}{2}\ln(2\pi)\mbox{\large )}\nonumber\\
                           +&\mbox{\large (}&\frac{2+nb-4b}{2nb}\,
                  \{\ln[1+(n-2)b]-\ln(nb)\}-\frac{1+(n-2)b}{nb}+
                  \frac{1}{2}\ln(2\pi)\mbox{\large )}\nonumber\\
                           -&\mbox{\large (}&\frac{2-nb-2b}{2nb}
                  \,[\ln(1-b)-\ln(nb)]-\frac{1-b}{nb}+\frac{1}{2}
                  \ln(2\pi)\mbox{\large )}\nonumber\\
                           -&\mbox{\large (}&\frac{2+nb-2b}{2nb}\,
                  \{\ln[1+(n-1)b]-\ln(nb)\}-\frac{1+(n-1)b}{nb}+
                  \frac{1}{2}\ln(2\pi)\mbox{\large )\}} \, . \nonumber
\end{eqnarray}\
We first consider the divergent terms with $\ln(nb)$.
The sum of their amplitudes appears to cancel exactly:
\begin{equation}
\frac{2-nb}{2nb}+\frac{2+nb-4b}{2nb}-\frac{2-nb-2b}{2nb}-\frac{2+nb-2b}{2nb}=0 \, .
\nonumber
\end{equation}\
Similarly, the sums of the amplitudes of terms with $\frac{1}{nb}$
and $\frac{1}{2}\ln(2\pi)$ vanish.
Therefore,
\begin{eqnarray*}
\hspace*{-21 mm}
f(n)=\lim_{b\to 0}&\mbox{\large\{}&\frac{2+nb-4b}{2nb}\,
 \ln[1+(n-2)b]\nonumber\\
       &-&\frac{2-nb-2b}{2nb}\, \ln(1-b)\nonumber\\
       &-&\frac{2+nb-2b}{2nb}\, \ln[1+(n-1)b]\mbox{\large\}} \, . \nonumber
\end{eqnarray*}
The prefactors depend linearly on $1/b$, and the logarithms are proportional
to $b$ in lowest order. It is therefore sufficient to keep the divergent
part of the prefactors and the terms with $b$ in the logarithms:
\begin{equation}
f(n)=\lim_{b\to 0}{\large\{}\frac{1}{nb}\,[(n-2)b+b-(n-1)b]{\large\}}=0 \, .
\label{b67}
\end{equation}
Thus, according to the Schultz solution, the free-energy density of
branches 6 and 7 vanishes in the thermodynamic limit $L \to \infty$. 

\subsection{Exact results for the universal parameters}
\subsubsection{Results for branch 1}
Although branch 1 can be mapped onto the critical Potts model, the
exact results for the temperature and magnetic scaling dimensions
of the Potts model do not apply to the completely packed system of
branch 1 and the associated dense O($n$) phase. Results for the
magnetic dimension  and the conformal anomaly of the dense O($n$)
phase have been obtained \cite{BB} from exact analysis of the model
on the honeycomb lattice. These results coincide with the Coulomb gas
results given below, and with an exact analysis of the
model on the square lattice \cite{BNW,3WBN}.

The Coulomb gas method, which offers a way to calculate some scaling 
dimensions, was explained in some detail in Ref.~\onlinecite{CG}. 
It considers an observable local density $p(r)$ on position $r$, 
which depends on the microstate at position $r$, and is conjugate
to the field $q$. In a critical state, we expect that the two-point
correlation behaves as $\langle p(0) p(r) \rangle \propto r^{-2X_q}$,
where the exponent $X_q$ is the scaling dimension of the density $p$.
Via the relation with the Coulomb gas one may now associate $p(0)$ with
a pair of charges, an electric charge $e_0$ and a magnetic one $m_0$.
Similarly we have a pair $e_r,m_r$ representing $p(r)$.
Then, the scaling dimension $X_q$ is 
given by \cite{CG} 
\begin{equation}
X_q= X(e,m) = - \frac{e_0 e_r}{2 g} - \frac{m_0 m_r g}{2} \, .
\label{Xem}
\end{equation}
The Coulomb gas coupling $g$ may be obtained if some exact information
about the universal properties is available. Its determination, as well
as that of the electric and magnetic charges, is a technical problem
that we leave aside. We shall copy their values from the literature,
and present only the result in terms of $X_q$ when needed.

For the critical O($n$) model, as well as for its analytic continuation
into the low-temperature O($n$) phase, it is well established how to
apply the Coulomb gas method Ref.~\onlinecite{CG}. In particular the
low-temperature O($n$) phase, which shares its universal properties
with the completely packed O($n$) loop model of branch 1, is important for
the present research.  The Coulomb gas results include the following
scaling dimensions of the critical O($n$) model and the
low-temperature phase
\begin{equation}
X_h =1-\frac{3g}{8}-\frac{1}{2g}\, ,~~~~~X_t=\frac{4}{ g}-2\, ,
\label{x_tx_h}
\end{equation}
where $X_t$ is the leading temperature dimension in the thermodynamics
of the critical O($n$) model. The parameter $g$, which is called the
Coulomb gas coupling constant, is a known function of $n$:
\begin{equation}
g = 1 \pm \frac{1}{\pi} \, \arccos\frac{n}{2} \, ,
\label{ccon}
\end{equation}
where the + sign applies to the critical O($n$) model and the $-$ sign
to the dense low-temperature phase, which applies to branch 1.
Furthermore, the introduction of the $x$- and $c$-type vertices into the
nonintersecting O($n$) loop model can be analyzed using the Coulomb 
gas \cite{dN,CG}. These perturbations are described by the cubic-crossover
exponent
\begin{equation}
X_c(g) = 1+\frac{3g}{2}-\frac{1}{2g} \, .
\label{xc}
\end{equation}
This perturbation is relevant in the dense phase, thus crossing bonds
and cubic vertices are expected to lead to different universal behavior
in the range $-2<n<2$.

Next we express the conformal anomaly $c_{\rm a}$ as a function of the
coupling constant $g$. From the definition of the parameter $y$ as a
function of $n$ in Ref.~\onlinecite{BCN}, one finds that it relates to
$g$ by $y=2-2g$ in our notation. Then, using Eqs.~(1) and (9) of
Ref.~\onlinecite{BCN}, one obtains the conformal anomaly as
\begin{equation}
c_{\rm a}(g) = 13 - 6g - \frac{6}{g} \, .
\label{cg}
\end{equation}

\subsubsection{Results for the other branches}
As far as we are aware, no exact results are available for the universal
parameters of branch 2 and 3 and equivalent models. However, as mentioned
in the preceding subsubsection, the cubic perturbation, i.e. the vertex
weight $c$, is expected to introduce new universal behavior for branch
2 and 3 with respect to branch 1. Numerical results \cite{onc} for the
dense phase (not completely packed) of the model with $z$- and $c$-type
vertices confirm this, and show the existence of a phase with a small
value of the magnetic dimension $X_h$, {\it i.e.}, a phase in which 
magnetic correlations persist over long distances.

The same Coulomb gas result applies to the introduction of crossing
bonds, which is, like the cubic perturbation, also described by the 
four-leg watermelon diagram, and one may thus expect new universal
behavior for branch 4.
A few results are available for a supersymmetric spin chain \cite{MNR}
related to branch 4, referred to as the Brauer model \cite{MN,dGN}.
Numerical as well as analytical arguments support, for $n \leq 2$,
the formula for the conformal anomaly 
\begin{equation}
c_{\rm a}(n) = n-1\, , \mbox{\hspace{20mm}} (n\leq 2) \, ,
\label{c4}
\end{equation}
and the magnetic dimension $X_h$ is reported to be very small, suggesting
anomalously slow decay of magnetic correlations, at least for $n=1$.
This behavior was confirmed, although with limited accuracy,
for a densely packed O($n$) model with crossing bonds, which is believed
to display similar universal behavior as branch 4 \cite{onc}. This model
was also studied by Jacobsen {\it et al.} \cite{jrs}, and recently,
correlation functions were obtained by Nahum {\it et al.} \cite{Serna}
for $n<2$, decaying as an inverse power of the logarithm of the distance.
As far as universal behavior is concerned, these findings for the
completely packed model apply as well in the dense O($n$) phase,
but not for the O($n$) transition to the high-temperature phase,
where the cubic perturbation is irrelevant for $|n|<2$. The  latter
point was numerically confirmed for the $n=0$ \cite{GBB} case which
describes intersecting trails.

\section{Transfer-matrix method}
\label{transfmat}
Consider a square lattice model, wrapped on the surface of a cylinder
with a circumference of $L$ lattice units.  The transfer-matrix method
is used for the calculation of the partition sum $Z$ of such systems.
The cylinder may be infinitely long but its circumference $L$ is finite.
We postpone the transfer-matrix construction to Appendix~\ref{tmtechnique}.
In this Section, we focus instead on the calculation of the free energy
and the universal quantities.

\subsection{Free energy and correlation lengths}
Using techniques described in Appendix~\ref{tmtechnique} and in
Ref.~\onlinecite{BN82}, we have computed a few of the leading eigenvalues
of the transfer matrix ${\bf T}$ for some relevant parameter choices.
We have restricted ourselves to eigenstates that are invariant under
rotations about the axis of the cylinder, and inversions.
It follows from Eq.~(\ref{Z-recursion}) that, in general, the reduced
free energy density for $M \to \infty$ is determined by the largest
eigenvalue $\Lambda_{0}(L)$ as
\begin{equation}
f(L)=L^{-1} \ln\Lambda_{0}(L) \, .
\label{flt}
\end{equation}
The transfer-matrix results for $f(L)$ can be used to estimate the
conformal anomaly $c_{\rm a}$ using the relation \cite{BCN,Affl}
\begin{equation}
f(L) \simeq f+ \frac{\pi c_{\rm a}}{ 6 L^{2} } \, .
\label{fcp}
\end{equation}
The subdominant eigenvalues $\Lambda_{k}(L)$ of ${\bf T}$ determine the
correlation
lengths $\xi_k$ belonging to the $k$-th correlation function. The gap
with respect to the largest eigenvalue determines the corresponding
correlation length along the cylinder as
\begin{equation}
\xi_{k}^{-1}(L) = \ln \frac{\Lambda_0 }{| \Lambda_k |} \, ,
\label{xkl}
\end{equation}
where it is usual to associate the label $k=1$ with the magnetic
correlation length $\xi_m$ and $k=2$ with the energy-energy correlation
length $\xi_{t}$.
For the purpose of numerical analysis, it is convenient
to define the corresponding scaled gaps $X_{k}(L)$ as
\begin{equation}
X_{k}(L)= \frac{L}{2\pi \xi_{k}(L)} \, ,
\label{x_k}
\end{equation}
In the presence of a temperature field $t$ and an irrelevant field $u$,
its scaling behavior is
\begin{equation}
X_k(t, u,L) \simeq X_k+ aL^{y_t} t+ bL^{y_u}u + \cdots \, ,
\label{xisce}
\end{equation}
where $X_k$ is the scaling dimension of the observable whose correlation
length is described by $\xi_k$ \cite{JCxi}. This formula provides a basis
to observe
the phase behavior as a function of a parameter, such as a vertex weight, 
that contributes to $t$. If $y_t>0$ and $u$ not too large, a set of
curves displaying $X_{k}(L)$ versus  that parameter for several values
of the system size $L$ will show intersections converging to the point
where the relevant scaling field $t$ vanishes, i.e., the point where a
phase transition occurs. According to Eq.~(\ref{xisce}), the slopes of
the $X_{k}(L)$ curves at the intersections increase with $L$ if $y_t>0$.
In the data analysis, we shall make use of this criterion for the relevance
of the scaling field $t$.

While the calculation of the temperature-like scaling dimension $X_2=X_t$
from $\Lambda_2$ is straightforward, that of the magnetic dimension $X_h$ 
needs some further comments.
Magnetic correlations between O($n$) spins are, in the equivalent O($n$)
loop model, represented by the insertion of a single loop segment between
these two points. In the present context of completely packed models, 
it is not possible to add another loop segment into the system, and we
use a method employed e.g., in Ref.~\onlinecite{onc}. It analyzes the 
difference between the leading eigenvalues of systems with odd system size
$L$ containing such a segment, and even systems without such a segment.
Thus, one may define scaled gaps using the average of two
consecutive even (or odd) systems  as
\begin{eqnarray}
X_{h}({\rm even}\;L)&=& \frac{L}{2\pi}\left[\ln{\Lambda_0(L)}-
\frac{1}{2}\left(\ln{|\Lambda_1(L-1)|}+
\ln{|\Lambda_1(L+1)|}\right)\right]\,,
\nonumber\\
{\rm or} \label{x_h}\\
X_{h}({\rm odd}\;L)&=& \frac{L}{2\pi}
\left[\frac{1}{2}\left(\ln{|\Lambda_0(L-1)|}+
\ln{|\Lambda_0(L+1)|}\right)-\ln{\Lambda_1(L)}\right]\,,
\nonumber
\end{eqnarray}
where $\Lambda_1$ denotes the largest eigenvalue of odd systems in the
transfer-matrix sector that includes odd connectivities.

\section{Numerical Results for Free Energy Density}
\label{numeranalf}
This section presents the finite-size analysis of the transfer-matrix
results for the free energy of the seven branches following from the
Schultz solutions \cite{S}, after transformation of the coloring model
into that of Eq.~(\ref{Zsql}) and (\ref{Zsqll}). The vertex weights
for these seven branches are listed in Table \ref{branches}.

\subsection{Branch 1}
Part of the numerical results for branch 1 has already appeared in
Ref.~\onlinecite{BWG}, together with an analytic derivation of the free
energy for $n<-2$. Here we summarize those results, and provide
some additional data.
The finite-size data for the free energy were extrapolated using
Eq.~(\ref{fcp}), thus yielding estimates of $f(n)$, which are listed in
Table \ref{fren}. For $n=-2$, the finite-size data for the free energy
did not obey Eq.~(\ref{fcp}), but were, up to numerical precision, 
precisely proportional to $1/L$.  Accordingly we quote the results
$f(-2)=0$ and, for the conformal anomaly, $c_{\rm a}=-\infty$.
For most values of $n$, these free energies agree satisfactorily with
the theoretical values given in Eqs.~(\ref{fgt2}) to (\ref{fltm2}).
Next, the free energies in Eq.~(\ref{fcp}) were fixed at their
theoretical values, in order to obtain improved estimates of the 
conformal anomaly. These results are also listed in Table \ref{fren},
and appear to agree well with the theoretical values, except for the
ranges where $|n|$ slightly exceeds 2, and where poor finite-size
convergence occurs.
\begin{table}
\vspace*{-2mm}
\caption {Fit results for the free
energy density and the conformal anomaly of the branch 1 model, compared
with the theoretical values. Estimated numerical
uncertainties in the last decimal place are given between parentheses.
The entries "0. (-)" indicate that the raw numerical data agree, up to
numerical precision, with a vanishing result.}
\label {fren}
\renewcommand{\arraystretch}{1.0}
\begin{tabular}{|c|c|c|c|c|}
\hline
$n$ & $f_{\rm exact}$&  $f_{\rm extr}$  &$c_{a,\rm exact}$&$c_{a,\rm extr}$ \\
\hline
$-20.$& 1.447952861454 & 1.4479528  (1)   &$  0        $&$0.000000   $ (1)\\
$-10.$& 1.052018311561 & 1.05202    (1)   &$  0        $&$0.001      $ (1)\\
$-4.0$& 0.456613026255 & 0.45       (5)   &$  0        $&$0          $ (50)\\
$-3.0$& 0.252039567005 & 0.26       (4)   &$  0        $&$20         $ (40)\\
$-2.0$&  0             & 0.005      (2)   &$-\infty    $&$-\infty    $ (-)\\
$-1.8$& 0.207751892795 & 0.2078     (3)   &$-29.6539374$&$-29.65     $ (2)\\
$-0.2$& 0.557322110937 & 0.557322   (1)   &$-2.62603787$&$-2.62604   $ (1)\\
 0.0  & 0.583121808062 & 0.582      (1)   &$-2         $&$-1.9998    $ (5)\\
 0.2  & 0.607404530379 & 0.60740453 (1)   &$-1.47195492$&$-1.47195500$ (5)\\
 0.4  & 0.630389998897 & 0.630389999 (5)  &$-1.02108633$&$-1.0210864$  (2)\\
 0.6  & 0.652252410906 & 0.652252411 (3)  &$-0.63239553$&$-0.6323956$  (2)\\
 0.8  & 0.673132748867 & 0.673132749 (2)  &$-0.29480810$&$-0.2948082$  (1)\\
 1.0  & 0.693147180560 & 0.69314718056 (1)&$  0        $&  $0.$        (-)\\
 1.2  & 0.712392984154 & 0.712392984 (1)  &$ 0.25834580$&  0.2583459   (1)\\
 1.4  & 0.730952859626 & 0.730952860 (2)  &$ 0.48499981$&  0.4850000   (1)\\
 1.6  & 0.748898172077 & 0.748898172 (2)  &$ 0.68341406$&  0.6834140   (1)\\
 1.8  & 0.766291499497 & 0.766291499 (2)  &$ 0.85560157$&  0.855610    (2)\\
 2.0  & 0.783188785414 & 0.78318875  (3)  &   1         &  1.002       (1)\\
 2.5  & 0.823597622499 & 0.823597    (1)  &   0         &  1.304       (?)\\
 3.0  & 0.861997334707 & 0.86205     (3)  &   0         &  1.6         (?)\\
 4.0  & 0.934112909108 & 0.9341      (1)  &   0         &  $-0.1$      (5)\\
 6.0  & 1.059762003273 & 1.05976     (1)  &   0         &  0.0000      (5)\\
 8.0  & 1.163519822868 & 1.1635195   (3)  &   0         &  0.00000     (3)\\
 10.  & 1.250668806419 & 1.25066880  (5)  &   0         &  0.00000     (1)\\
 15.  & 1.420503142656 & 1.4205031427 (2) &   0         &  0.000000    (1)\\
 20.  & 1.547785693447 & 1.5477856934 (1) &   0         &  0.00000000  (1)\\
\hline
\end{tabular}
\end{table}

\subsection{Branch 2 and 3}
The finite-size data for the free energy of branch 2 were fitted by
Eq.~(\ref{fcp}). Fits with two iteration steps, as described e.g. in
Ref.~\onlinecite{BN82}, were employed, using various combinations of exponents
that were left free, or fixed at expected integer values. A comparison   
between the different fits, and between fits using even and odd system
sizes, thus yielded error estimates. The best  estimates of $f(n)$ 
are listed in Table \ref{fren_b2b3}. 

\begin{table}
\caption {Numerical results for the free energy density of the branch 2
and 3 models, compared with the theoretical values. Fit results for the
conformal anomaly of branch 2 are also shown.  Estimated numerical
uncertainties in the last decimal places are given between parentheses.
The entries "0. (-)" indicate that the raw numerical data agree, up to
numerical precision, with a vanishing result.}
\label {fren_b2b3}
\renewcommand{\arraystretch}{0.9}
\begin{tabular}{|c|c|c|c||c|c|}
\hline
    \multirow{2}{*}{$n$} &\multicolumn{3}{c}{Branch 2}  \vline 
\hspace{-0.57mm}  \vline &
    \multicolumn{2}{c}{Branch 3} \vline\\ \cline{2-6}
$ $& $f_{\rm exact}$&$f_{\rm extr}$ &$c_{a,\rm extr}$&$f_{\rm exact}$&
$f_{\rm extr}$ \\
\hline
1.0&0           &0.        (-)&0.    (-)&0            &0.0      (-)\\
1.1&-	        &0.321623  (2)&0.00  (1)&-            &0.113    (2)\\
1.2&-	        &0.435512  (5)&0.0   (1)&-     	      &0.140    (2)\\
1.4&-	        &0.571672  (5)&0.3   (2)&-            &0.158    (3)\\
1.6&-	        &0.660925  (1)&0.4   (2)&-            &0.152    (5)\\
1.8&-	        &0.7284404 (5)&0.85  (2)&-            &0.126    (5)\\
1.9&-	        &0.7570799 (2)&0.930 (2)&-            &0.10     (1)\\
2.0&0.7831887854&0.7831888 (1)&1.002 (1)&0            &0.0      (-)\\
2.2&0.8294617947&0.8294618 (2)&1.12  (1)&0.0476594124 &0.04766  (2)\\
2.4&0.8696665810&0.8696665 (5)&1.2   (1)&0.0912111100 &0.09121  (2)\\
2.6&0.9052961420&0.905296  (1)&-        &0.1313749544 &0.13175  (5)\\
2.8&0.9373438162&0.937344  (2)&-        &0.1687122292 &0.16872  (1)\\
3.0&0.9665056811&0.966506  (2)&-        &0.2036655317 &0.20365  (2)\\
3.2&0.9932894055&0.993290  (3)&-        &0.2365893173 &0.23655  (5)\\
3.4&1.0180772108&1.018078  (2)&-        &0.2677686298 &0.2677   (1)\\
3.6&1.0411644293&1.041165  (2)&-        &0.2974314708 &0.2972   (2)\\
3.8&1.0627842309&1.062785  (2)&-        &0.3257593222 &0.3254   (2)\\
4.0&1.0831240913&1.083125  (1)&-        &0.3528968320 &0.3525   (2)\\
5.0&1.1701712128&1.170166  (1)&-        &0.4741476927 &0.473    (1)\\
10.&1.4360209233&1.4362    (3)&-        &0.8762747795 &0.88     (1)\\
15.&1.5944925888&1.5945    (3)&-        &1.1179474170 &1.118    (3)\\
20.&1.7097376056&1.7098    (3)&0.0   (5)&1.2884777913 &1.2885   (5)\\
25.&1.8009364248&1.80093   (5)&0.01  (5)&1.4195325895 &1.41954  (5)\\
30.&1.8766478021&1.87665  (2)&0.01   (2)&1.5256574392 &1.52567  (2)\\
50 &2.0943380303&2.094337 (1)&0.001  (2)&1.8180845623 &1.818084 (5)\\
100&2.4014692469&2.401692 (1)&0.000  (1)&2.2038009127 &2.203800 (2)\\
\hline
\end{tabular}
\end{table}
One observes that the bulk free energy for $n>2$ is in agreement with
the Schultz solution \cite{S}. Since branches 1 and 2 intersect at $n=2$,
we took the $n=2$ exact result for branch 1 in the second column of
Table \ref{fren_b2b3}. For $n=1$, branches 2 and 3 are connected and
the partition sum allows independent summation on the vertex states,
which yields a factor 1 per vertex. This yields the exact results
$f(1)=0$ and $c_{\rm a}=0$, also shown in Table \ref{fren_b2b3}.
For $n=2$, branch 3, the largest eigenvalue of the transfer matrix is
$2$ for all even system sizes, therefore the bulk free energy 
and the conformal anomaly also vanish in this case.

While the bulk free energy is well resolved in most cases, complications
arise for the part of branch 3 with small $n$. The free energy for small
system seems to converge well to a limiting value as
\begin{equation}
f(n)=\pm\sqrt{n-1}+\frac{1}{2}(n-1)\mp \frac{5}{3}(n-1)^{3/2}+\cdots \,,
\label{br23sn}  
\end{equation}
where the upper signs apply to branch 2 and the lower sign to branch 3.
This behavior agrees well with the numerical data for larger $n$ in
the case of branch 2, but not with those for branch 3. There appears
to be an eigenvalue crossing for branch 3, which, for $n=1.05$, occurs
at $L=10$, near the middle of the range of accessible system sizes.
The eigenvalue crossings shift to smaller $L$ for larger values of $n$.
Thus, we believe that Eq.~(\ref{br23sn}) does not apply to the bulk free 
energy of branch 3, not even for $n$ close to 1.  In addition to the
level crossing, the free-energy data display oscillations with a period
4 in the system size for branch 3.  For these reasons the free energies
of branch 3 could not be accurately determined in the interval $1<n<2$.

The conformal anomaly 
for branch 2 was estimated by least-squares fits on the basis
Eq.~(\ref{fcp}), with the finite-size exponent fixed at $-2$.
These fits do not show the type of fast convergence as that of branch 1.
Especially for $n<2$ we observe that strong crossover effects play
a role, so that the errors are difficult to estimate.
For $n>2$, there
exists a range of $n$ where the numerical results seem to suggest, as
indicated in the Table, a conformal anomaly $c_{\rm a}>1$, but iterated
fits display a diverging trend, which becomes progressively stronger with
increasing $n$. Thus, the entries for $c_{\rm a}$ in Table \ref{fren_b2b3}
for $2<n<2.6$ should not be taken too seriously. 
For larger $n$, the data  are no longer suggestive of convergence to
a value of $c_{\rm a}>0$.  Only for $n \gae 20$ do we observe the
exponential convergence of the free energy with $L$ that is expected
in a non-critical phase, corresponding with $c_{\rm a}=0$.

For branch 3, the finite-size data for $f$ are, remarkably, behaving
more like $f(L)=f(\infty)+a/L$, which does not suggest a finite 
conformal anomaly. For $n \gae 5$, the absolute value of the effective
exponent becomes significantly larger than 1, and tends to increase
with $L$, in accordance with the expected crossover to exponential
behavior, which is indeed seen for $L \gae 25$. Convergence is poor 
in the crossover range around $n=10$, and the extrapolated values of
the free energy are relatively inaccurate in that range.

An investigation how branches 2 and 3 are embedded in
the $c_n$ versus $n$ phase diagram will be reported in Sec.~\ref{embed}.

\subsection{Branch 4}
\label{freeb4}
The branch-4 system contains crossing bonds instead of the cubic
vertices considered in the preceding subsection. The largest eigenvalues
of the transfer matrix were computed for a number of values of the loop
weight $n$, for system sizes up to $L=16$.
The extrapolated values of the free-energy density for $n\le2$ agree 
accurately with the exact expression given by Eq.~(\ref{fbr4R}), as 
shown in  Table \ref{fren_b4}. That expression does however no longer
agree with the numerical results listed for the range $n>2$.

\begin{table}
\caption {Fit results for the bulk free energy density of branch 4 and
branch 5, compared with the theoretical values $f_{\rm R}$ given by
Rietman \cite{RPhD}.  Estimated numerical
uncertainties in the last decimal place are given between parentheses.
Error margins quoted as "(-)" indicate that the raw finite-size data
agree, with a numerical precision determined only by rounding errors,
with the listed result. This table is organized such as to display the
symmetry $f(n+x)=f(n-x)$ of the free energy. Results for the conformal
anomaly, estimated from the finite-size dependence of the free-energy
data, are also listed.
}
\label {fren_b4}
\renewcommand{\arraystretch}{1.0}
\begin{tabular}{|c|c|c|c||c|c|c|}
\hline
$n$ & $f_{\rm R}$&$f_{\rm extr}$&$c_{a,\rm extr}$& $n$&$f_{\rm extr}$&
$c_{a,\rm extr}$  \\
\hline
 2.0  &0.783188785414& 0.783190 (2)&1.002   (2)&2.0&0.783190 (2)&1.002 (2)\\
 1.6  &0.788072581927& 0.788072 (2)&0.66    (1)&2.4&0.788070 (4)&1.32  (2)\\
 1.2  &0.801609709369& 0.801610 (2)&0.22    (1)&2.8&0.801607 (5)&1.40  (1)\\
 1.0  &$\ln(9/4)$   &$\ln(9/4)$ (-)& 0      (-)&3.0&0.810929 (2)&1.50  (1)\\ 
 0.8  &0.821583452525& 0.82158  (1)&$-0.22$ (2)&3.2&0.821583 (1)&1.60  (1)\\
 0.6  &0.833330017842& 0.83333  (1)&$-0.44$ (3)&3.4&0.833329 (1)&1.70  (1)\\
 0.4  &0.845964458726& 0.84596  (1)&$-0.65$ (5)&3.6&0.845964 (1)&1.80  (1)\\
 0.2  &0.859313113225& 0.85931  (1)&$-0.92$ (8)&3.8&0.859313 (1)&1.90  (1)\\
 0.0  &0.873230390267& 0.87323  (1)&$-1.1$  (1)&4.0&0.873230 (1)&2.003 (5)\\
$-0.2$&0.887594745620& 0.88760  (1)&$-1.3$  (2)&4.2&0.887594 (1)&2.103 (5)\\
$-0.4$&0.902304904172& 0.90230  (1)&$-1.6$  (2)&4.4&0.902304 (1)&2.202 (5)\\
$-0.6$&0.917276530696& 0.91727  (1)&$-1.8$  (2)&4.6&0.917276 (1)&2.302 (5)\\
$-0.8$&0.932439389367& 0.93243  (2)&$-2.0$  (3)&4.8&0.932439 (1)&2.402 (5)\\
$-1.0$&0.947734962298& 0.94773  (1)&$-2.2$  (3)&5.0&0.947735 (1)&2.502 (5)\\
$-1.2$&0.963114471587& 0.96311  (1)&$-2.5$  (3)&5.2&0.963114 (1)&2.603 (5)\\
$-1.6$&0.993969362786& 0.99397  (1)&$-2.9$  (4)&5.6&0.993969 (1)&2.802 (5)\\
$-2.0$&1.024753260684& 1.0247   (1)&$-3.4$  (5)&6.0&1.024752 (1)&3.002 (5)\\
$-2.5$&1.062873680798& 1.0629   (1)&$-4 $   (1)&6.5&1.06287  (1)&3.25  (1)\\ 
$-3.0$&1.100390077368& 1.1004   (2)&$-5 $   (1)&7.0&1.10038  (1)&3.51  (1)\\
$-4.0$&1.173116860698& 1.1731   (5)&$-6 $   (2)&8.0&1.17311  (2)&4.00  (2)\\ 
$-6.0$&1.308199777002& 1.308    (3)&$-10$   (2)&10.0&1.3082  (2)&5.0   (1)\\ 
$-8.0$&1.429801657071& 1.43     (3)&-          &12.0&1.4298  (2)&6.0   (2)\\
\hline
\end{tabular}
\end{table}
However, the free-energy data listed in Table \ref{fren_b4} 
accurately display a symmetry with respect to the point $n=2$.
It is thus straightforward to conjecture an exact expression for the
free-energy density along branch 4 for all $n$, by
replacing $n-2$ with $|n-2|$ in  Eq.~(\ref{fbr4R}):
\belowdisplayskip=1mm
\begin{displaymath}
 f(n)  =\ln\left(\frac{|n-2|}{4}\right)+
 2\ln\Gamma\left(\frac{1}{4}\right)-2\ln\Gamma\left(\frac{3}{4}\right)
\hspace{40mm}
\end{displaymath}
\begin{equation}
\hspace*{40mm}
-2\ln\Gamma\left(\frac{1}{4}+\frac{1}{|n-2|}\right)+
 2\ln\Gamma\left(\frac{3}{4}+\frac{1}{|n-2|}\right) \label{Rm_ge} \, .
\end{equation}
Whereas the bulk free energy displays a clear symmetry with respect to
the point $n=2$, this is not the case for the finite-size results for the
free energy. Accordingly, the estimated values of the conformal anomaly,
also included in Table \ref{fren_b4}, do not obey the symmetry.
These estimates of $c_{\rm a}$ were obtained by fits according to
Eq.~(\ref{fcp}), with  the bulk free energy fixed according to
Eq.~(\ref{Rm_ge}). Our confidence in this procedure is based on the
degree of accuracy found above for the agreement between the extrapolated
values of the free energy and Eq.~(\ref{Rm_ge}).

In the range $n\leq 2$, the results for the conformal anomaly are 
suggestive of behavior according to Eq.~(\ref{c4}).
While the finite-size dependence of the estimates of $c_{\rm a}$ is
quite small, their apparent convergence is very slow in this range.
This makes it difficult to estimate the error margins, so that our
new evidence supporting Eq.~(\ref{c4}) may not be considered as
very convincing.
The fits for $c_{\rm a}$ in the range $n\geq 2$ are better behaved,
and the numerical results in Table \ref{fren_b4} allow the conjecture
\begin{equation}
c_{\rm a}=n/2 \mbox{\hspace{20mm}} (n \geq 2) \, .
\label{fb4ng2}
\end{equation}
This type of behavior is already strongly suggested by first estimates
of $c_{\rm a}$ as $6 \pi^{-1} L^2[f(L)-f(\infty)]$. Such estimates 
in the range $3\leq n \leq 8$ differ less than $10^{-2}$ from $n/2$
for $L=16$.  However, apparent convergence is slow, and we were unable
to reduce the estimated uncertainty margins much below the $10^{-2}$ 
level by means of iterated fits.

\subsection{Branch 5}
The definition of branch 5 specifies that both crossing bonds and cubic
vertices occur in addition to the original O($n$)-type vertices.
In the representation of the coloring model, the vertex weights of
branch 5 according to Eq.~(\ref{map}) are equal to those for branch 4,
except for a change of sign of the weight $W^0$ describing a color
crossing. In an infinite system, such crossings occur in pairs, so that
the free-energy density for branch 5 must be equal to that for branch 4.
This may be expected to hold also for finite systems with an even system
size, which is confirmed by our transfer-matrix results for the largest
eigenvalue, at least for $n>-2$. A level crossing occurs at $n=-2$, and
for $n<-2$ the largest eigenvalue of a system with a size $L$ divisible
by 4 has an eigenvector that is antisymmetric under translations. Such 
eigenvalues do not contribute to the free energy of a translationally
invariant system.  For systems with a size equal to an odd multiple of
2, the largest eigenvalues of branch 4 coincide with those of branch 5,
also for $n<-2$.

Another point of interest is that, in the
representation of the generalized loop model, the size of the transfer
matrix for the branch-5 model is larger than that for branch 4, due to
the  larger number of connectivities in the presence of cubic vertices.
The larger vector space leads to the occurrence of additional eigenvalues,
so that we may expect additional scaling dimensions for branch 5.

After verification that the leading transfer-matrix eigenvalues for
branches 4 and 5 coincide, there is no reason for a separate analysis
of the free energy of branch 5 besides that of branch 4.

\subsection{Branch 6 and 7}
The transfer-matrix results for the free-energy density  $f(n,L)$ of
finite branch-6 systems are found to behave precisely as
\begin{equation}
f(n,L)=\frac{\ln(n)}{L} \, ,
\label{fbr6eo}
\end{equation}
and those for branch 7 as
\begin{equation}
f(n,L)=
\begin{cases}
L^{-1}  \ln(n)   \text { ~~~~~~~~ ($L$ even)}\\
L^{-1}  \ln(n-2) \text { ~~~~($L$ odd) .}\\
\end{cases}
\label{fbr7eo}
\end{equation}
The apparent simplicity of these results is due to the conservation of
colors along lines of vertices, or the absence of $z$-type vertices for
branches 6 and 7. This condition is imposed by the symmetry requirement
Eq.~(\ref{lrsym}).  For branch 6, there are only $x$-type vertices,
and every layer of vertices trivially contributes a weight $n$ for a 
loop closing around the cylinder, thus explaining Eq.~(\ref{fbr6eo}).
The coloring-model parameters
of branch 7 are $W^0=1$, $W^{\rm d}=-1$. The leading eigenvalue of the
transfer matrix occurs in the sector in which the colors on the lines
parallel to the axis of the cylinder are the same. Summation on the $n$
colors of a newly added layer thus contributes $n-1+(-1)^L$, which
yields $n$ for even systems and $n-2$ for odd ones, in agreement with
Eq.~(\ref{fbr7eo}).

These results imply that the bulk free energy vanishes for branch 6 and
7. This agrees with the Schultz solution which, in the symmetric case
described by Eq.~(\ref{lrsym}), becomes trivial  as expressed by
Eq.~(\ref{b67}).

\section{Evaluation of scaling dimensions}
\label{numericanx}
In view of the trivial nature of branches 6 and 7 described in the
preceding subsection, branches 6 and 7 do not require further analysis.
This section will therefore focus on the transfer-matrix results for the
scaling dimensions of branches 1 to 5.

\subsection{Branch 1}
\begin{figure}[h]
\begin{center}
\includegraphics[scale=1.0]{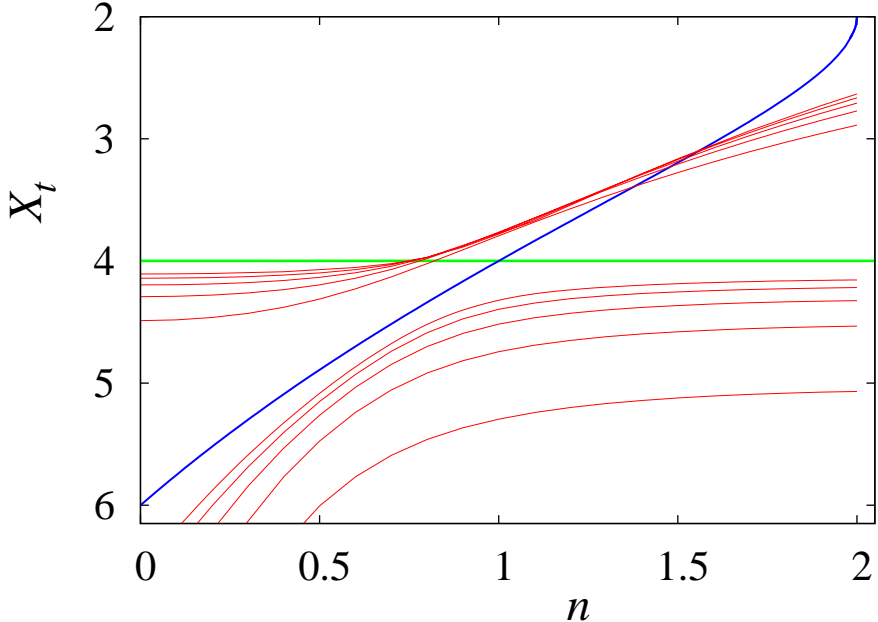}
\end{center}
\caption{(Color online) The two leading thermal scaled gaps of the
branch-1 model versus loop weight $n$, for even system sizes $L=8$ to 16.
The scaled gaps are shown as thin lines, smoothly connecting a series
of data points. The scaled gaps increase with $L$ in most of the
range of $n$.  Also shown
are two thicker lines, of which one represents a constant scaling
dimension $X=4$, and the other the Coulomb gas result for $X_t$.
Extrapolation 
of the finite-size data indicates that the leading gap (upper set of
curves) converges to $X_{t}=4$ for $n<1$ and to the Coulomb gas result
for $X_t$ for $n>1$. The second gap behaves similarly but with the
intervals of $n$ interchanged.
}
\label{fig:sgaps}
\end{figure}

The extrapolated results for the temperature dimension $X_t$, and those
for the magnetic dimension $X_h$ are, together with the exact Coulomb gas
predictions \cite{CG}, listed in Table \ref{scad} for several values of
the loop weight $n$. These results supplement earlier data for the
temperature dimension listed in Ref.~\onlinecite{BWG}, and data for the dense
(not completely packed) phase of the O($n$) model~\cite{BN}, which is
related by universality. For $n \geq 1$ the extrapolated transfer-matrix
results for the leading temperature-like dimension agree with the 
Coulomb gas result for $X_t$, but this is no longer the case for $n<1$, 
where the extrapolations seem to converge to the exact value 4.
The Coulomb gas values for $X_t$ are omitted in most of the range $n<1$,
where they no longer match the numerical results. Calculations of the
three leading eigenvalues whose eigenvectors satisfy the translational
and inversion symmetries of the lattice, indicate that the scaling dimension
$X_t=4$ still exist for $n>1$, as well as that predicted by the Coulomb gas
theory in the range $n<1$. This is illustrated in Fig.~\ref{fig:sgaps},
which shows the two leading temperature-like scaled gaps
for system sizes $L=8$, 10, 12, 14 and 16. 

The data for $X_{h}$ in Table \ref{scad} agree well with the Coulomb gas 
results, except near $|n|=2$. Poor finite-size convergence occurs near $n=2$,
and for $n=-2$, the whole eigenvalue spectrum of finite systems collapses
to $|\Lambda_i|=2$, which would correspond with $X_t=X_h=0$. But one may
expect that the result for these scaling dimensions will be different if
the order of the limits $n\to -2$ and $L \to \infty$ is reversed.

Our numerical data for $|n|\ge 10$ show a divergent behavior of the 
scaled gaps, in agreement with the expected absence of criticality for
large $|n|$. Extrapolations in the ranges $|n| \gae 2$ (not shown in
Table \ref{scad}), while unsatisfactory in accuracy, are consistent with 
the presence of a marginally relevant operator at $|n| = 2$. The ranges
$|n|> 2$ of branch 1 have earlier been identified \cite{BWG} as lines 
of phase coexistence separating two lattice-gas-like ordered phases.
The associated vanishing scaling dimension corresponds with an
eigenvector that is not invariant under lattice translations.

\begin{table}
\caption {Fit results for the temperature dimension $X_t$ and the magnetic
dimension $X_h$ of the branch 1 model,
compared with the theoretical values. Estimated numerical
uncertainties in the last decimal place are given between parentheses.
The entries "0. (-)" indicate that the raw numerical data for finite
systems  agree, up to numerical precision, with a vanishing result.}
\label {scad}
\vspace{-5mm}
\renewcommand{\arraystretch}{1.0} 
\begin{center}
\begin{tabular}{|c|c|c|c|c|}
\hline
$n$&$X_{t,\rm extr}$&$X_{t,\rm exact}$&$X_{h,\rm extr}$&$X_{h,\rm exact}$\\
\hline
$-2.0$& 0        (-)& --              &  0          (-)&$-\infty$           \\
$-1.8$& 4.1      (1)& --              &$-2.5360$    (5)&$-2.53654900\cdots$ \\
$-1.6$& 4.01     (1)& --              &$-1.51782$   (2)&$-1.51782801\cdots$ \\
$-1.4$& 4.002    (2)& --              &$-1.06797$   (1)&$-1.06979745\cdots$ \\
$-1.2$& 4.000    (1)& --              &$-0.804642$  (1)&$-0.80464267\cdots$ \\
$-1.0$& 4.0000   (1)& --              &$-0.6250000$ (1)&$-0.62500000\cdots$ \\
$-0.8$& 4.0000   (1)& --              &$-0.4933552$ (2)&$-0.49335519\cdots$ \\
$-0.6$& 4.0000   (1)& --              &$-0.3917838$ (1)&$-0.39178379\cdots$ \\
$-0.4$& 4.00000  (1)& --              &$-0.3105015$ (1)&$-0.31050153\cdots$ \\
$-0.2$& 4.000000 (2)& --              &$-0.2436554$ (1)&$-0.24365536\cdots$ \\
  0.0 & 4.000001 (1)& --              &$-0.1875000$ (1)&$-0.18750000\cdots$ \\
  0.2 & 4.000000 (1)& --              &$-0.1395107$ (1)&$-0.13951071\cdots$ \\
  0.4 & 4.00001  (2)& 5.0910$\cdots$  &$-0.0979121$ (1)&$-0.09791208\cdots$ \\
  0.6 & 3.99999  (2)& 4.7003$\cdots$  &$-0.0614096$ (1)&$-0.06140963\cdots$ \\
  0.8 & 4.000    (1)& 4.3392$\cdots$  &$-0.0290269$ (1)&$-0.02902694\cdots$ \\
  1.0 & 4.000    (2)& 4.0000$\cdots$  &  0          (-)& 0.00000000$\cdots$ \\
  1.2 & 3.68     (2)& 3.6751$\cdots$  &  0.0262995  (1)& 0.02629958$\cdots$ \\
  1.4 & 3.357    (2)& 3.3561$\cdots$  &  0.0504353  (1)& 0.05043540$\cdots$ \\
  1.6 & 3.029    (2)& 3.0304$\cdots$  &  0.073015   (1)& 0.07301374$\cdots$ \\
  1.8 & 2.67     (1)& 2.6705$\cdots$  &  0.095032   (5)& 0.09502101$\cdots$ \\
  2.0 & 2.1      (1)& 2.0000$\cdots$  &  0.122      (1)& 0.12500000$\cdots$ \\
\hline
\end{tabular}
\end{center}
\end{table}

\subsection{Branch 2 and 3}
We followed a similar procedure in order to obtain the scaling dimensions
$X_t$ and $X_h$ as for branch 1. The extrapolated results are shown in
Table \ref{scad23}. The entries for $X_t$ at $n=1$ are shown to indicate
that the temperature-like energy gaps of finite systems diverge for
$n \to 1$. However, this is due to another eigenvalue of the transfer
matrix that obscures the true scaling behavior for small system sizes.
If one would first take the limit $L \to \infty$,
and then the limit $n \to 1$, a result $X_t \approx 2$ is expected.
For $n=1$, the finite-size results for the scaled magnetic gaps vanish,
and the corresponding entry $X_h=0$  is in line with the entries for
branch 2 with $n > 1$. 

\begin{table}
\caption {Numerical results for the temperature dimension $X_t$ and the
magnetic dimension $X_h$ of the branch-2 and branch-3 models.
Estimated numerical
uncertainties in the last decimal place are given between parentheses.}
\label {scad23}
\renewcommand{\arraystretch}{1.5}
\begin{center}
\begin{tabular}{|c|c|c|c|}
\hline
$n$&$X_t({\rm branch~2})$&$X_h({\rm branch~2})$&$X_t({\rm branch~3})$\\
\hline
  1.0  & $\infty$ & 0 (-)   & $\infty$  \\ 
  1.2  & 2.0  (?) & 0.00 (2)&  0.8  (1) \\ 
  1.4  & 1.9  (1) & 0.04 (2)&  1.0  (2) \\ 
  1.6  & 1.90 (5) & 0.08 (1)&  1.4  (?) \\ 
  1.8  & 2.00 (5) &0.105 (5)&  1.6  (2) \\ 
  2.0  & 1.9  (1) &0.122 (2)&  0    (-) \\ 

  10   &$-$0.2 (2)& 0.0  (1)&$-$0.1 (3) \\ 
  20   & 0.0  (1) & 0.0  (2)&  0.0  (1) \\ 
  30   & 0.00 (1) &0.000 (2)&  0.00 (1) \\ 
\hline
\end{tabular}
\end{center}
\end{table}
For branch 2, a range $n\gae 2$ exists where the scaled temperature-like
gaps decrease slowly with increasing $L$, but power-law fits in the
range of accessible values of $L$ do not suggest convergence. Only at
much larger values of $n$ does it become clear (see Table \ref{scad23})
that crossover occurs to a fixed point with a vanishing $X_t$.

A similar result is found for $X_t$ on branch 3 at large $n$. But for
$n\to 2$ the behavior is different and the thermal scaled gaps of finite
systems vanish in this limit.

The finite-size data for $X_h$ on branch 2 with $n<1.5$ could not be
satisfactorily fitted with a power law. The assumption that
$X_h(L) \simeq X_h+a/\ln L$ gave somewhat better behaved results, 
but the errors are hard to estimate. In Table \ref{scad23} we base the
error estimates on the differences between the above logarithmic fits
and fits with a fixed power $-1$. Also for $n=2$ we used logarithmic
fits, which yielded a best estimate not far from the exact value
$X_h=1/8$. 
In the case of branch 3, the free-energy data appear to oscillate not
only between even and odd systems, but there is also a period four, and
we are unable to produce any meaningful estimates of $X_h$.

\subsection{Branch 4 and 5}
As noted in Sec.~\ref{numeranalf}, the leading eigenvalues of the
transfer matrices of finite branch-4 and branch-5 systems are equal
for $n>-2$.
However, this does not hold for the rest of the eigenvalue spectra,
and we perform separate analyses for the two branches.
Unfortunately the convergence of the scaled gaps is very poor, and we
are unable to find accurate results. Power-law fits tend to yield
finite-size exponents that vary considerably with system size, often
assuming positive values. Logarithmic fits $X_t(L)\simeq X_t+a/\ln L$
were not very satisfactory either, because the finite-size data display
an extremum as a function of the finite size for some values of $n$.
Under these circumstances, we take the branch-4 scaled gaps at system 
size $L=16$ as our final estimates. They are shown in Table \ref{scad45}.
The difference with the result of the logarithmic fit, or 10 times the
difference between the $L=14$ and 16 results, is quoted as a rough 
estimate of the error margin. 

\begin{table}
\caption {Fit results for the temperature dimension $X_t$ and the magnetic
dimension $X_h$ of the branch 4 and branch 5 models. Estimated numerical
uncertainties in the last decimal place are given between parentheses.
The entries "0. (-)" indicate that the raw numerical data agree, up to
numerical precision, with a vanishing result.}
\label {scad45}
\renewcommand{\arraystretch}{0.9}
\begin{center}
\begin{tabular}{|c|c|c||c|c|}
\hline
$n$ &$X_t({\rm branch 4})$ &$X_h({\rm branch 4})$ &
$X_t({\rm branch 5})$ &$X_h({\rm branch 5})$\\
\hline
$-8.0$ &2.3  (5) &  --        & --      &--          \\
$-4.0$ &2.3  (2) &$-0.12$ (5) &--       &$-0.8$  (2) \\
$-2.0$ &2.3  (2) &$-0.09$ (5) &0    (-) &$-0.47$ (5) \\
$-1.8$ &2.3  (2) &$-0.08$ (5) &0.0  (1) &$-0.43$ (5) \\
$-1.6$ &2.3  (2) &$-0.08$ (5) &0.0  (1) &$-0.40$ (5) \\
$-1.4$ &2.3  (2) &$-0.08$ (4) &0.0  (1) &$-0.37$ (5) \\
$-1.2$ &2.3  (2) &$-0.07$ (5) &0.0  (1) &$-0.33$ (5) \\
$-1.0$ &2.3  (2) &$-0.07$ (5) &0.0  (1) &$-0.30$ (5) \\
$-0.8$ &2.3  (2) &$-0.06$ (5) &0.1  (1) &$-0.27$ (5) \\
$-0.6$ &2.2  (2) &$-0.06$ (4) &0.1  (2) &$-0.24$ (4) \\
$-0.4$ &2.2  (1) &$-0.05$ (4) &0.1  (2) &$-0.21$ (4) \\
$-0.2$ &2.2  (1) &$-0.05$ (3) &0.1  (2) &$-0.17$ (4) \\
  0.0  &2.2  (1) &$-0.04$ (3) &0.1  (3) &--          \\
  0.2  &2.2  (1) &$-0.03$ (3) &0.2  (3) &$-0.11$ (5) \\
  0.4  &2.2  (1) &$-0.03$ (3) &0.3  (3) &$-0.08$ (2) \\
  0.6  &2.2  (1) &$-0.02$ (2) &0.3  (3) &$-0.05$ (2) \\
  0.8  &2.2  (2) &$-0.01$ (2) &0.4  (3) &$-0.02$ (2) \\
  1.0  &2.1  (2) &0. (-)      &0.5  (3) &0       (-) \\
  1.2  &2.0  (3) &0.013   (3) &0.7  (3) &0.02    (2) \\
  1.4  &2.0  (3) &0.03    (4) &0.9  (2) &0.054   (2) \\
  1.6  &1.9  (3) &0.05    (3) &1.1  (2) &0.07    (2) \\
  1.8  &1.8  (3) &0.07    (3) &1.5  (2) &0.09    (2) \\
  2.0  &1.7  (2) &0.11    (2) &1.9  (2) &0.11    (2) \\
  4.0  &1.4  (4) &0.5     (1) &1.3  (2) &0.125   (3) \\
  8.0  &1.5  (5) &--          &1.4  (2) &0.12    (4) \\
\hline
\end{tabular}
\end{center}
\end{table}

A similarly slow convergence is observed for the branch-5 scaled gaps.
For $n<-2$ we have the additional problem that the largest eigenvalues
display a finite-size dependence not only with an odd-even alternation,      
but also with an effect of period 4.       
But  some observations can still be made: for large
negative $n$ the scaled gaps tend to become very small, and for $n$
closer to $-2$ they are at most a few tenths, and tend to decrease with
increasing $L$. For $n=-2$ the largest eigenvalues become degenerate,
which corresponds with $X_t=0$. The final estimates of $X_t$ shown in
Table \ref{scad45} for $n>-2$ are taken from logarithmic fits, and the
error estimates are taken as their differences with the scaled gaps at
$L=14$. The entry at $n=0$ is obtained from interpolation between small
negative and positive values of $n$, because the vertex weight $c/n$ in
Eq.~(\ref{Zsqll}) diverges at $n=0$.
Similar numerical problems appear during analysis of the magnetic gaps
as defined in Eq.~(\ref{x_h}). Thus also the results  for $X_h$ in
Table \ref{scad45}, and their error estimates, are somewhat uncertain.

\newpage
\section{Location of phase transitions}
\label{embed}
In order to explore the physical properties of the seven branches of
solvable models described in Table \ref{branches}, we performed some 
further numerical work. Without aiming at a complete coverage of the
phase diagram, we wish to investigate the possible association  of the
solvable branches with lines of phase transitions, or the location of
these branches with respect to such phase transitions.

For this purpose, we have calculated finite-size data for the scaled
temperature-like gap, using Eq.~(\ref{x_k}), and for the magnetic gap
using Eq.~(\ref{x_h}), along lines in the phase diagram that intersect 
with the branches of interest.

\subsection{Branch 1}
The completely packed nonintersecting O($n$) loop model with $|n|<2$
on the square lattice belongs to the same
universality classes as the dense phase of the  O($n$) model.
For the latter model, the introduction of crossing bonds, as well as
that of cubic vertices, leads to crossover to different universal behavior.
Both of these perturbations are described by the cubic-crossover exponent
given by Eq.~(\ref{xc}), which is relevant in the dense O($n$) phase.
Thus branch 1 is a locus of phase transitions in the $(n,x,c_n)$ 
parameter space, at least for $|n|<2$. This was already illustrated  for the
dense O($n$) phase by transfer-matrix calculations in Ref.~\onlinecite{onc}.
For the present completely packed case, a few instances of the effect of a
variation of the weights of the cubic and crossing-bond vertices on branch
1 will be included in the following subsections treating branches 2-5.

\subsection{Branch 2 and 3}
\begin{figure}
\hspace*{-15mm}
\includegraphics[scale=0.75]{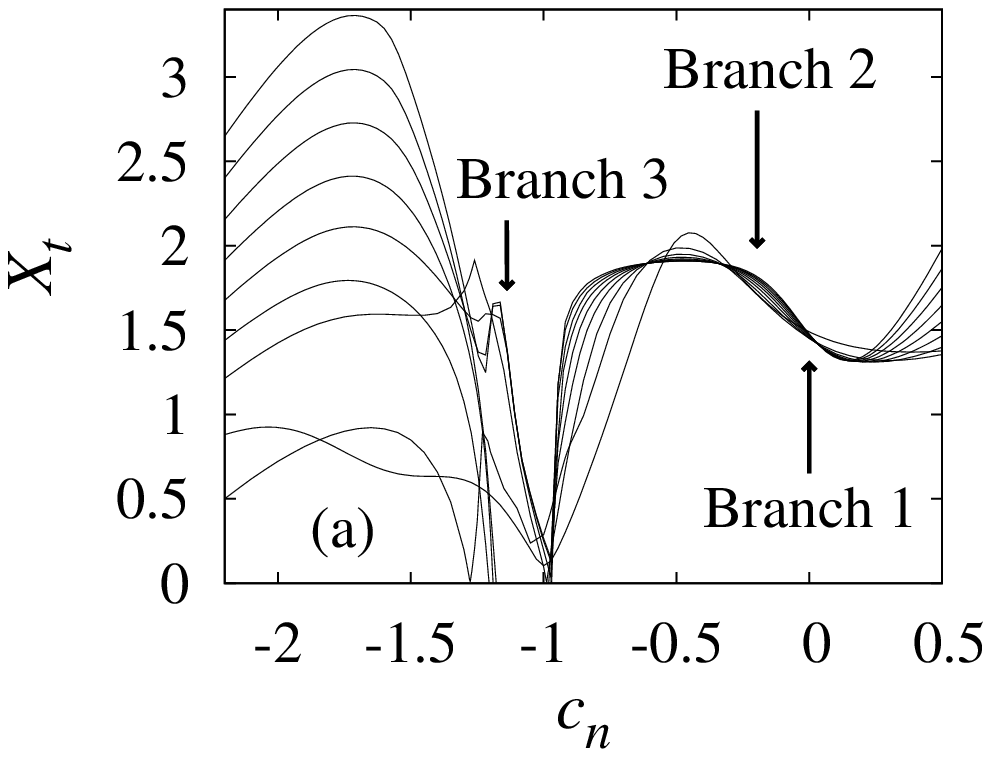}
\hspace*{-20mm}
\includegraphics[scale=0.75]{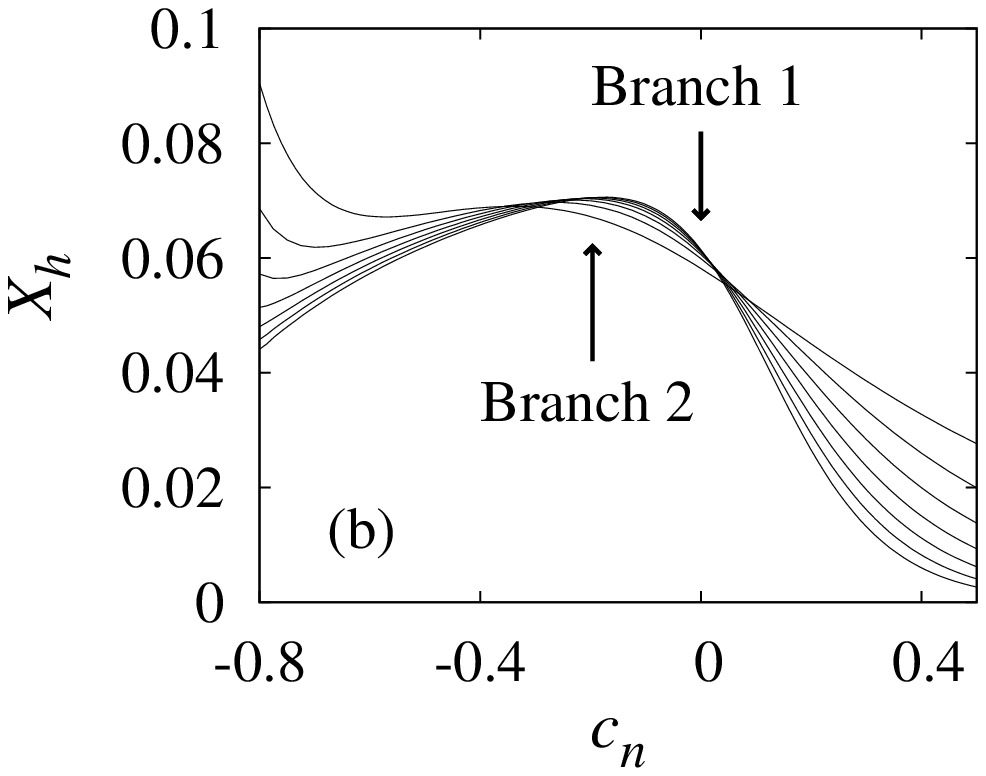}
\caption{Scaled thermal (a) and magnetic (b) gaps versus $c_n$ covering
branches 1, 2 and 3 of the completely packed O($n$) loop model with $n=1.5$.
Results are shown for even system sizes $L=4$ to 20 for the thermal case,
and for $L=4$ to $18$ for the magnetic case. In figure (a) the scaled
thermal gaps increase with $L$, both on the left and the right side of 
the scale. Instead, in figure (b) the scaled magnetic gaps decrease on both
sides. The data for $X_t$ display cusps near $c_n \approx -1.2$, which are
due to intersections between transfer-matrix eigenvalues.
Complex pairs of eigenvalues then appear in a range of $c_n$ for system sizes
equal to odd multiples of 2. The corresponding data for these ranges are
not shown in this figure.
}
\label{fig:qn1_5_t}
\end{figure}
\begin{figure}
\hspace*{-15mm}
\includegraphics[scale=0.75]{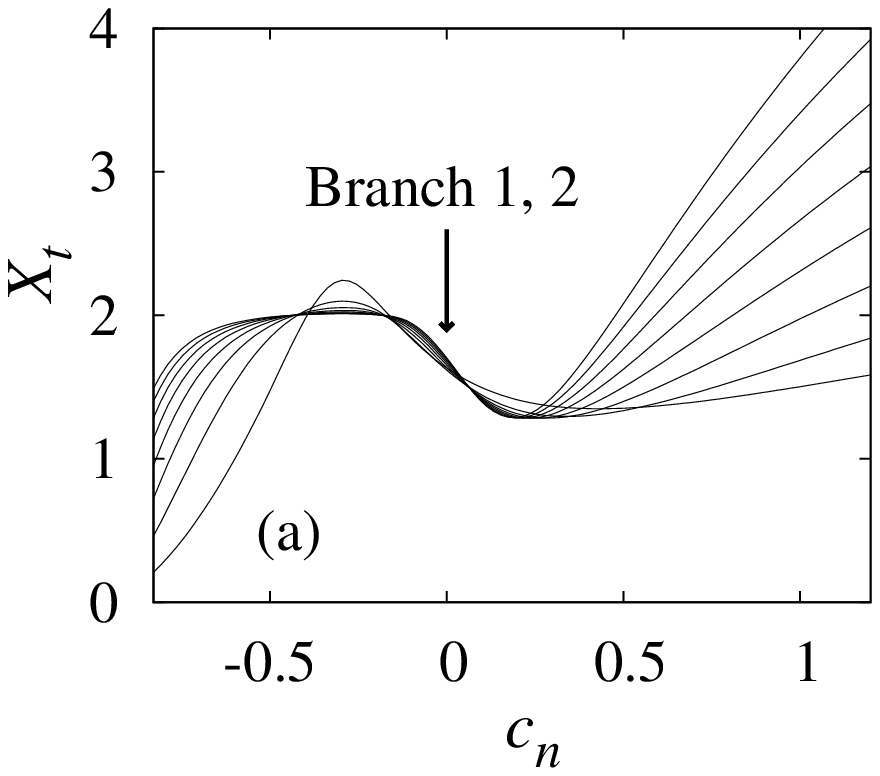}
\hspace*{-20mm}
\includegraphics[scale=0.75]{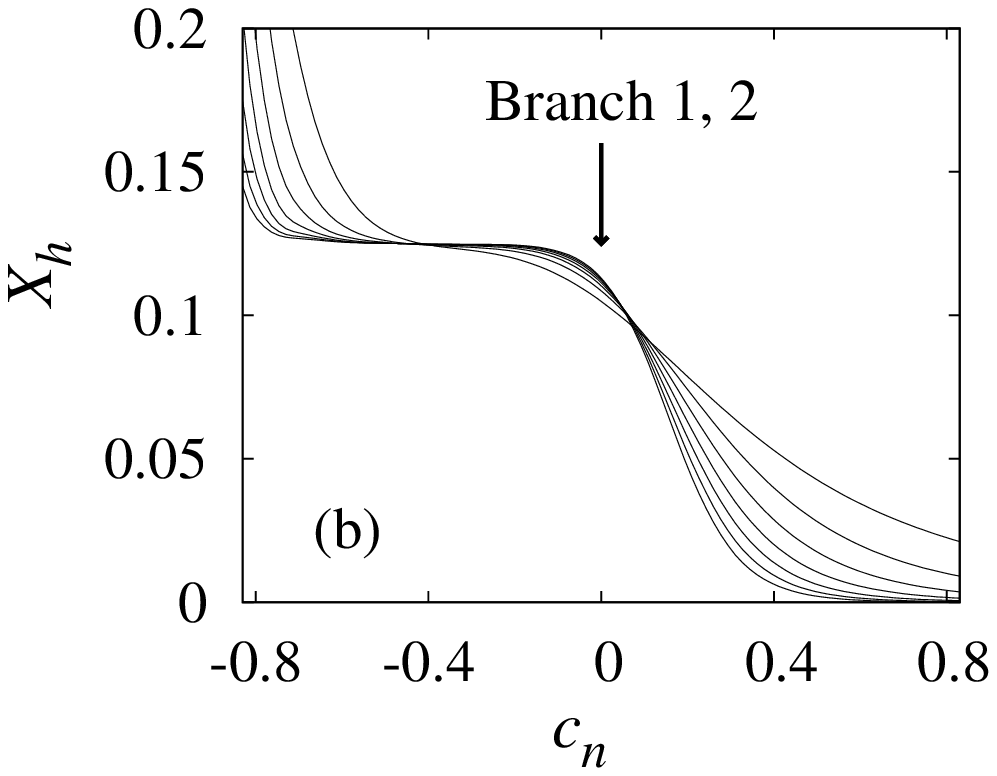}
\caption{Scaled thermal (a) and magnetic (b) gaps versus $c_n$ covering 
branches 1 and 2 of the completely packed O($n$) loop model with $n=2$.
Results are shown for even system sizes $L=4$ to $18$ for the thermal case,
and for $L=4$ to $16$ for the magnetic case. In figure (a) the scaled
thermal gaps increase with $L$ near $c_n=1$, while the scaled magnetic gaps 
instead decrease on the right-hand end of the scale.
}
\label{fig:xb2n2}
\end{figure}

\begin{figure}
\hspace*{-15mm}
\includegraphics[scale=0.75]{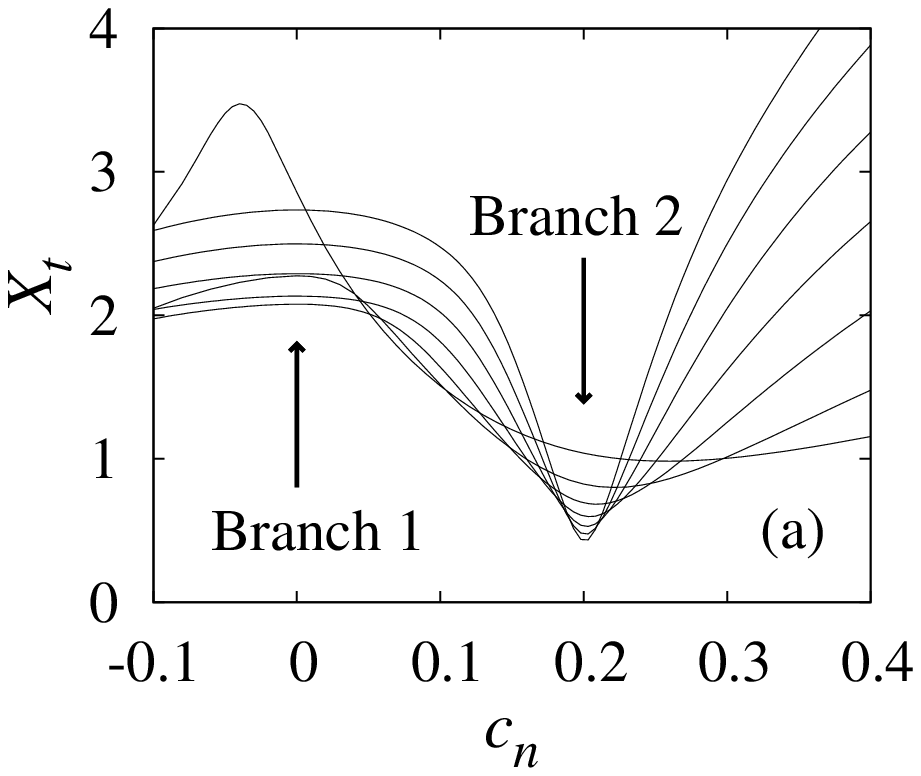}
\hspace*{-20mm}
\includegraphics[scale=0.75]{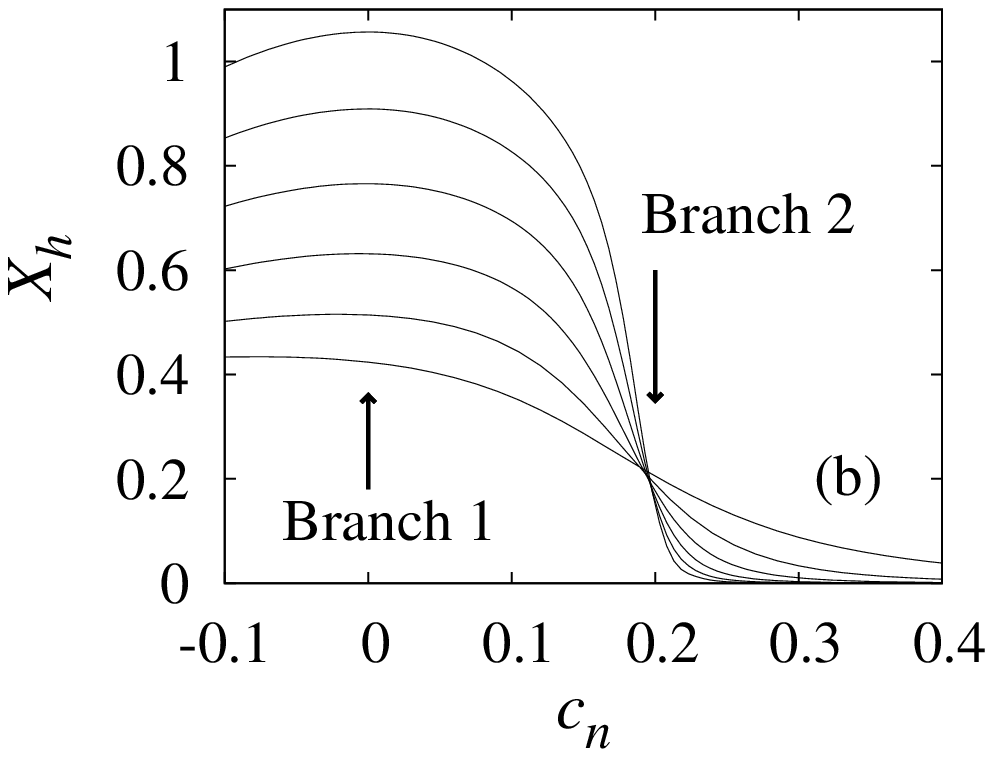}

\caption{Scaled thermal (a) and magnetic (b) gaps versus $c_n=c/n$ of
the cubic-O($n$) loop model with $n=10$.
The scaled gaps are represented by means of smooth curves connecting the
numerical data points. Different system sizes correspond with separate
curves.  The thermal gaps are shown for even system sizes $L=4$ to 16,
and the magnetic ones for $L=4$ to 14. 
The correspondence is such that, in the neighborhood of branch 2,
 steeper curves belong to larger $L$.
The intersections, as well as the data taken at branch 2, seem to
converge to a vanishing value of the scaled gaps, which indicates that a
first-order transition occurs at branch 2. }
\label{fig:qn10b2}
\end{figure}

\begin{figure}
\hspace*{-15mm}
\includegraphics[scale=0.75]{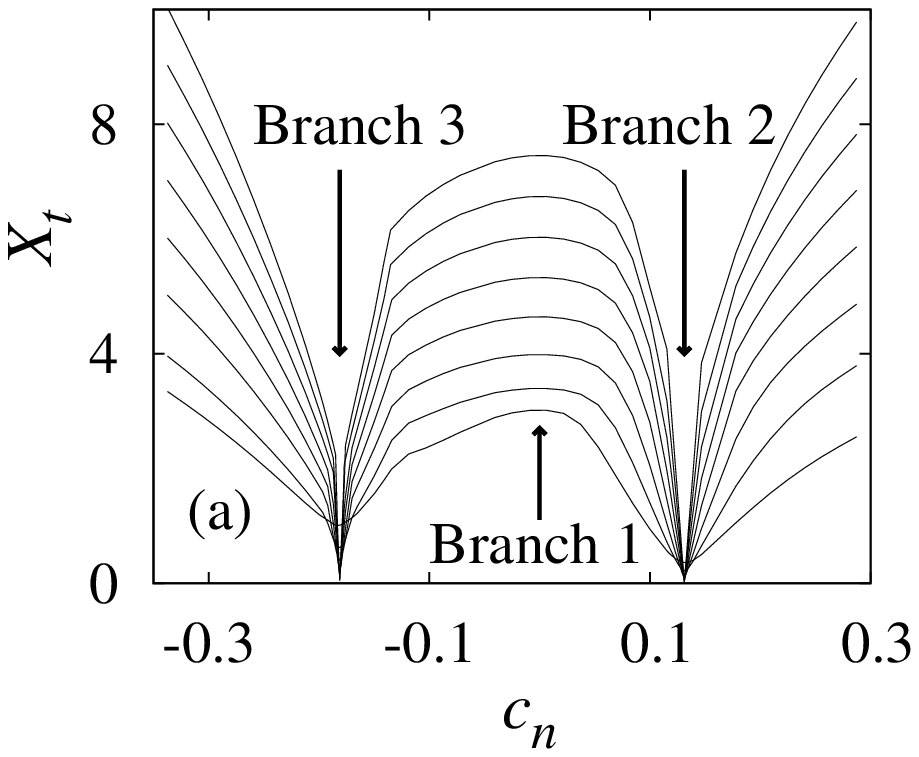}
\hspace*{-20mm}
\includegraphics[scale=0.75]{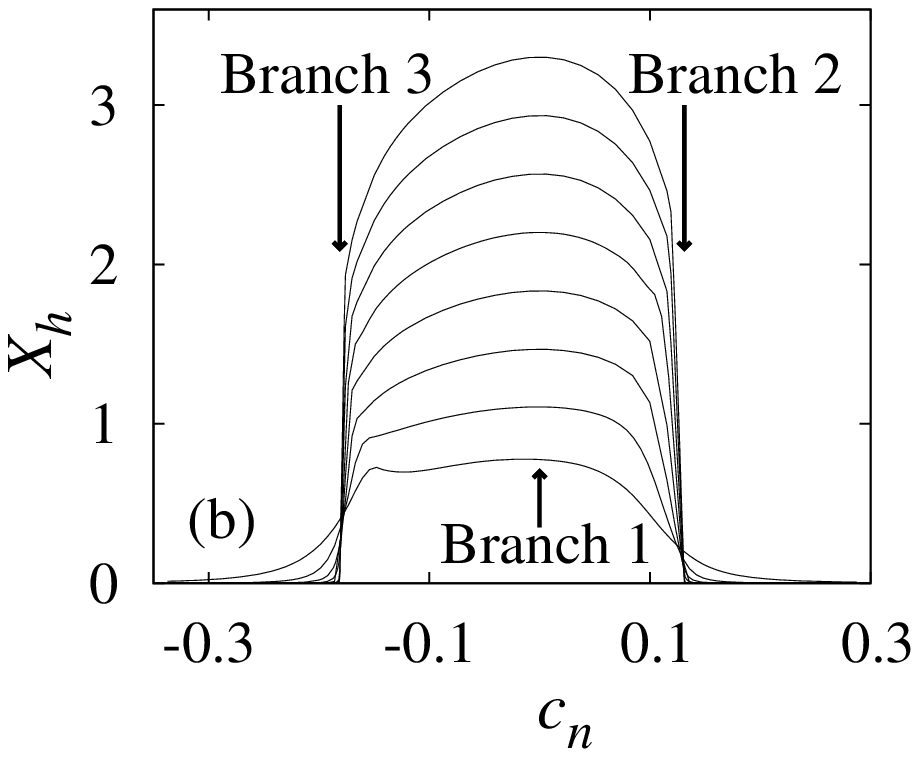}
\caption{Scaled thermal (a) and magnetic (b) gaps versus $c_n$ covering
branches 1, 2 and 3 of the completely packed O($n$) loop model with $n=40$.
Results are shown for even system sizes $L=6$ to 20 for the thermal case, 
and for $L=4$ to 18 for the magnetic case.
In the middle part of the figures the scaled gaps increase with $L$.}
\label{b2b3n40}
\end{figure}

\begin{figure}
\hspace*{-15mm}
\includegraphics[scale=0.75]{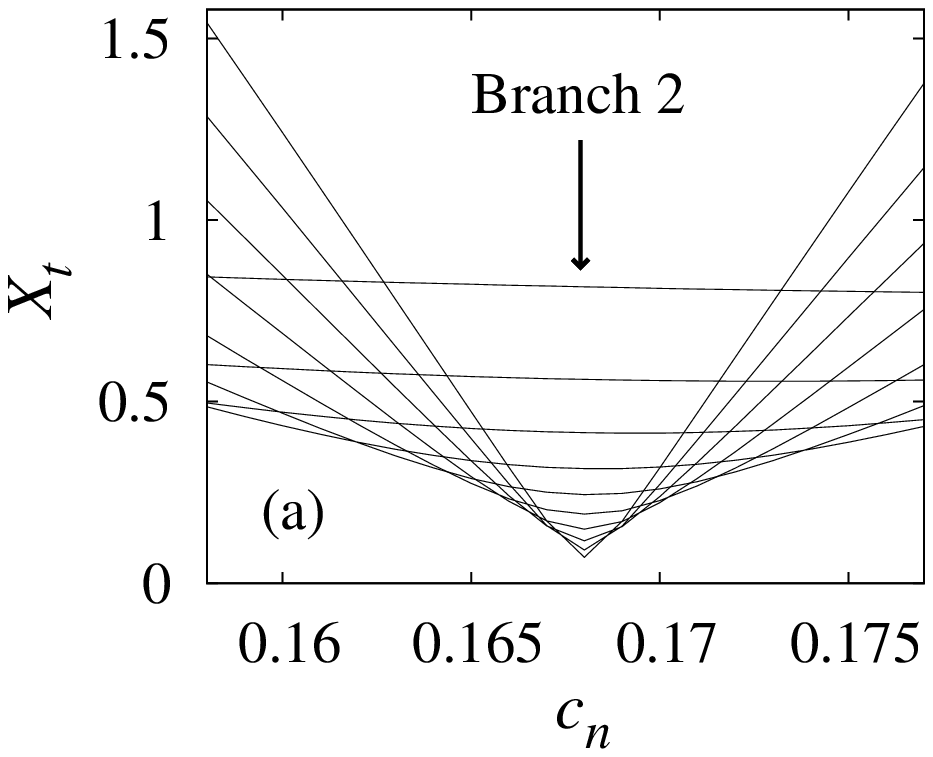}
\hspace*{-20mm}
\includegraphics[scale=0.75]{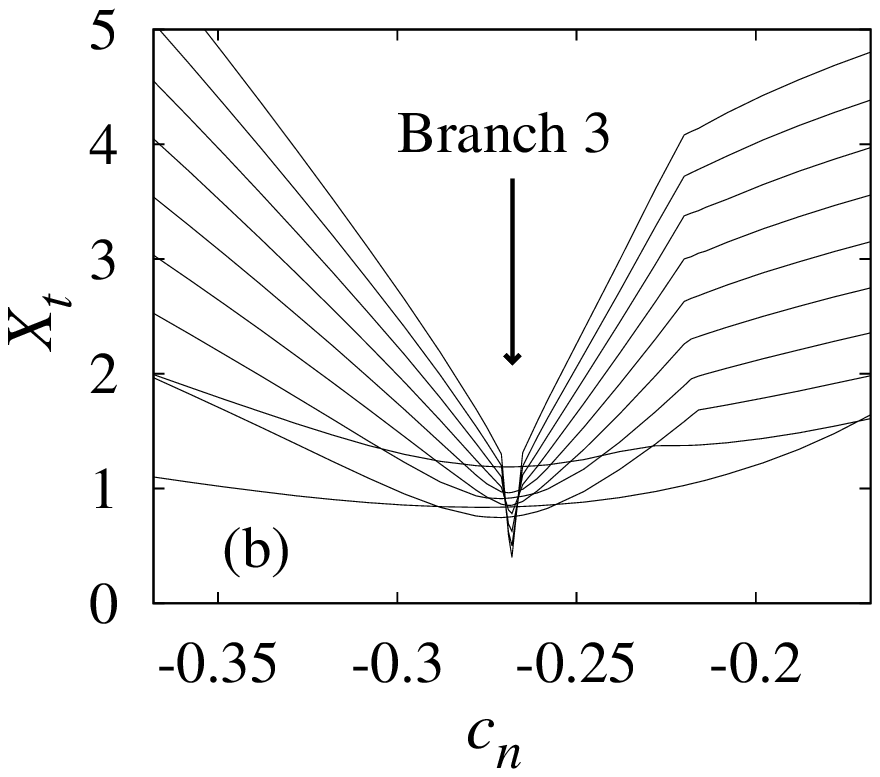}
\caption{Scaled thermal gaps versus $c_n$ of the completely packed O($n$)
loop model with $n=20$ around branch 2 in figure (a) and branch 3 in figure
(b).  Results are shown for even system sizes $L=4$ to 22. The scaled gaps
display, at branch 2 and 3, apparent convergence to 0 when $L$ increases.
Away from these transition points, they tend to diverge instead.}
\label{b23n20}
\end{figure}

For branch 2 and 3, only $z$-type and $c$-type vertices are present.
These two branches exist only for $n \geq 1$.
They merge at the end point $n=1$, where the system reduces to a
trivial case with effective weight 1 for each loop and each vertex.
We first consider the thermal and magnetic scaled gaps $X_t$ and $X_h$
of a system with $n=1.5$ as a function of $c_n$. Results are shown in
Figs.~\ref{fig:qn1_5_t}. Several details can be noted.
At $c_n=0$, which is the location of branch 1, the  scaled gaps are
nicely approaching the values given by Eqs.~(\ref{x_tx_h}).
Furthermore, the curves for $X_t$ show intersections close to 
the branch-1 point, with slopes that increase with $L$.
Then, a comparison with the scaling behavior expressed by
Eq.~(\ref{xisce}), with $y_t$ playing the role of the exponent of the
cubic perturbation $c_n$, shows that the cubic perturbation is
relevant on branch 1 at $n=1.5$, because the slopes increase with $L$.
Slightly to the left of branch 2, intersections occur as well, but here
they seem to indicate that the cubic weight is irrelevant in that range.
Indeed for $c_n<0$ there exists a range about branch 2 where the $X_t$
data are consistent with slow  convergence to a
value independent of $c_n$. This limiting value may be close to 2.
The data in the range $c_n<-1$ appear to behave irregularly due to
finite-size effects with a period exceeding 2. But the data for system
sizes restricted to multiples of 4 may still suggest convergence at the
branch-3 point. The data  in the range  with $c_n$ smaller than the 
branch-3 value indicate that scaled gaps diverge with increasing $L$.

The results for $X_h$ in Fig.~\ref{fig:qn1_5_t}b display a similar scaling
behavior near branch 1 and 2. At $c_n=0$ (branch 1) the data agree with
convergence to the theoretical value given by Eq.~(\ref{x_tx_h}). 
The slow apparent convergence for $c_n<0$ indicates the existence
of a marginal or almost marginal temperature dimension $X_t \approx 2$.
In the neighborhood of branch 3, the $X_h$ data (not shown) lose
transparency because of the irregular finite-size dependence.  For $c_n$
significantly less than the branch-3 value, as well as for  $c_n$
significantly exceeding the branch-2 value, the data are consistent
with convergence to $X_h=0$, as expected for a phase dominated by cubic
vertices.

For $n=2$, $c_n>-2$, again one finds divergent behavior of the gaps $X_t$,
corresponding with a non-critical phase dominated by $c$-type vertices.
The same observation applies to the range where $c_n$ considerably exceeds
the branch-1 value. Again, complex eigenvalues occur near branch 3.
The behavior of $X_t$ and $X_h$ in the neighborhood of branch 1, which
coincides with branch 2 for $n=2$, is shown in Figs.~\ref{fig:xb2n2}.
These data indicate that there exists a range $c_n<0$ where the cubic
weight is marginal, for which $X_h=1/8$ and $X_t=2$.

Next, we consider the thermal and magnetic scaled gaps in the range $n>2$.
Fig.~\ref{fig:qn10b2} shows these quantities for $n=10$ as a function of the
cubic weight $c_n=c/n$, for a range of system sizes, with $z$ fixed at $z=1$.
The $X_t$ curves are seen to display minima, which become increasingly
pronounced for larger system sizes, and whose location rapidly converges
to the branch-2 value $c_n=0.2$. 
The $X_h$ curves instead monotonically decrease as a function of $c_n$,
and they intersect at points that rapidly approach the branch-2 value of
$c_n$. Furthermore, extrapolation of the two types of scaled gaps at the
minima or at the intersections leads to values close to 0 
(see also Table \ref{scad23}), strongly suggesting a first-order phase
transition at branch-2. The scaled magnetic gaps for $c_n$ smaller
than the branch-2 value in this figure seem to diverge, as expected for
a disordered phase. Instead, for $c_n$ exceeding the branch-2 value,
the magnetic gaps rapidly approach zero, indicating a long-range-ordered
phase in which the cubic vertices percolate.
The divergent behavior of the scaled thermal gaps on either side of
the branch-2 value indicates a finite energy-energy correlation
length, consistent with this phase behavior.
 
Similar data were computed for other values of $n$. For $n>10$ the
minima and intersections display even more rapid convergence to 
the branch-2 values, and the gaps tend to vanish more rapidly with
increasing $L$. For $n<10$ the picture becomes less clear, and for
$n\gae 2$ we are unable to see clear signs of a first-order transition
from the available data. But these results do not exclude a weak
first-order transition, and one may expect that branch 2 is the locus
of a first-order transition for all $n>2$.

For larger values of $n$ also the behavior of the scaled gaps near
branch 3 can be resolved. This is illustrated by the $X_t$ and $X_h$
plots for $n=40$ shown in Fig.~\ref{b2b3n40}. It shows the scaled gaps
as a function of $c_n$. The scaled gaps  extrapolate to a value close 
to 0 at the branch 2 and 3 points. For other values of $c_n$, the
thermal scaled gaps display a divergent behavior. So do the magnetic
scaled gaps in the range between branch 2 and 3. Outside this range,
the magnetic gaps rapidly approach the value $X_h=0$, which is as
expected for a phase in which the $c$-type vertices dominate.

More detailed pictures of the scaled thermal gaps in the regions near
the locations of branches 2 and 3 are shown in Fig.~\ref{b23n20}. 
These figures show data for $n=20$, and include system sizes up to $L=22$.

\subsection{Branch 4 }
In the absence of cubic vertices for branch 4, we investigate
of the phase behavior as a function of the crossing-bond weight, i.e.,
crossover phenomena between branch 1 and branch 4.
The case of branch 5 involves all three vertex types and will therefore
be treated separately. For the interpretation of the results for the
scaled thermal gaps still denoted $X_t$, it should be realized that these 
gaps are obtained from a transfer matrix in an extended connectivity
space in comparison respect to that used for branch 1, thus allowing
for additional eigenvalues and associated scaling dimensions.

\begin{figure}
\begin{center}
\hspace*{-18mm}
\hspace*{-8mm}
\includegraphics[scale=0.7]{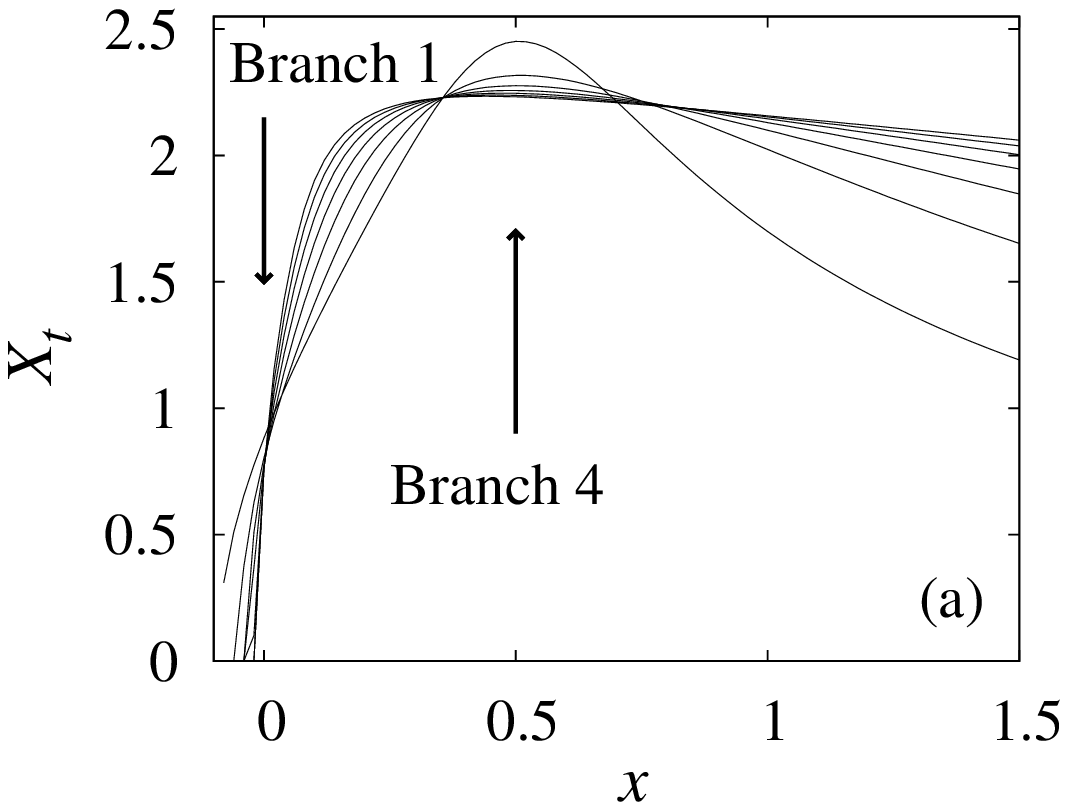}
\hspace{-18mm}
\includegraphics[scale=0.7]{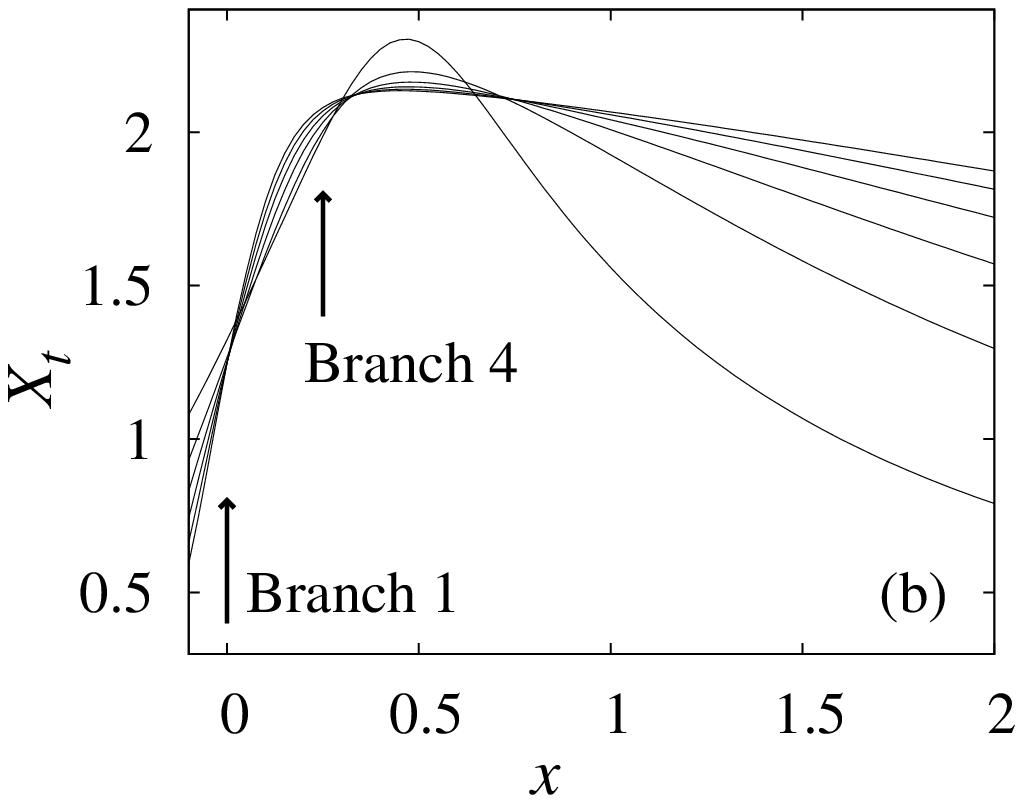}
\end{center}
\caption{Scaled thermal gaps versus the crossing bond weight of
the completely packed O($n$) loop model around branch 4, for $n=0$ in
figure (a) and for $n=1$ in figure (b). The location of branch 4 is
indicated.  Results are shown for
even system sizes $L=4$ to 16. The scaled gaps increase as a function of
$L$ for large $x$. The intersections of the curves near $x=0$ agree well
with the value $X_t=3/4$ for cubic crossover in the dense phase of the 
$n=0$ loop model, and with $X_t=5/4$ for the $n=1$ model.}
\label{fig:b4_01}
\end{figure}

\begin{figure}
\begin{center}
\hspace*{-8mm}
\includegraphics[scale=0.70]{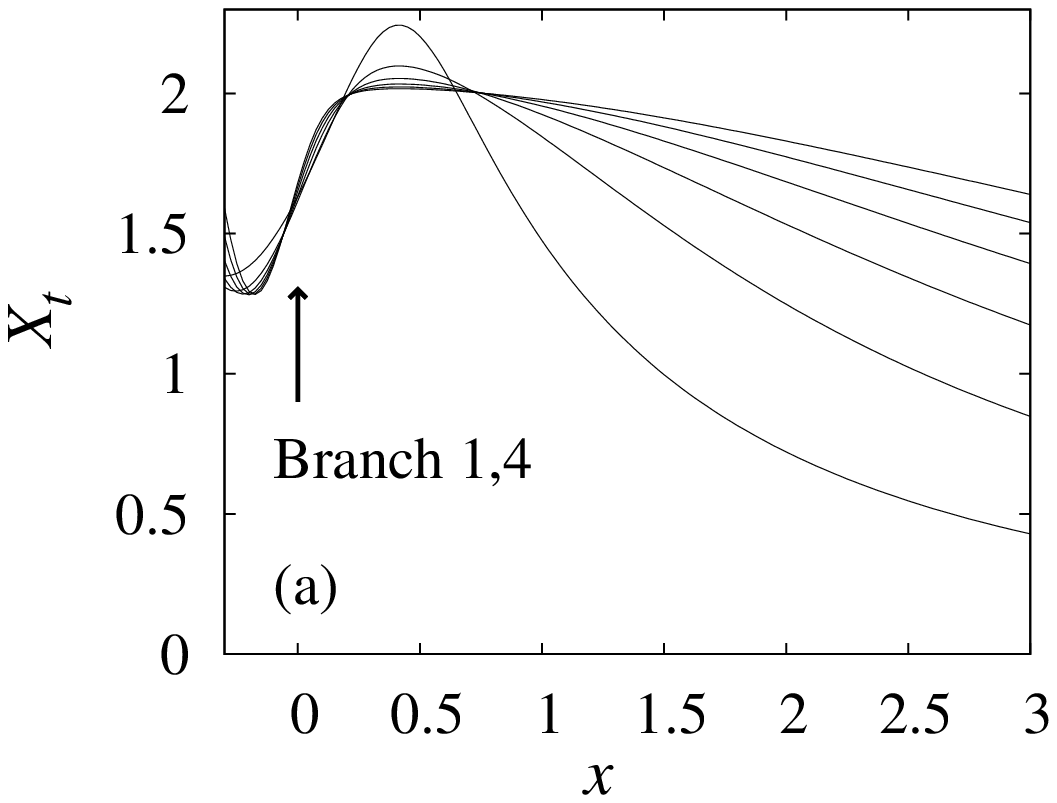}
\hspace{-18mm}
\includegraphics[scale=0.70]{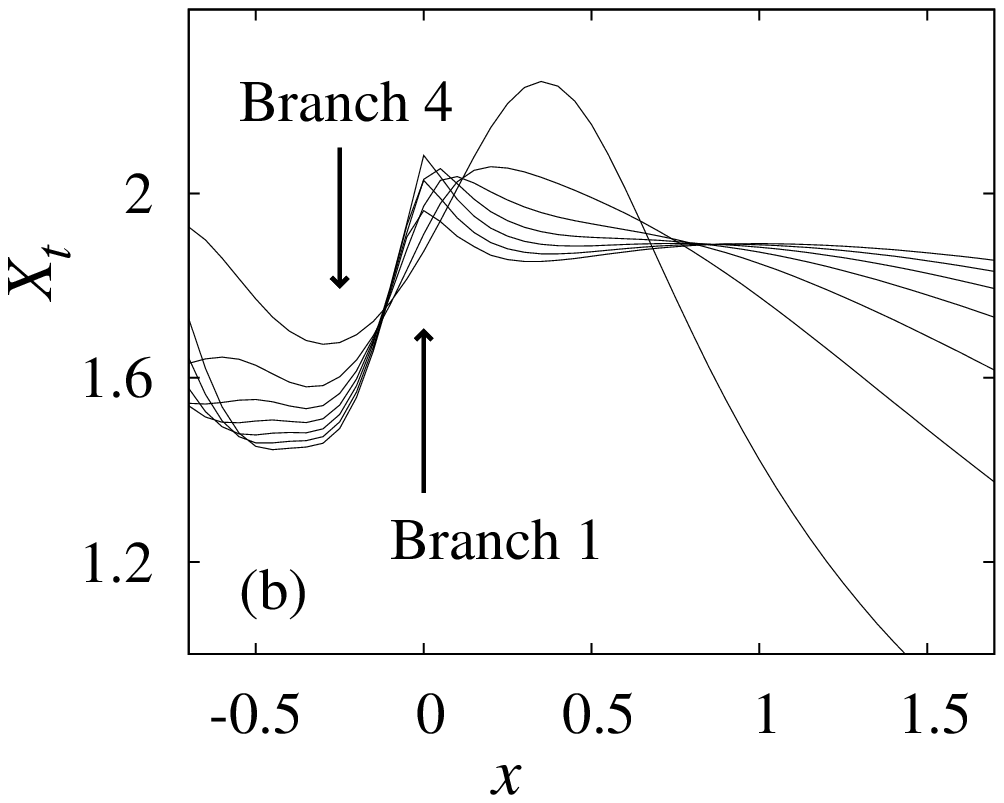}
\end{center}
\caption{Scaled thermal gaps versus the crossing bond weight of
the completely packed O($n$) loop model around branch 4, for $n=2$ in
figure (a) and for $n=3$ in figure (b). The location of branch 4 is
indicated.  Results are shown for even system sizes $L=4$ to 16.
The scaled gaps increase as a function of $L$ for large $x$.}
\label{fig:b4_23}
\end{figure}

In Figs.~\ref{fig:b4_01} and \ref{fig:b4_23}, we present diagrams
describing the scaling behavior of the thermal gaps as a function of
$x$ near branch 4 for $n=0$ and 1, and  $n=2$ and 3 respectively.
For $n=0$ and 1, there are intersections close to $x=0$, and the 
behavior of the slopes confirms that $x$ is relevant, which tells us
that a continuous phase transition takes place here. It is noteworthy
that, for $n=1$, the free energy is a trivial nonsingular function of
the summed vertex weights. Thus the phase transition at $x=0$
can, for $n=1$, only apply to the geometric properties of the loop
configurations. Indeed, the intersections indicate that $X_t=5/4$ at
the transition, corresponding with the thermal scaling dimension of the
percolation critical point. For $n=2$, the scaled gaps in the range $x>0$
display a behavior consistent with marginal behavior as a function of $x$. 
Also in Fig.~\ref{fig:b4_23}b for $n=3$ one observes  hints of marginal
behavior for $x>0$. For $x<0$ there is a range where the scaled gaps
are suggestive of another critical phase with a smaller dimension $X_t$.

\begin{figure}
\begin{center}
\hspace*{-8mm}
\includegraphics[scale=0.70]{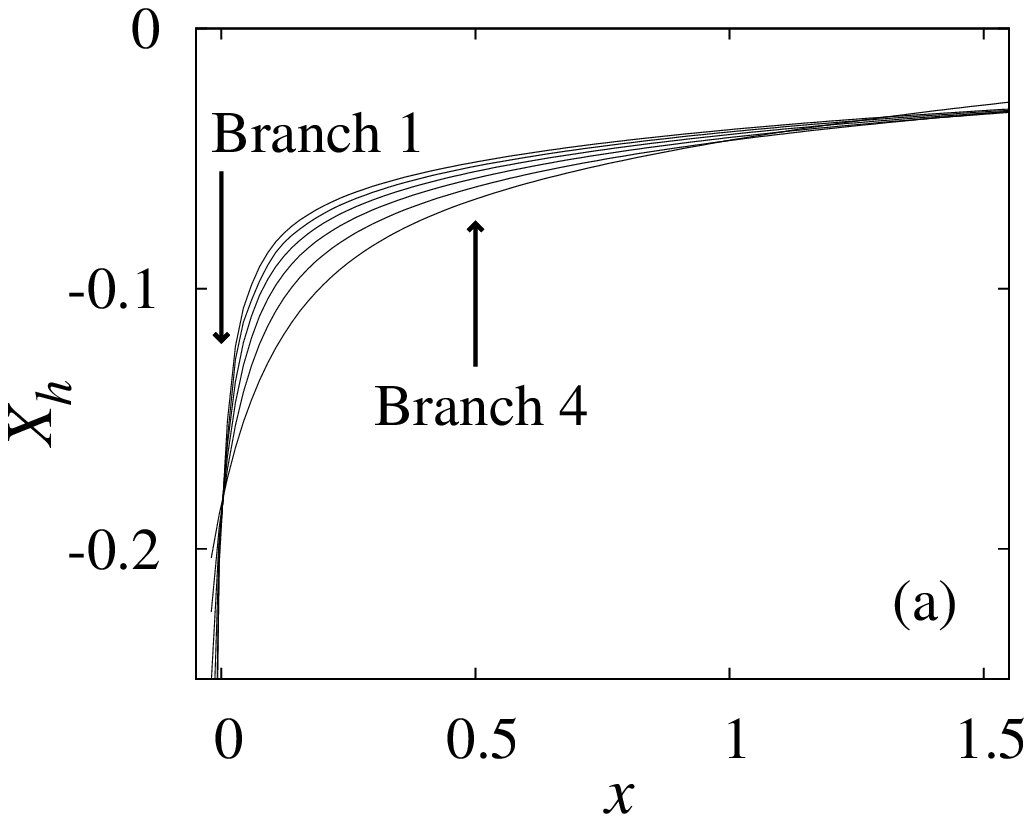}
\hspace{-18mm}
\includegraphics[scale=0.70]{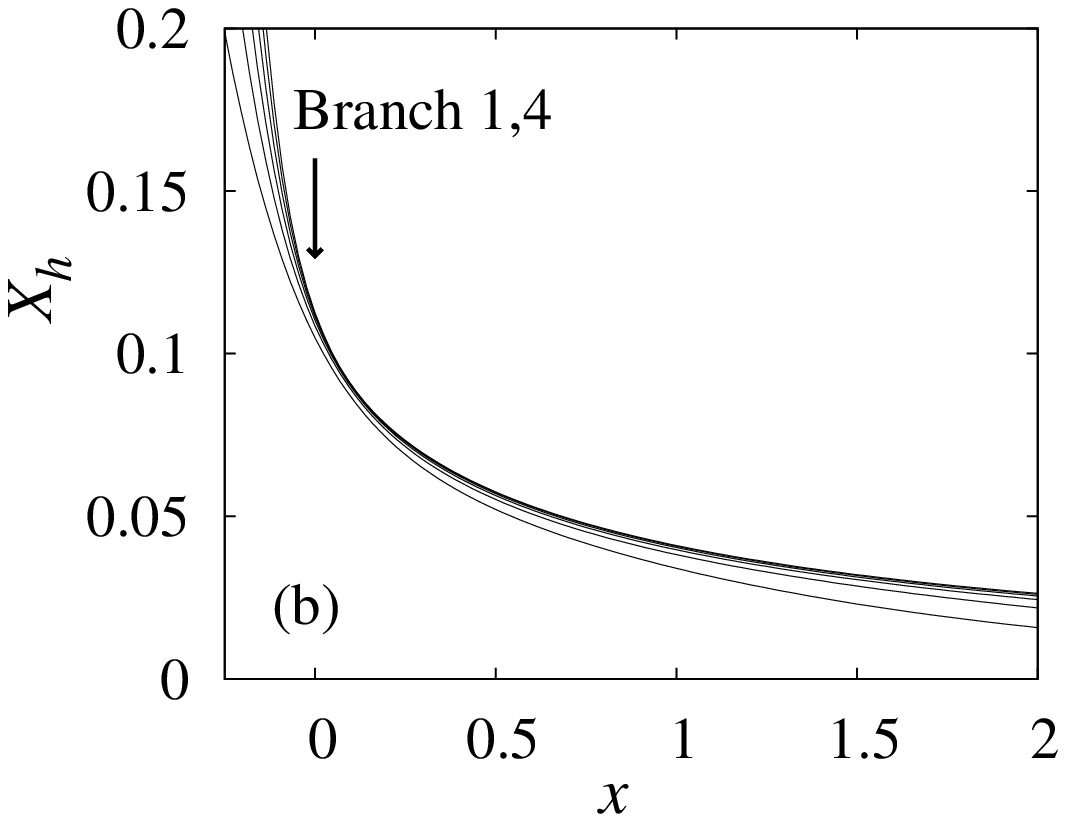}
\end{center}
\caption{Scaled magnetic gaps versus the crossing bond weight of
the completely packed O($n$) loop model around branch 4, for $n=0$ in
figure (a) and for $n=2$ in figure (b). The location of branch 4 is
indicated.  Results are shown for even system sizes $L=4$ to 14.
The scaled gaps increase as a function of $L$ at branch 4.}
\label{fig:b4h02}
\end{figure}

The behavior of the scaled magnetic gaps, shown in Figs.~\ref{fig:b4h02}
for $n=0$ and 2, is consistent with  that of $X_t$. Intersections are
found for $n=0$ near branch 1, rapidly converging to the expected value
$X_h=-3/16$. Crossover to much smaller absolute values of $X_h$ occurs 
for $x>0$. For $n=2$ the crossing-bond weight seems marginal, and in a 
range $x>0$ one observes apparent convergence to an $x$-dependent value,
thus indicating ``nonuniversal'' behavior.

\begin{figure}
\begin{center}
\hspace*{-8mm}
\includegraphics[scale=0.70]{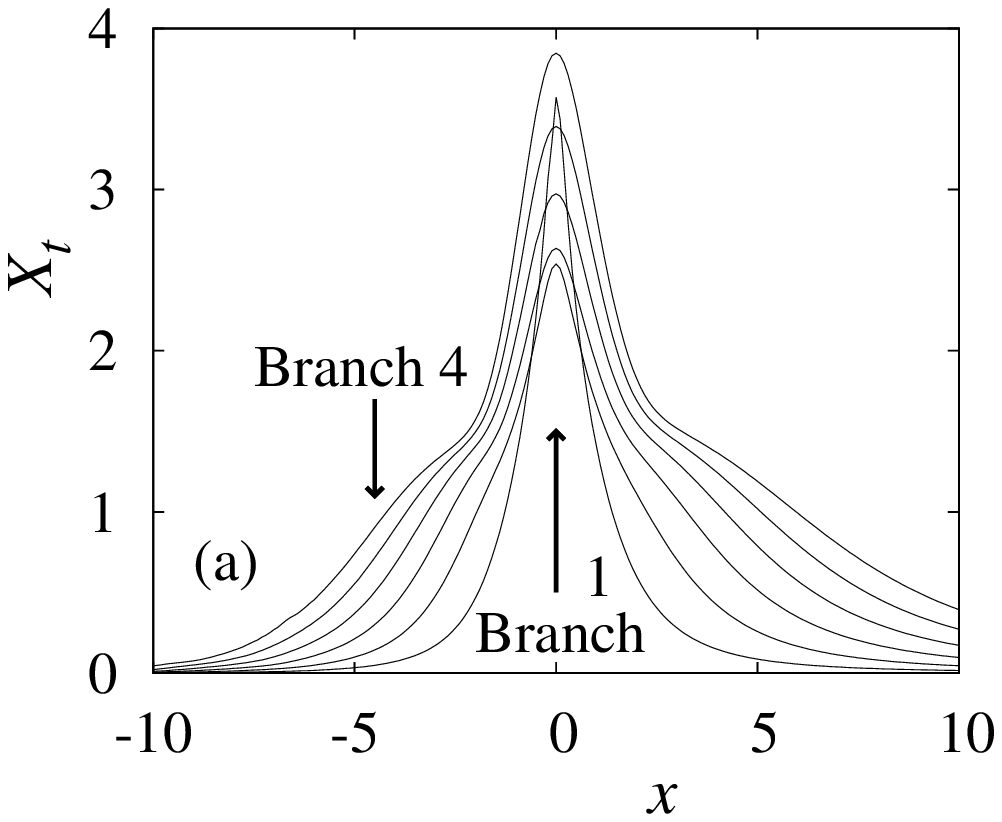}
\hspace{-18mm}
\includegraphics[scale=0.70]{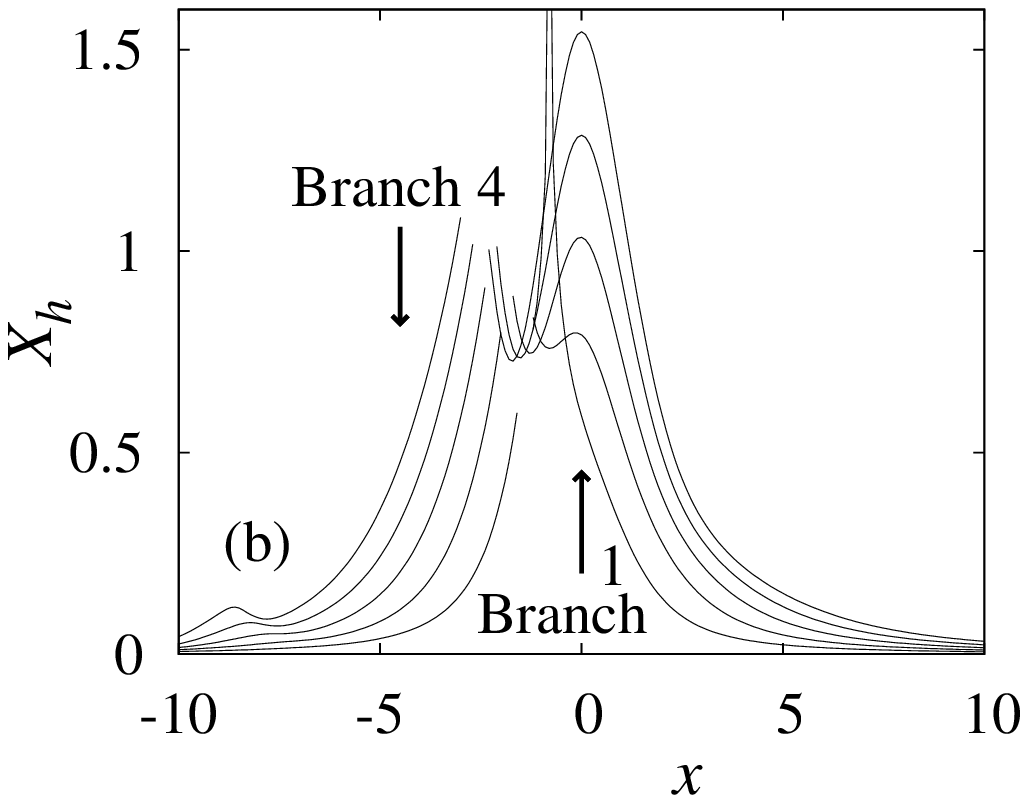}
\label{fig:qnx20}
\end{center}
\caption{Scaled thermal (a) and magnetic (b) gaps versus the crossing
bond weight of the completely packed O($n$) loop model with $n=20$. 
The thermal gaps are shown as separate curves for even system sizes
$L=4$ to 14, and the magnetic gaps for $L=4$ to 12. Larger scaled gaps
correspond with larger $L$ in most of the range of the $x$ scale.
The missing parts in the $X_h$ curves correspond with a level crossing
of the leading eigenvalues of the odd systems that appear in
Eq.~(\ref{x_h}).
The intersections in the $X_h$ curves occur close to the level crossing.
The absence of intersections between the $X_t$ curves for $L>4$ indicates
that there is at most a weakly singular phase transition as a function 
of $x$.
}
\label{qnx20}
\end{figure}

One may expect that the introduction of crossing-bond type vertices in
the completely packed non-intersecting loop model with large $n$ will
affect the checkerboard-like ordering of the elementary loops. Thus, we
numerically investigate the scaled gaps as a function of $x$ for $n=20$,
in order to address the question whether a phase transition occurs as a
function of the crossing-bond weight $x$.
The results for $X_t$ and $X_h$ are plotted in Figs.~\ref{qnx20}.
The $X_t$ curves do have some intersections, but only involving the
smallest system size. They do not provide evidence for an Ising-like
transition where the checkerboard-like order of the $x=0$ model vanishes.
But if, as the results in Sec.~\ref{freeb4} suggest, the branch-4
model is in a critical state at $n=20$, while the branch-1 model (at $x=0$)
is off-critical, then there must be a transition of some kind. Perhaps
the "shoulder" that develops in the curves near $x=-3$ is a sign of a
weak transition. A similar shoulder is present in the physical range $x>0$.
We note that, while the results at $x=0$ in Fig.~\ref{qnx20} increase
rapidly with $L$, the results near the shoulders seem consistent with
convergence to a finite value of $X_t$.
The data for $X_h$ do display intersections for $x<0$, close to a crossing
of the leading transfer-matrix eigenvalues of odd systems, involving a
doublet and a singlet. For positive $x$ the singlet is the largest eigenvalue.

\subsection{Branch 5}
The analysis of the phase behavior in the neighborhood of  branch 5 is
somewhat more involved in the sense that we now have all three types of
vertices in the system. Due to the larger number of connectivities for
a given system size, the calculations for branch 5 are restricted to
smaller systems than those for branch 4. We investigate the influence
of a variation of $x$ as well as of $c_n$.

\subsubsection{Variation of the crossing-bond weight}
\begin{figure}
\begin{center}
\hspace*{-8mm}
\includegraphics[scale=0.7]{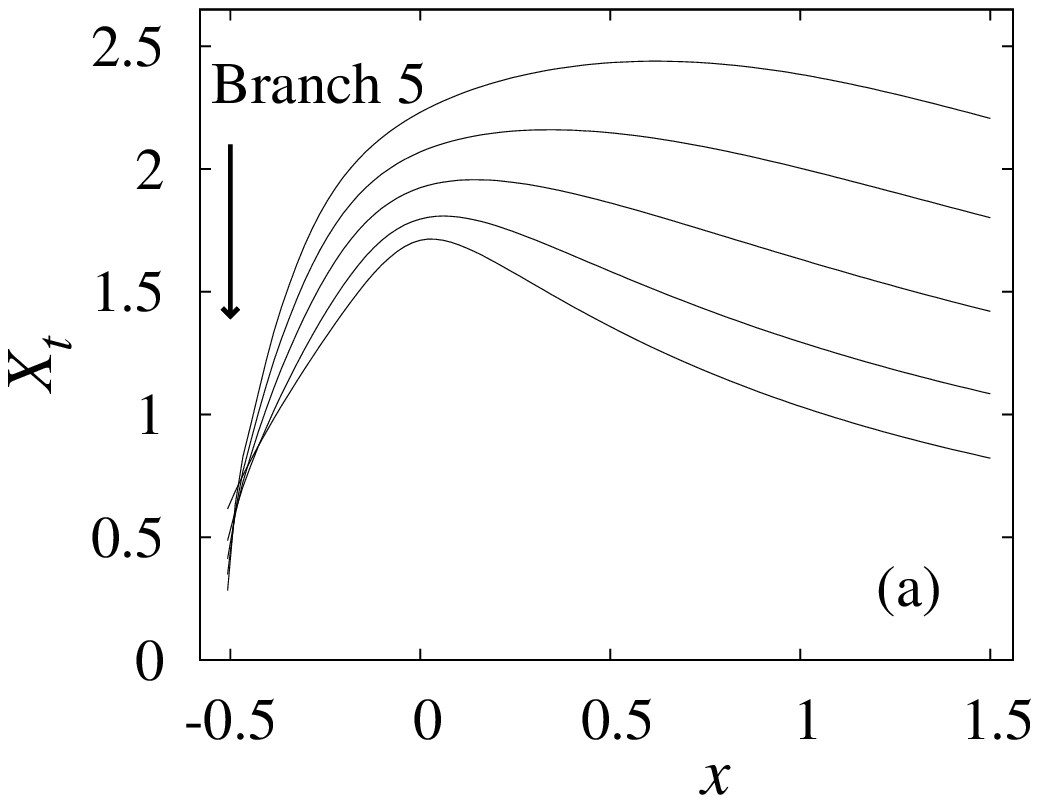}
\hspace{-18mm}
\includegraphics[scale=0.7]{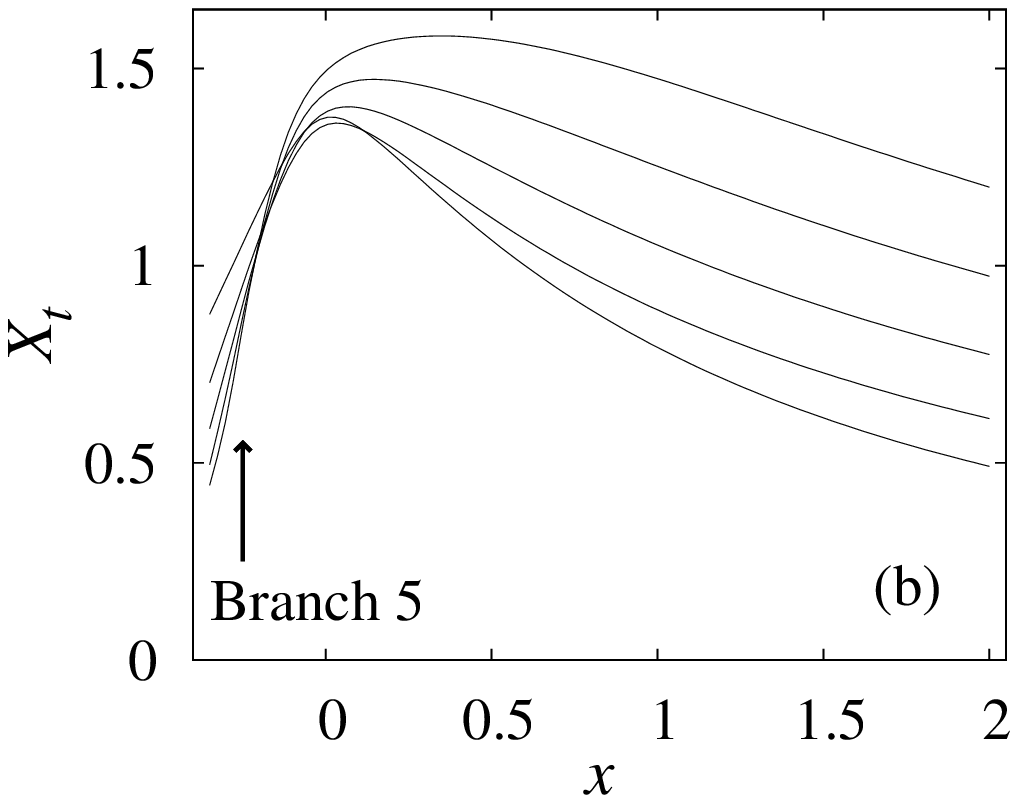}
\label{fig:b5_01x}
\end{center}
\caption{Scaled thermal gaps versus the crossing bond weight of
the completely packed O($n$) loop model around branch 5, for $n=0$ in
figure (a) and for $n=1$ in figure (b). The cubic weight is fixed at its
value at branch 5.  Results are shown for even system sizes $L=4$ to 12.
The scaled gaps increase as a function of $L$ on the right hand side.}
\label{b5c0}
\end{figure}

The scaled thermal gaps for $n=0$ and 1 are shown in Figs.~\ref{b5c0},
as a function of the crossing-bond weight $x$, while the cubic weight
$c_n$ is kept at its branch-5 value. Although the partition sum remains
well-behaved, the cubic weight as used in the transfer-matrix
calculations diverges at $n=0$. Therefore the thermal gaps for $n=0$
were obtained by averaging those for $n=\pm 0.05$. The resulting thermal
gaps for $n=0$ display intersections near branch 5, and the two
corresponding eigenvalues of the transfer matrix merge into a complex
pair at values of $x$ that are only slightly smaller.
For $n=1$, there are also intersections near branch 5, approaching
the branch-5 point when $L$ increases.

At the intersections, the slopes of the
curves increase with $L$, which indicates that the crossing bonds
are relevant at branch 5, and thus induce a phase transition.
For $n=1$, this transition may describe some
geometric property of the graph configurations.
  
\begin{figure}
\begin{center}
\hspace*{-8mm}
\includegraphics[scale=0.7]{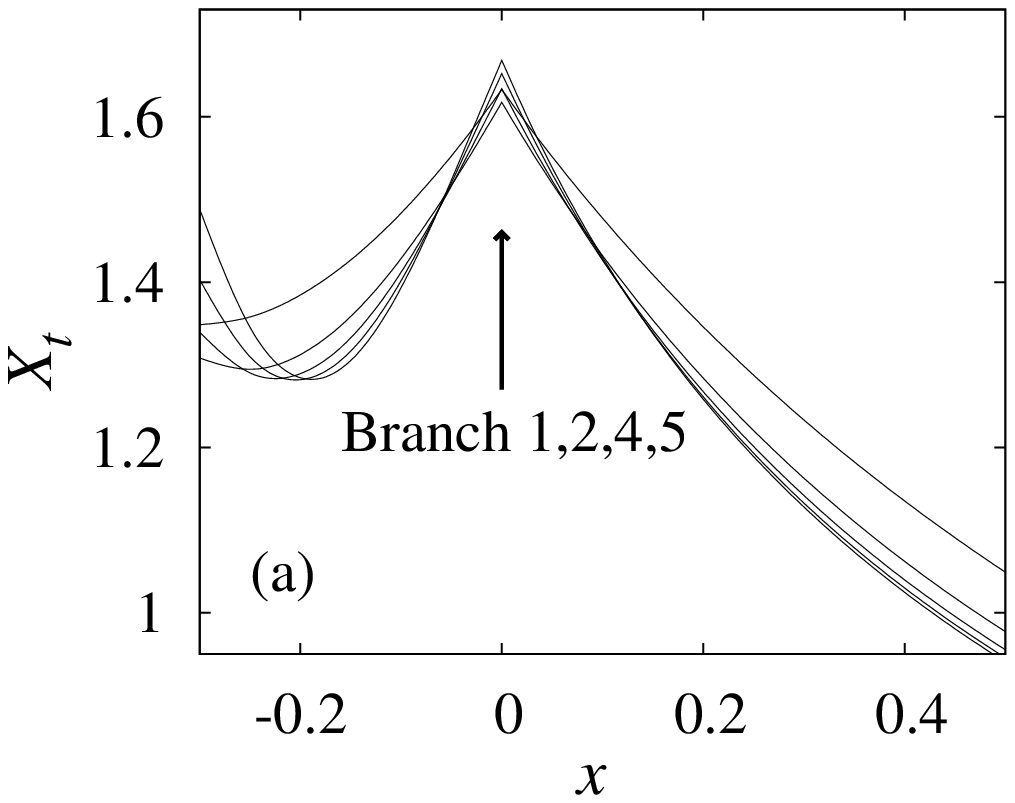}
\hspace{-18mm}
\includegraphics[scale=0.7]{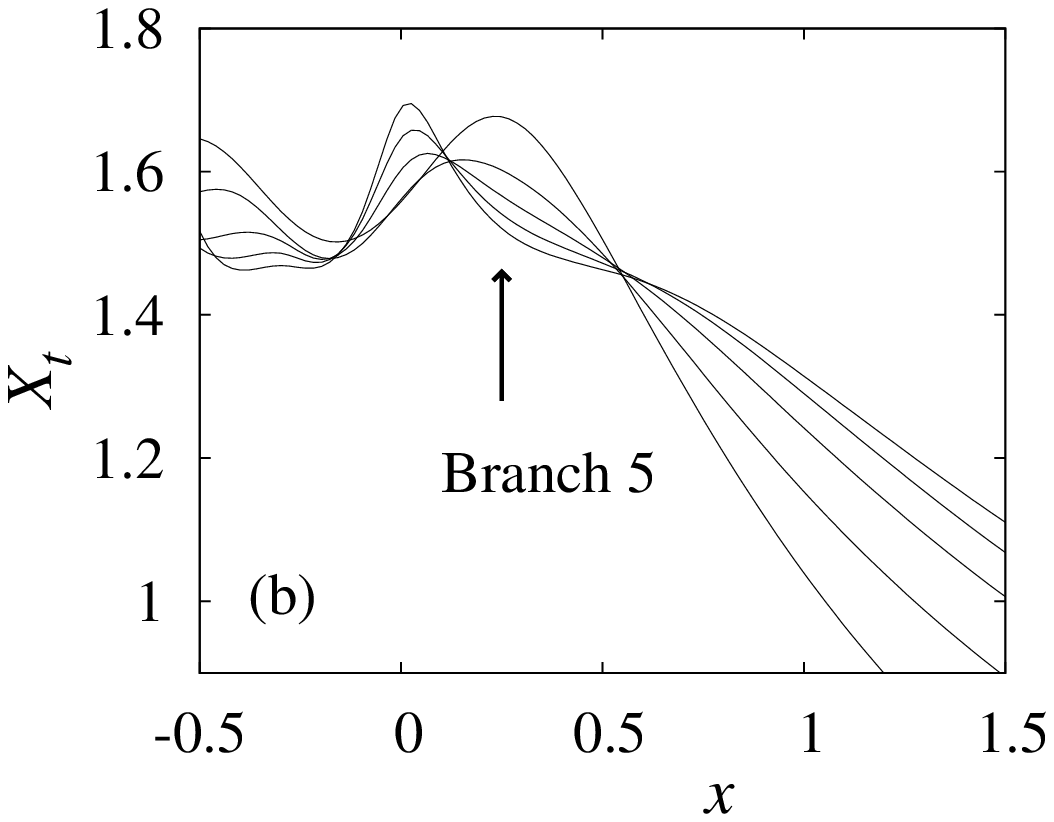}
\end{center}
\caption{Scaled thermal gaps versus the crossing bond weight of
the completely packed O($n$) loop model around branch 5, for $n=2$ in
figure (a) and for $n=3$ in figure (b).
The cubic weight vanishes for $n=2$ at branch 5.
This is precisely the intersection point of branches 1, 2, 4 and 5.
Results are shown for even system sizes $L=4$ to 12. The cusps at
$x=0$ are due to intersections between the second and third
eigenvalues of the transfer matrix.
For $n=2$, the scaled gaps decrease as a function of $L$ on the right
hand side; for $n=3$, they increase.}
\label{fig:b5n23x}
\end{figure}

In Figs.~\ref{fig:b5n23x} we show the scaled gaps as a function of $x$ for
$n=2$ and 3. For $n=2$, a cusp appears at the branch-5 point, which
is due to an intersection of the second and third eigenvalues of the
transfer matrix. The curves are suggestive of ``nonuniversal'' behavior
of $X_t$ when $x$ is varied. Note that, 
although branches 4 and 5 coincide at $n=2$, the cusps are absent in
Fig.~\ref{fig:b4_23}a. This is due to the fact that, for $x>0$, the
subleading thermal eigenvalue for branch 5 is absent for branch 4,
whose transfer matrix acts in a smaller configuration space. For
$x \le 0$, Figs.~\ref{fig:b4_23}a and \ref{fig:b5n23x}a match exactly.
The curves for $n=3$ in Fig.~\ref{fig:b5n23x}b display some structure
superimposed on marginal-like behavior, which may however be due to
slow crossover effects as may be expected for $n\gae 2$.

\begin{figure}
\begin{center}
\hspace*{-8mm}
\includegraphics[scale=0.70]{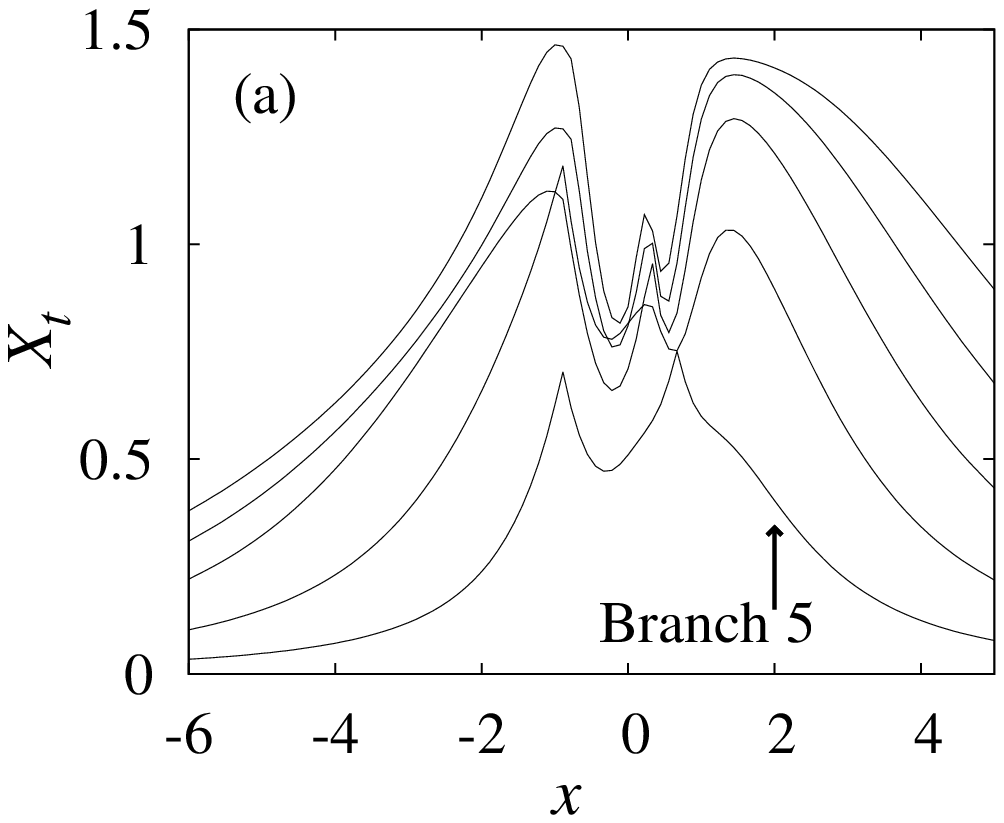}
\hspace{-12mm}
\includegraphics[scale=0.70]{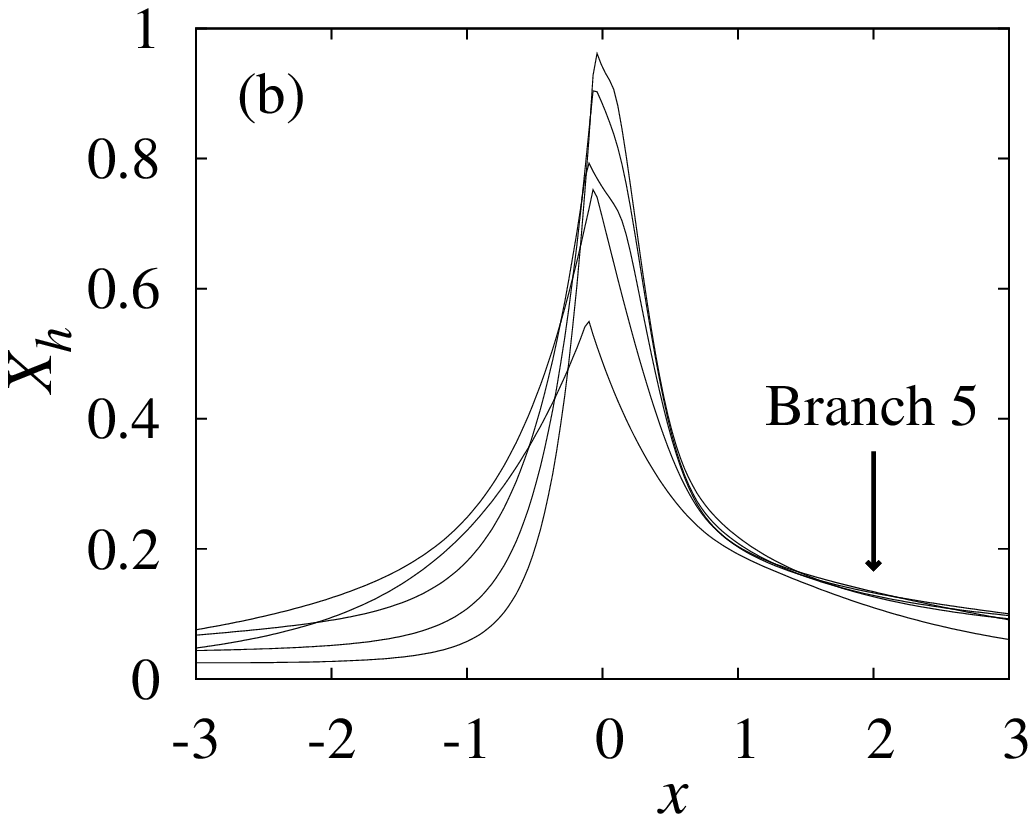}
\end{center}
\caption{Scaled gaps versus the crossing bond weight of the completely
packed O($n$) loop model around branch 5 for  $n=10$. The thermal  gaps
are shown in figure (a) for even $4 \leq L \leq 12$, and the magnetic
gaps in figure (b) for odd $3 \leq L \leq 11$.
The cubic weight is fixed at its branch-5 value.
The scaled gaps increase as a function of $L$ in the neighborhood of
branch 5. But for $X_h$, $x<0$ they tend to become smaller instead. }
\label{fig:b5nl0x}
\end{figure}

\begin{figure}
\begin{center}
\hspace*{-8mm}
\includegraphics[scale=0.70]{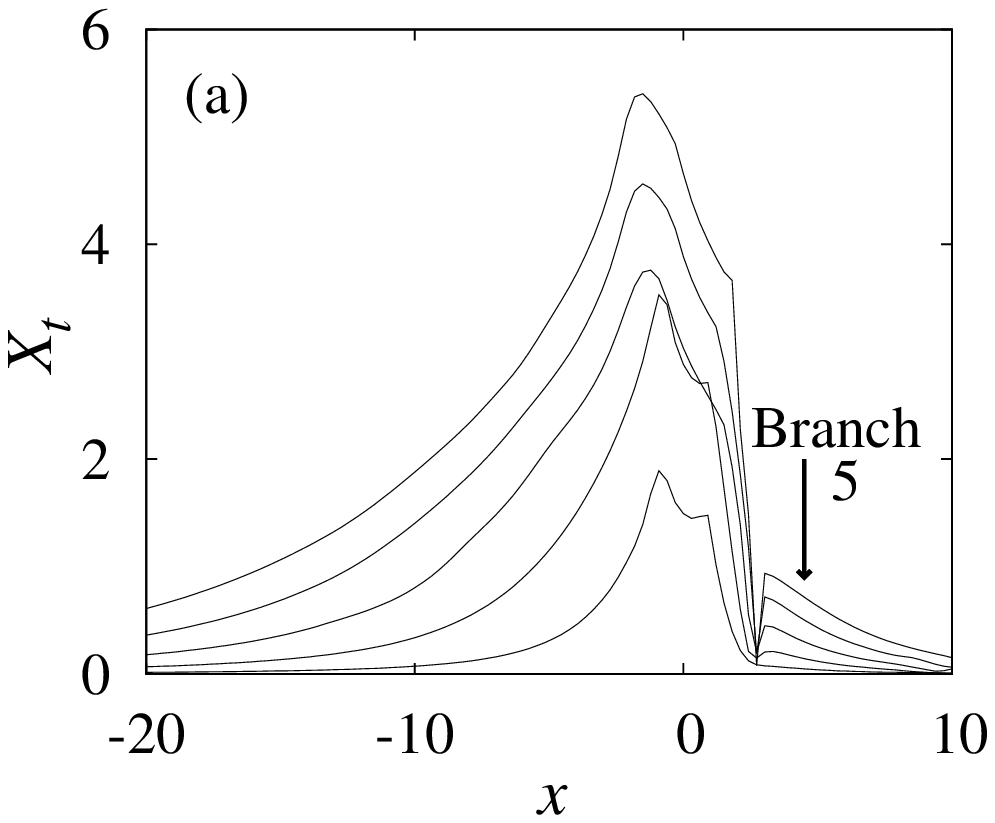}
\hspace{-12mm}
\includegraphics[scale=0.70]{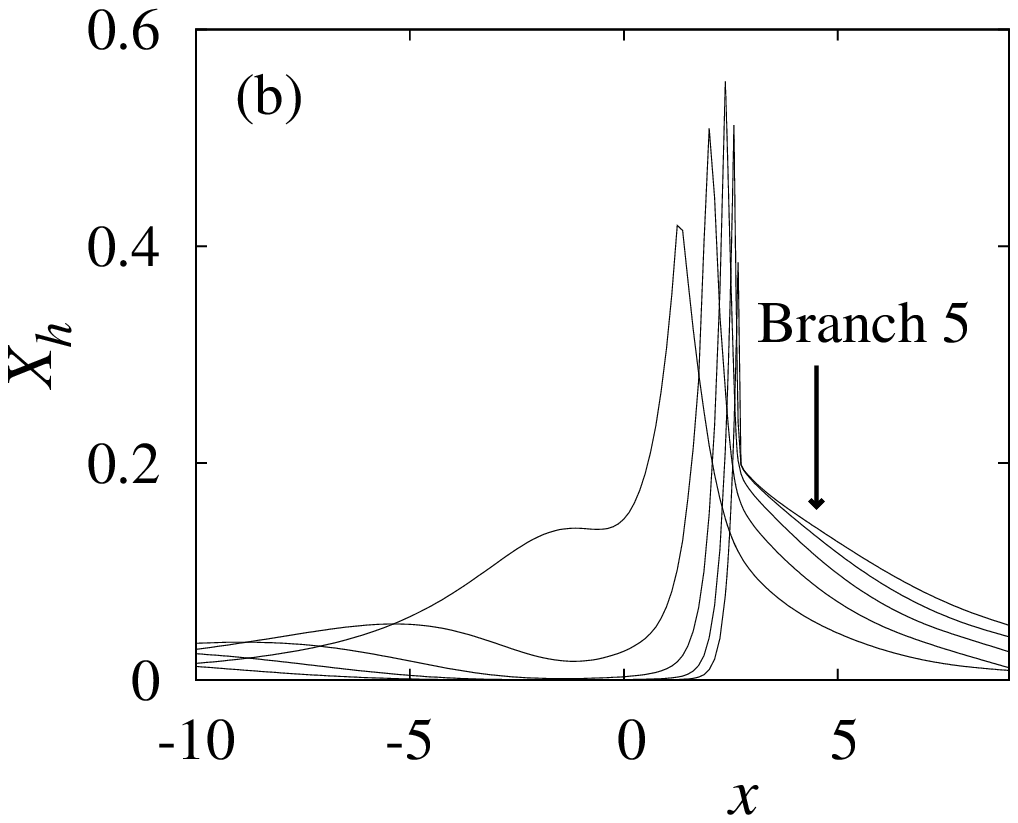}
\end{center}
\caption{Scaled gaps versus the crossing bond weight of the completely 
packed O($n$) loop model around branch 5 for  $n=20$. The thermal  gaps
are shown in figure (a) for even $4 \leq L \leq 12$, and the magnetic 
gaps in figure (b) for odd $3 \leq L \leq 11$.
The cubic weight is fixed at its branch-5 value.
The scaled gaps increase as a function of $L$ near branch 5. But for
$X_h$, $x<0$ there exists a range where they tend to 0 instead.}
\label{fig:b5n20x}
\end{figure}

We also include results for the thermal and magnetic gaps for $n=10$
and 20 in Figs.~\ref{fig:b5nl0x} and \ref{fig:b5n20x}. The magnetic gaps
were calculated  on the basis of Eq.~(\ref{x_h}). The scaling behavior
of the results for even $L$ is consistent with that for odd $L$, but there
is some alternation effect. We show the magnetic gaps only for odd $L$.
The results for $n=10$ still seem consistent with convergence to
nontrivial values $X_t \approx 3/2$ and $X_h \approx 1/8$. For $n=20$
this is even less clear. A pronounced difference between $n=10$ and
$20$ is seen in the $X_h$ plots near $x=0$, where the magnetic gaps for
$n=20$ rapidly approach 0 with increasing $L$, thereby revealing a phase
dominated by cubic vertices. The sharp extrema for $n=20$ near $x=2$
may be associated with a transition between a phase with mainly $c$-type
vertices and one with $x$-type vertices.

\subsubsection{Variation of the cubic vertex weight}
\begin{figure}
\begin{center}
\hspace*{-8mm}
\includegraphics[scale=0.7]{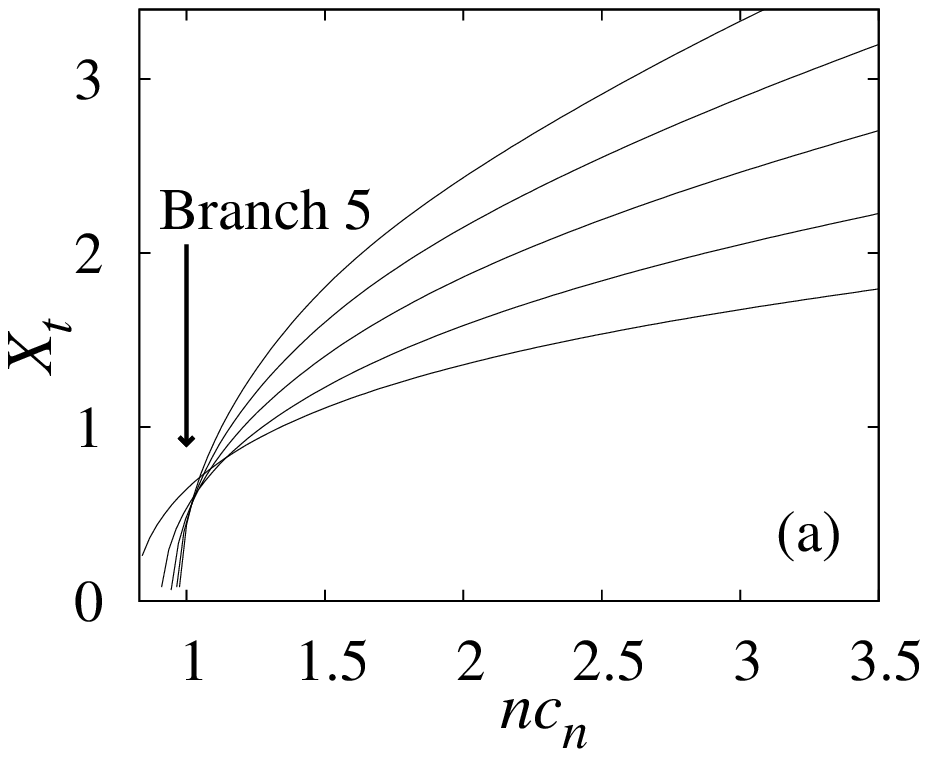}
\hspace{-18mm}
\includegraphics[scale=0.7]{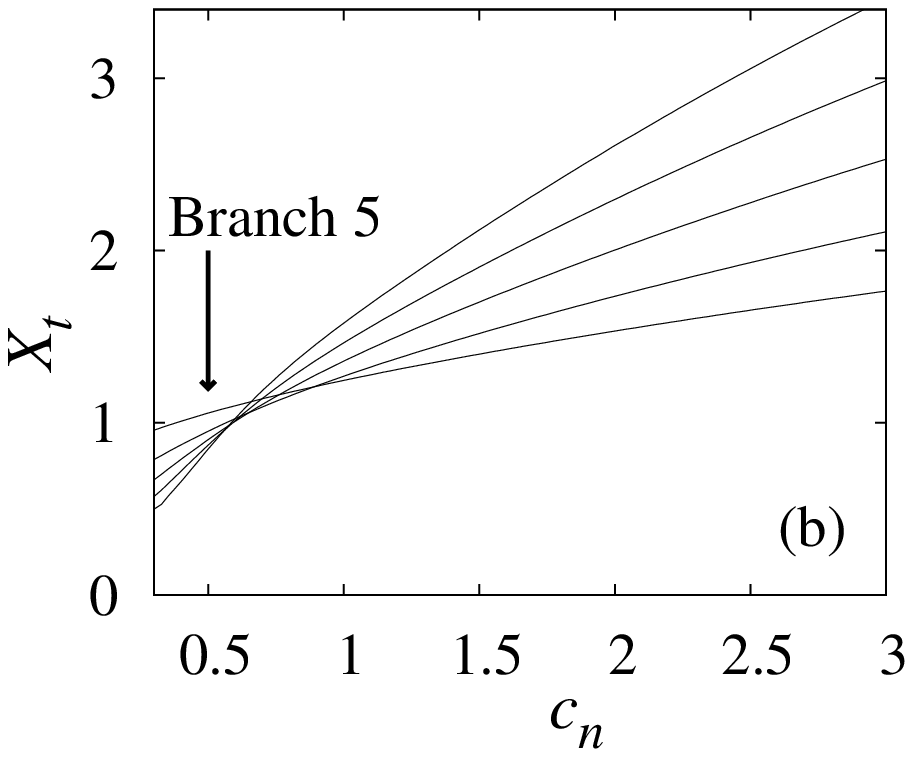}
\end{center}
\caption{Scaled thermal gaps versus the cubic vertex weight of
the completely packed O($n$) loop model around branch 5, for $n=0$ in
figure (a) and for $n=1$ in figure (b). The crossing-bond weight $x$ is fixed
at its branch-5 value.  Results are shown for even system sizes $L=4$ to 12.
The scaled gaps increase as a function of $L$ on the right hand side. The
branch-5 cubic vertex weight $c_n$ diverges at $n=0$, but $c=n c_n$
(horizontal scale) remains finite.}
\label{fig:b5c01}
\end{figure}

We also investigated the behavior of the system near branch 5 under a change
of $c_n$. Figs.~\ref{fig:b5c01} display the results for the scaled thermal
gap for $n=0$ and $n=1$. Those for $n=0$ are again obtained by interpolation
between $n=-0.05$ and 0.05. The intersections indicate that a continuous
phase transition takes place at branch 5. At cubic vertex weights somewhat
smaller than the branch-5 value one finds complex eigenvalues, similar to
the situation found when $x$ becomes smaller with respect to its branch-5
value, see under Figs.~\ref{b5c0}. The apparent divergence of the scaled
gaps for larger values of the cubic vertex weight indicates a non-critical
state dominated by $c$-type vertices.

\begin{figure}
\begin{center}
\hspace*{-8mm}
\includegraphics[scale=0.7]{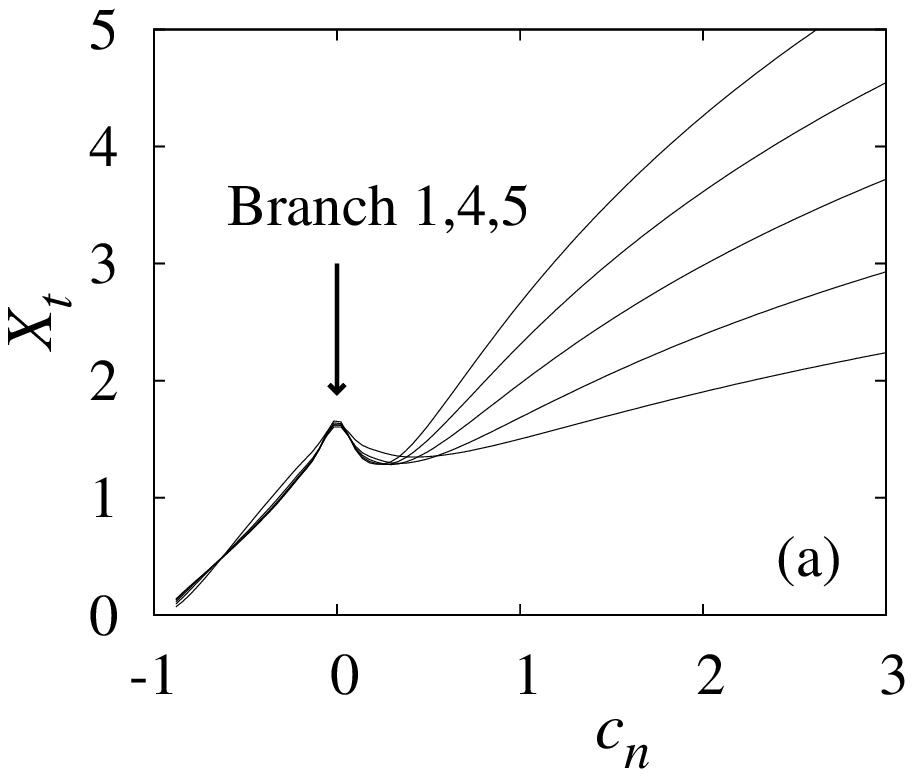}
\hspace{-18mm}
\includegraphics[scale=0.7]{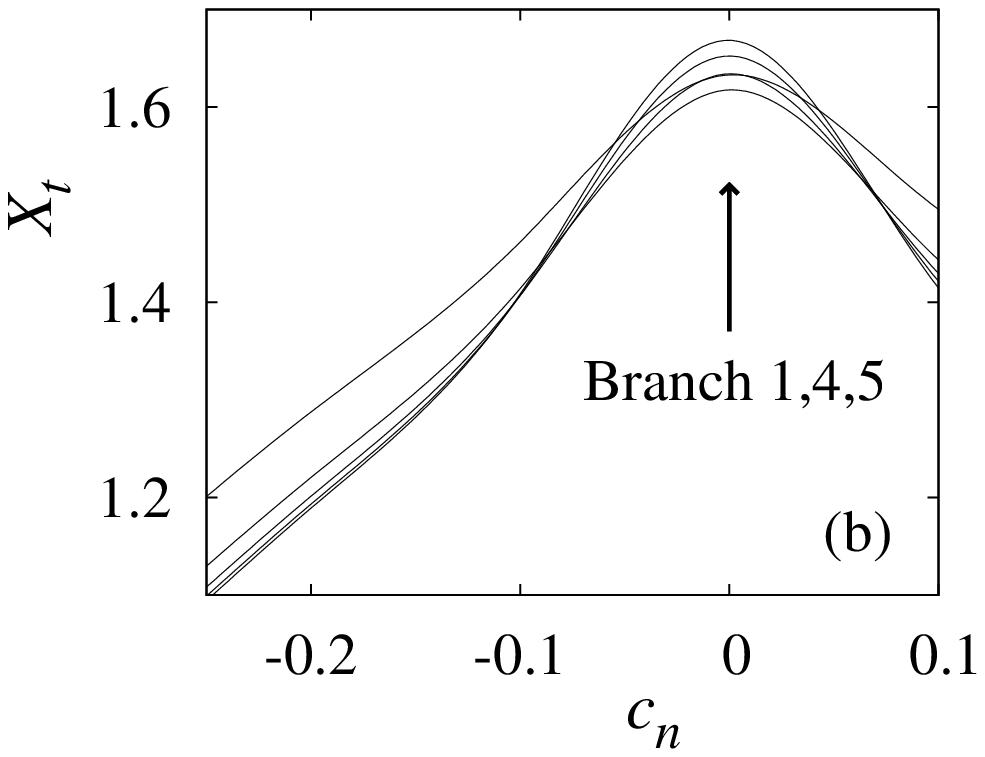}
\end{center}
\caption{Scaled thermal gaps versus the cubic vertex weight of
the completely packed O($n$) loop model around branch 5 for $n=2$. The
scaled gaps for $c_n<0$ are hard to distinguish in figure (a); an
enlarged view is shown in figure (b). The cubic weight is fixed at its
value at branch 5.  Results are shown for even system sizes $L=4$ to 12.
The scaled gaps increase as a function of $L$ on the right-hand side of
figure (a), and they decrease on the left-hand side of figure (b).}
\label{fig:b5cn2}
\end{figure}

The $X_t$ data for $n=2$ are shown in Figs.~\ref{fig:b5cn2}. The point 
$c_n=0$ is the intersection of branches 1, 2, 4 and 5.  For $c_n>0$, the
scaled gaps again display a divergent behavior. Although the vector space
of the transfer matrix for branch 5 is larger than for branch 2, the $X_t$
data coincide with those in Fig.~\ref{fig:xb2n2}a in this range. 
Different behavior occurs for $c_n>0$. The $X_t$ data no longer agree
with those in Fig.~\ref{fig:xb2n2}a and display a ``nonuniversal'' range
$c_n<0$
where the thermal scaling dimension $X_t$ depends continuously on $c_n$.
The difference with the corresponding branch-2 data for $X_t$ is due to
the second largest eigenvalue for branch 5. The largest eigenvalues are
the same for both branches at $n=2$; thus, the behavior of the magnetic
gaps for branch 5 is the same as shown in Fig.~\ref{fig:xb2n2}b.

\begin{figure}
\begin{center}
\hspace*{-8mm}
\includegraphics[scale=0.73]{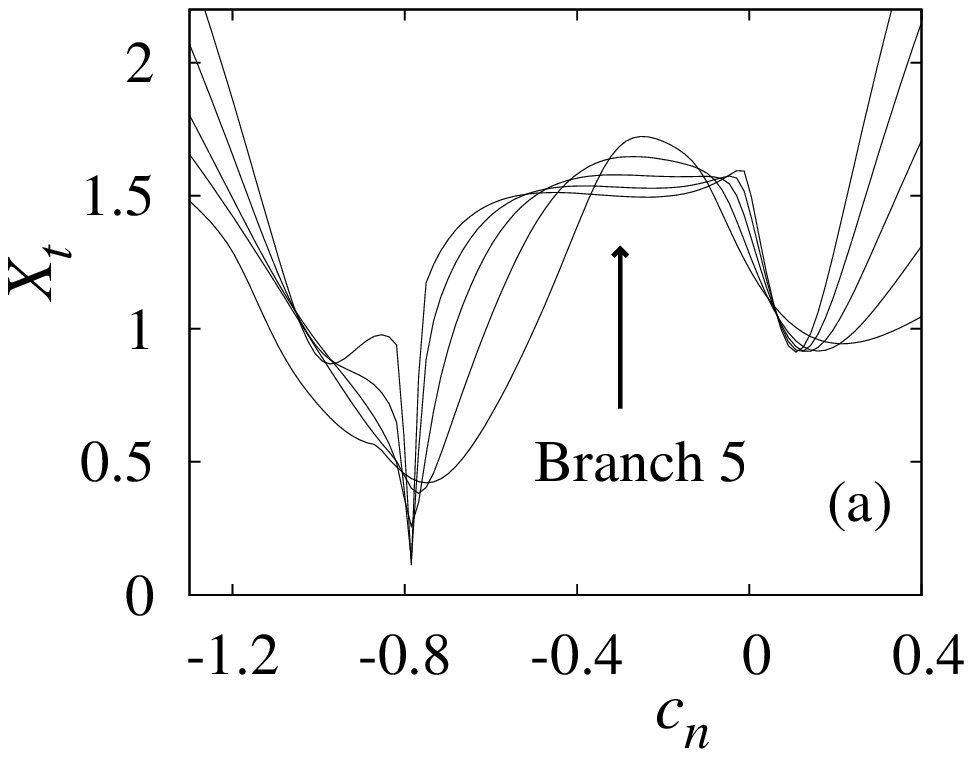}
\hspace{-18mm}
\includegraphics[scale=0.73]{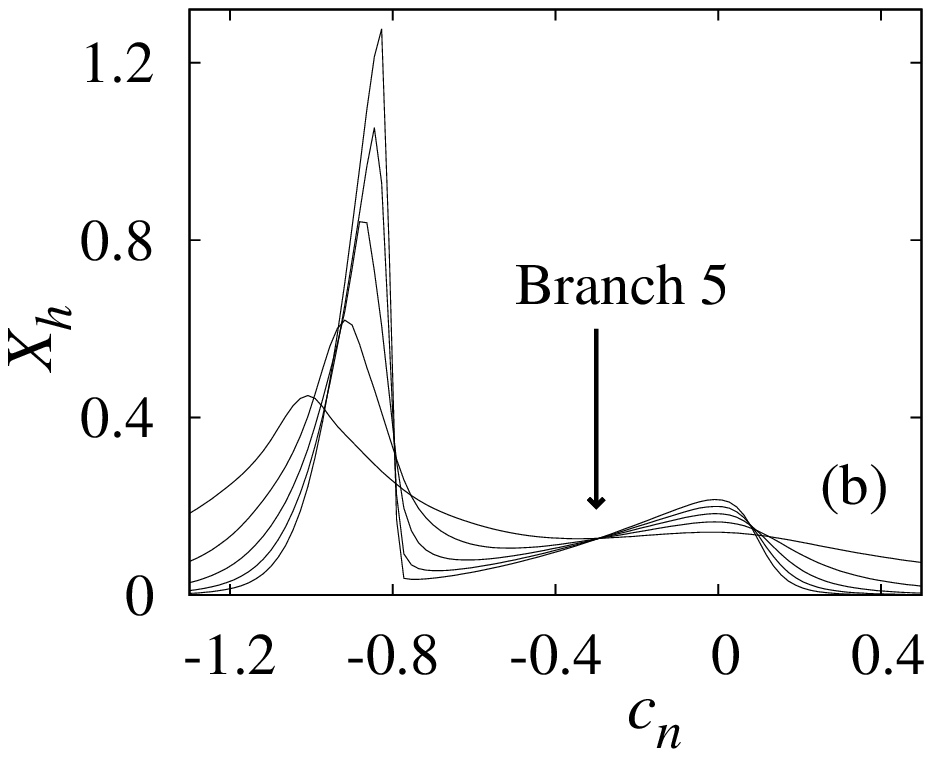}
\end{center}
\caption{Scaled gaps versus $c_n$ of the completely packed O($n$)
loop model around branch 5 with $n=5$. Figure (a) shows the data for 
$X_t$ and figure (b) for $X_h$.
The crossing bond weight is fixed at its value at branch 5.
Results are shown for odd $L$ in the range $3 \leq L \leq 11$.
The scaled thermal gaps at branch 5 decrease as a function of $L$, 
while the magnetic gaps display intersections near branch 5;
there, the steeper curves correspond with larger $L$.}
\label{fig:b5n5c}
\end{figure}

\begin{figure}
\begin{center}
\hspace*{-8mm}
\includegraphics[scale=0.73]{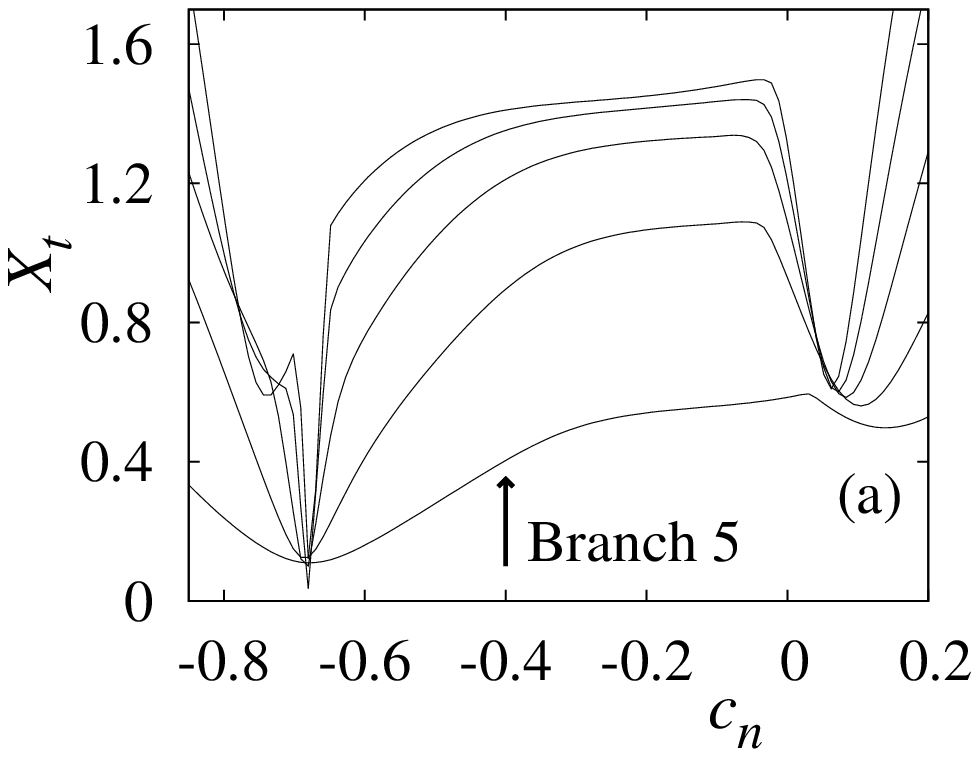}
\hspace{-18mm}
\includegraphics[scale=0.73]{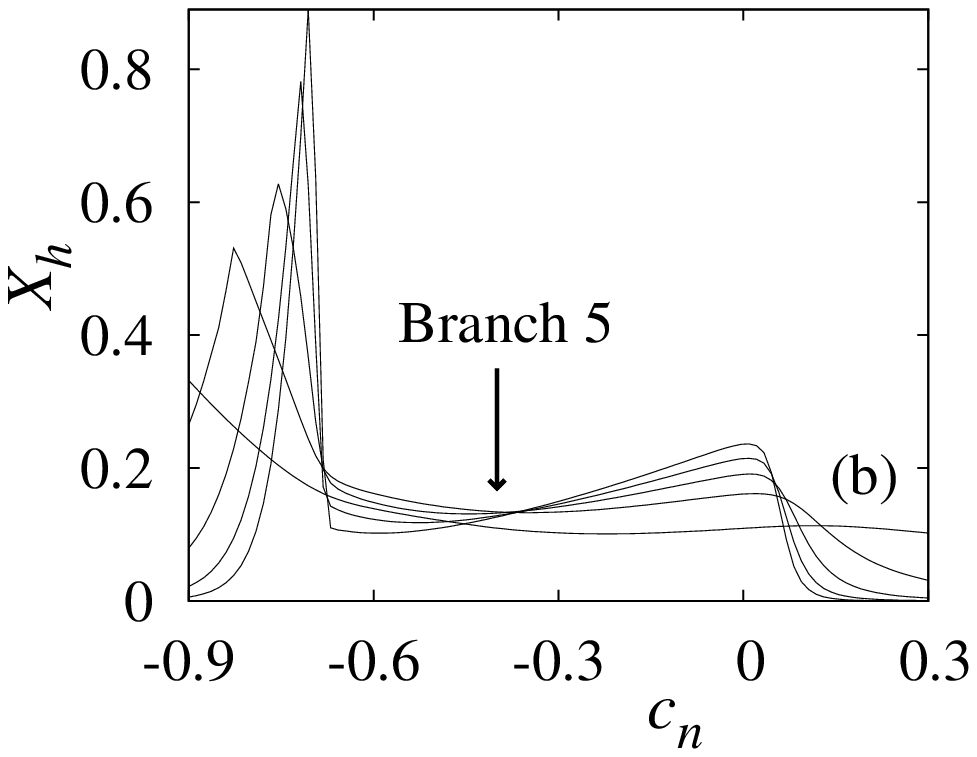}
\end{center}
\caption{Scaled gaps versus $c_n$ of the completely packed O($n$)
loop model around branch 5 with $n=10$. Figure (a) shows the data for 
$X_t$ and figure (b) for $X_h$.
The crossing bond weight is fixed at its value at branch 5.
Results are shown for odd $L$ in the range $3 \leq L \leq 11$.
The scaled thermal gaps at branch 5 increase as a function of $L$,          
while the magnetic gaps display intersections near branch 5;  
there, the steeper curves correspond with larger $L$.}
\label{fig:b5n10c}
\end{figure}

Finally we display the effect of a change of the cubic vertex weight
on the branch-5 systems with $n=5$ and 10. The corresponding scaling
plots are shown in Figs.~\ref{fig:b5n5c} and \ref{fig:b5n10c}.
These results may suggest convergence to a nontrivial temperature
dimension $X_t \approx 1.5$, in a range of $c_n$ about branch 5, but again
we have to consider the possibility of strong crossover phenomena.
The sharp minima near $c_n=-0.7$ tend to $X_t=0$, and thus suggest a
first-order transition to a state dominated by $c$-type vertices.
Instead, the intersections of the curves near $c_n=0$ tend to a nonzero
value, corresponding to a continuous transition to a $c$-dominated phase.

Our numerical results for $n=30$ (not shown) are entirely consistent with
this picture. There are sharp minima in the $X_t$ curves near $c_n=-0.6$ 
and $c_n=0$. In between, the data increase slowly with $L$, up to a maximum
of about $X_t=0.6$ for $L=12$, but do not allow a firm conclusion about 
their convergence. Outside this range, the scaled gaps clearly display 
divergent behavior as expected for off-critical phases dominated by the
$c$-type vertices.

\section{Discussion}
\label{discus}
We have investigated the completely packed Eulerian graph model on the
square lattice with three types of vertices.  In particular we focused
on the symmetric case with vertex weights satisfying the fourfold
rotational symmetry of the lattice. We explored 
the physics associated with the exactly solvable cases of the
equivalent Perk-Schultz coloring model. We have checked and extended
the results for the bulk free energy given by Schultz \cite{S} for the
symmetric cases of the coloring model. 
We also explored the phase diagram in the neighborhood of branches 1 to 5,
which revealed many details concerning the nature of the relevant phases
and their transitions.

{\bf Branch 1:}
This model (or its equivalent models) has already been studied extensively
and there are well-established results.
We investigated the equivalence of two seemingly different solutions
applying to the case $n>2$ case analytically. Indeed we found that the
Schultz solution for the partition sum per site for $n>2$ is exactly
equivalent with the Baxter solution for the $q>4$-state Potts model,
and thus with the corresponding range of the Lieb solution of the
6-vertex model.

We also checked the consistency between the analytic solutions and
our numerical results for the free energy. We find accurate agreement
with the Baxter solution for the Potts model, which applies to the
range $n>0$, and with the Lieb solution \cite{EL} which is specified
for all $n>-1$. The latter solution also agrees with our results when
continued down to $n=-2$.  Furthermore, we continued the Lieb solution
to complex parameters \cite{BWG}, which then covers the range $n<-2$.
We checked that the resulting formula agrees precisely with our
numerical free energy results for that range of $n$.

We also compared our numerical results for the  scaling dimensions with
the Coulomb gas predictions for $|n|<2$, and found  a satisfactory
agreement.  Our numerical
results show that a temperature-like scaling dimension $X_{t1}=4$ exists,
which is the leading dimension of that type for $n<1$.

{\bf Branch 2:}
The Schultz solution for the partition sum of this branch proved to be
consistent with our transfer-matrix analysis. The latter calculations
benefited from an improved coding algorithm (see Appendix \ref{tmtechnique})
that allowed us to reach larger system sizes in comparison with
Ref.~\onlinecite{onc}.

The physical character of this model, which includes $z$-type as well
as $c$-type vertices, depends on the range on $n$.
For $n>2$, branch 2 is physical in the sense of positive Boltzmann
weights. For large system sizes, the scaled gaps approach a value
consistent with $0$, indicative of a first-order phase transition.
Furthermore, the exploration of the phase diagram of the model as a function
of the cubic vertex weight, reported in Sec.~\ref{embed}, indeed shows that, 
at least for $n>>2$, branch 2 corresponds with a locus of first-order
phase transitions. As for the nature of this transition, we recall that
for $x=c=0$ (branch 1) the system displays a checkerboard-like order.
This order can break down when vacant vertices are introduced \cite{FGB},
and it is plausible that the introduction of cubic vertices will yield
a similar result. Concerning the physical reason behind the first-order
nature, we mention that the  transition is located at a cubic weight
$c\approx \sqrt n$ for large $n$. That is where the Boltzmann weights
of the checkerboard O($n$) phase and that of the fully ordered cubic phase
coincide. There, the introduction of a cubic vertex in the checkerboard
background of $z$-type vertices increases the Boltzmann weight
by a factor $\sqrt n$, but the number of components in Eq.~(\ref{Zsql})
decreases by one, which costs a factor $n$.
Furthermore, two or three cubic vertices do not interact. Only when
four cubic vertices form a square, there is no factor $1/n$ involved
in the addition of the last vertex.
For this reason, there is no appreciable attraction between the cubic
vertices when their density $\rho=N_c/N$ is low. As long as the density
is small, it is thus mainly governed by the fugacity $c_n$ of the cubic
vertices, since we have set $z=1$ in Eq.~(\ref{Zsql}).

\begin{figure}
\begin{center}
\includegraphics[scale=0.75]{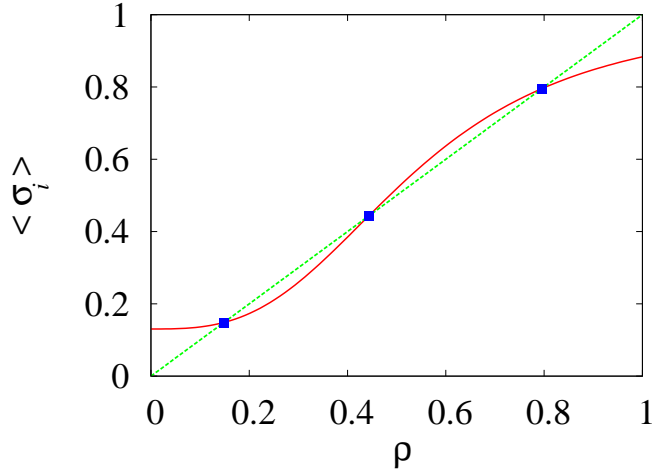}
\end{center}
\caption{(Color online) Mean-field approximation for the density of the type-$c$
vertices, for $n=100$ and $c_n=3/40$. The existence of three
solutions (square symbols) for this density is a sign of a first-order
transition.}
\label{fig:sceq}
\end{figure}

Let us next
consider the attraction between the cubic vertices when their density
is no longer negligible, using a mean-field type approximation.
For this purpose we denote the absence or presence of a cubic vertex
on site $i$ by means of a site variable $\sigma_i$ with corresponding
values $\sigma_i=0$ and 1 respectively.  Due to the absence of a factor
$1/n$ when four cubic vertices form a square, the above-mentioned weight
$c_n$ of a cubic vertex has to be replaced by $(c_n)[n\rho^3+1-\rho^3]$.
Thus, the mean-field self-consistency equation at low densities of the
cubic vertices becomes
\begin{equation}
<\sigma_i> = \frac{(n-1)\rho^3+1}{(n-1)\rho^3+1+1/c_n} \, .
\label{mf}
\end{equation}
A self-consistent solution of the equation $\rho=<\sigma_i>$ for large
$n$ exists with $\rho$ not exceeding a value of order $1/\sqrt n$, as long
as $c_n$ does not exceed $\sqrt n$, near the locus of the phase transition.
This smallness of $\rho$ is already a sign that the phase transition
for large $n$ is first order. Numerical evaluation of $<\sigma_i> $ for
large $n$, $c_n\approx \sqrt n$ indeed shows three solutions of the equation
$<\sigma_i> =\rho$, corresponding with a jump in $\rho$ when $c_n$ is varied.
An example is given in Fig.~\ref{fig:sceq} for $n=100$ and $c_n=3/40$.
The curve shows Eq.~(\ref{mf}), and the straight line the self-consistency
condition $<\sigma_i> = \rho$.

For $n<2$, the cubic weight at branch 2 becomes negative, but the model 
is not necessarily unphysical, since its weights in the equivalent
coloring-model representation are non-negative. Furthermore, the sign of
a cubic perturbation is important in the context of the universal behavior
of the O($n$) spin model. Depending on this sign, crossover will occur to
the face-cubic or to the corner-cubic phase. From the association of the
face-cubic model with four-leg vertices \cite{BNcub} one may interpret
a negative cubic vertex weight with crossover to a corner-cubic state.
Indeed, the cubic perturbation is relevant in the dense O($n$) loop 
phase, as is clear from the Coulomb gas theory \cite{CG}, and confirmed
by numerical work \cite{onc}. The fact that the cubic weight is rather
limited for branch 2 with $1<n<2$, supports its physical association
with a low-temperature corner-cubic state.

{\bf Branch 3:} 
Also our numerical results for the free energy of branch 3 are in a
good agreement with the exact result of Schultz \cite{S}. 
The data presented in Sec.~\ref{embed} indicate that also branch
3 is the locus of a first-order transition line in the $n$ versus
$c_n$ phase diagram for sufficiently large $n$. Just as for branch 2, 
the transition can be interpreted as the frontier of the long-range
ordered lattice-gas-like state that occurs when the $z$-type vertex
dominates. The scaling behavior of the gaps is, for large $n$, 
similar to that of branch 2. The loop-model version of the branch-3
model is, however, unlike branch 2, unphysical for all $n$, because the
cubic weight $c_n$ is negative. But the coloring-model weights are still
positive for $1<n<2$. In that range, the numerical results for the
conformal anomaly seem to diverge, and the scaled gaps at branch 3 display 
poor convergence with $L$, and do not allow a satisfactory estimation
of the scaling dimensions. But it is clear that a range of $c_n$ exists 
between branches 2 and 3 where $X_t$ tends to converge to a $c_n$-independent
value close to 2, which suggests that an algebraic phase exists. It does
still seem well possible that branch 3 defines the boundary of that phase.

{\bf Branch 4:}
In this case, both $z$-type and $x$-type vertices are present.
An analysis of the difference between the expressions given by
Schultz \cite{S} and Rietman \cite{RPhD} in Sec.~\ref{RvsS} showed that
the Schultz result has to be modified with a factor ${\rm ctg}(\alpha\pi)$,
after which it becomes equivalent with the Rietman result. This factor
specifies the periodic function $p(s)$  mentioned in Ref.~\onlinecite{PS},
which refers to Ref.~\onlinecite{S} in its footnote 64. Indeed, the
numerical analysis presented in Sec.~\ref{numeranalf} is in a good
agreement with the Rietman solution \cite{RPhD} for $n\le2$. The latter
solution does not apply to the range $n>2$. On the basis of our
transfer-matrix results we conjecture that the free energy is symmetric
with respect to $n=2$, i.e., $f(n)$ satisfies $f(2+x)=f(2-x)$. This
generalizes the Rietman result to an expression for the
free energy per site for all $n$, listed in Eq.~(\ref{Rm_ge}).

Our numerical estimates of the conformal anomaly $c_{\rm a}$ for
$n\le 2$ are, although not accurate,  confirm the existing
result \cite{MNR} $c_{\rm a}=n-1$. For $n\ge 2$ our results allow the
conjecture $c_{\rm a}=n/2$. The results for the scaling dimensions
for branch 4 with $n<2$ in Table \ref{scad} are mildly suggestive of 
$X_t=2$ and $X_h=0$,

Concerning the phase diagram of the intersecting O($n$) loop model, thus
the system described by Eq.~(\ref{Zsql}) with $c=0$ and $x/z$ and $n$ as
variable parameters, we find different types of behavior in the ranges $n>2$
and $n<2$. For $n>>2$ we see no evidence  of a strongly singular transition
as a function of $x$. However, the lattice-gas-like order that exists
for $x=0$ should dissolve when the $z$-type vertices become sparse for
larger $x$, and thus a phase transition of a weak signature seems very
likely. 

For $n<2$, the dense phase of the nonintersecting loop model still
displays Ising-like ordering, but here the introduction of crossing
bond ($x$-type) vertices is a relevant perturbation. It is, just as the
cubic perturbation, described by a 4-leg vertex, for which the Coulomb
gas analysis \cite{CG} can be applied, which then yields the exact
scaling dimension of this perturbation.  Its relevance for $n<2$       
leads to different universal properties of the dense phase
of the model with crossing bonds in comparison with those of the
nonintersecting loop model \cite{MNR,onc}. Indeed, we observed such
crossover to different universal behavior in Sec.~\ref{embed}
when a nonzero weight $x$ is introduced.
It is noteworthy that, for $x=0$, the leading temperature-like dimension
$X_t$ assumes a value corresponding with the cubic-crossover exponent
given by Eq.~(\ref{xc}), different from the corresponding result in 
Table \ref{scad}. It thus reflects that the enlargement of the set of
connectivities, caused by the introduction of crossing bonds, allows
the coding of more correlation functions in comparison with the
nonintersecting subset.  

Our analysis of the scaled gaps and the associated scaling dimensions
as a function of the crossing-bond weight,
while the weight of the nonintersecting vertices is kept constant,
confirmed that a phase transition takes place at $x=0$ for $-2<n<2$.
For $x>0$, we did not find clear signs of a phase transition for any value
of $n$, in contrast with the findings reported in Ref.~\onlinecite{onc}.
The latter work does, however, not concern completely packed systems. 

{\bf Branch 5:}
The identity of the free energy of branches 4 and 5 can be understood by
translating the generalized loop model back into coloring model language.
As according to Eq.~(\ref{map}) and Table \ref{branches}, both $W^d$ and
$W^r$ take the same values for branch 4 as for branch 5.
The weight $W^0$ has different signs for the two branches,
but the absolute values are the same.
Furthermore the weight $W^0$ describes the crossing of loops of a different
color, and the number of such intersections must be even in the even
systems that we are considering. Therefore, the free energy, and the largest
eigenvalue of the transfer matrix as well, must be the same for the two
branches. Thus, we may still use the generalized Rietman result
as the theoretical prediction of the free energy of the branch-5 model.

The transfer-matrix analysis for branch 5 is somewhat more involved than
in the previous cases, since there are now three different types of 
vertices in the system. New eigenvalues appear in the configuration space
of the transfer matrix, and the leading scaling dimensions of branch 5 are
different from branch 4. The temperature exponent appears to be very
small for $n<2$, and the magnetic exponent seems to be negative.

A complete analysis of the phase behavior in a vicinity of branch 5 would
involve the scanning of the 3D phase diagram parametrized by $(x,c_n,n)$.
Concerning this matter, we only performed superficial exploration in
the $x-n$ diagram with the cubic weight fixed at its branch-5 value,
and in the $c_n-n$ diagram at a similarly fixed crossing-bond weight.
In the $x-n$ diagram, no signs of phase transitions emerged in the
immediate neighborhood of branch 5 for sufficiently large $n$ ($n>10$), 
but a transition to a $c_n$-dominated phase is seen at small $x$.
For $n=2$ there is a clear change of behavior at $x=0$, where several
branches intersect. On both sides of this point there is a ``nonuniversal''
range of $x$. For $n<2$ the data show a transition at or near branch 5.

Similarly, in the $c_n-n$ diagram, a critical transition occurs at the
location of branch 5 for $n<2$. Our results indicate the existence 
of cubic long-range order when $c_n$ exceeds its branch-5 value.
For $n=2$ there is again a clear ``nonuniversal'' range, but only for 
$c_n<0$. For $c_n>0$, the divergent behavior of $X_t$ indicates the
existence of a phase where the cubic vertices percolate, except perhaps
in a small range close to $c_n=0$. 
Also for $n>>2$ we observe phase transitions to the cubic phase,
both at positive and larger negative value of $c_n$. In between, there
seems to exist a phase where the scaling dimensions depend continuously
on $c_n$.

{\bf Branch 6:}
Branch 6 is a very simple case with only one nonzero vertex weight,
namely the crossing-bond or $x$-type. Indeed the Schultz solution
predicts a trivial free-energy density. This agrees well with the
largest transfer-matrix eigenvalues which are, according to
Eq.~(\ref{fbr6eo}), equal to
\begin{equation}
\Lambda_0(n,L)=n \, ,
\label{fit2}
\end{equation}
This result follows immediately from the vertex weights $n_z=n_c=0$,
$x=1$ and Eq.~(\ref{Tdef}). Every new row added by the transfer matrix 
forms a loop closed about the cylinder, and thus contributes a factor
$n$ to the partition sum. The transfer matrix is diagonal, with 
all elements and eigenvalues equal to $n$.

{\bf Branch 7:}
This case is similar compared to branch 6, and again the Schultz solution
predicts a trivial free-energy density, in agreement with our
transfer-matrix results. Eq.~(\ref{fbr7eo}) implies the following
eigenvalues
\begin{equation}
\Lambda_0(n,L)=
\begin{cases}
~~n, \text { if $L$ is even.}\\
~~n-2, \text { if $L$ is odd.}\\
\end{cases}
\label{fit4}
\end{equation}
This result can be explained using the language of the coloring model,
for which we only have nonzero vertex weights $W^{0}=1$ and $W^{\rm d}=-1$.
Thus all edges on a line in the transfer direction have the same color,
and the same holds for edges on lines in the perpendicular direction.
The weight of a newly added row depends on the colors of the lines in the 
transfer direction. The maximum weight is realized for $L$ lines of the
same color. The weight of a newly added row is 1 if it has one of the
$n-1$ different colors, and $(-1)^L$ if it has the same color. These
weights indeed sum up to the multiplicities $n$ and $n-2$ appearing
in Eq.~(\ref{fit4}).

Finally we remark that the present explorations, although yielding
a lot of new information, are necessarily far from complete.
Furthermore, the limited ranges of accessible finite sizes in
our transfer-matrix analyses did, in several cases, not allow the
derivation of satisfactorily accurate results. Perhaps Monte Carlo
methods will appear to be helpful to resolve some of these issues.

\acknowledgments
It is a pleasure to acknowledge valuable comments by and enlightening
discussions with Profs.~B. Nienhuis and J.~H.~H. Perk.
This work was supported in part by the NSFC under Grant 11175018 (WG)
and by the Lorentz Fund.

\appendix
\section{Transfer-matrix technique}
\label{tmtechnique}
A crucial piece of information for the construction of the transfer matrix
for the generalized loop model is the way in which
the dangling bonds at the end of the cylinder are mutually connected
via some path of bonds in the cylinder.
This information is called ``connectivity'', denoted by Greek symbols
$\alpha$ or $\beta$ etc.. Let there be in total $C_L$ possible
connectivities for $L$ dangling bonds.
The partition sum $Z^{(M)}$ of a cylinder consisting of $M$ circular 
rows of $L$ vertices is divided into $C_L$ restricted sums, according
to the connectivity $\beta$ of the dangling bonds. The restricted sums
for the model of Eq.~(\ref{Zsqll}) are formally expressed as
\begin{equation}
Z^{(M)}_\beta=\sum_{{\mathcal G}_M}\delta_{\beta\varphi({\mathcal G}_M)}z^{N_z}
c_n^{N_{c}} x^{N_{x}} n^{N_{l}} \, ,
\label{res_sum}
\end{equation}
where $\varphi$ is the connectivity implied by the Eulerian graph
 ${\mathcal G}_M$.
Let us now add another row of $L$ vertices and rewrite the restricted
partition sums of the $(M+1)$-row system as follows:
\begin{displaymath}
Z^{(M+1)}_\beta=\sum_{{\mathcal G}_{M+1}}\delta_{\beta\varphi
({\mathcal G}_{M+1})}z^{N'_z} c_n^{N'_{c}} x^{N'_{x}} n^{N'_{l}}
\end{displaymath}
\begin{equation}
=\sum_{{\mathcal G}_M}z^{N_z} c_n^{N_{c}} x^{N_{x}} n^{N_{l}}\sum_{g_{M+1}}
\delta_{\beta\varphi({{\mathcal G}_M,g}_{M+1})}z^{n_z} c_n^{n_{c}}
x^{n_{x}} n^{n_{l}(\alpha)}\,.
\label{res_sum_plus2}
\end{equation}
where the primed quantities refer to the $M+1$-row system, 
${\mathcal G}_{M+1}={\mathcal G}_M \cup g_{M+1}$ where $g_{M+1}$ is the
vertex configuration on the $M+1$-th row, and the
lower-case symbols $n_z$, $n_c$, $n_x$ and $n_l$ denote the increase of
the numbers of vertices and loops caused by the addition of the $(M+1)$th 
row. All these numbers depend on $g_{M+1}$, but only $n_l$ depends also on
$\alpha$, which dependence is explicitly shown.
Next, we note that the connectivity $\varphi({\mathcal G}_{M+1})$ depends
only on $\varphi({\mathcal G}_{M})$ and $g_{M+1}$, and insert an innocent
factor $\sum_{\alpha=1}^{C_L} \delta_{\alpha\varphi({\mathcal G}_{M})}$
which yields
\begin{equation}
Z^{(M+1)}_\beta=\sum_{\alpha=1}^{C_L} \sum_{{\mathcal G}_M}
\delta_{\alpha\varphi({\mathcal G}_{M})}z^{N_z} c_n^{N_{c}} x^{N_{x}}
n^{N_{l}}\sum_{g_{M+1}}\delta_{\beta\varphi(\alpha,{g}_{M+1})}z^{n_z}
c_n^{n_{c}} x^{n_{x}} n^{n_{l}(\alpha)}\,.
\label{rec1}
\end{equation}
With the definition of the transfer-matrix elements by
\begin{equation}
T_{\beta \alpha}=\sum_{g_{M+1}}\delta_{\beta\varphi(\alpha,{g}_{M+1})}
z^{n_z} c_n^{n_{c}} x^{n_{x}} n^{n_{l}(\alpha)} \, ,
\label{Tdef}
\end{equation}
Eq.~(\ref{rec1}) assumes the recursive form 
\begin{equation}
Z^{(M+1)}_\beta=\sum_{\alpha=1}^{C_L} T_{\beta\alpha} Z^{(M)}_{\alpha} \, .
\label{Z-recursion}
\end{equation}
Repeated application yields that, in the large-$M$ limit, the largest
eigenvalue $\Lambda_0$ of the transfer matrix determines the free-energy
density. In actual calculations, we do not explicitly compute the elements
$T_{\beta \alpha}$, but the transfer-matrix is decomposed \cite{BN82,BN}
instead in $L$ sparse matrices, for which the required memory is only 
proportional to the number of connectivities, instead of quadratic.

\subsection{Coding and decoding of the connectivities}
For actual calculations one needs to determine the number $C_L$ of 
$L$-point connectivities of the model, and to code each of these by
consecutive and unique integers $1, 2, 3, ..., C_L$. A decoding
algorithm is needed as well. The numbers $C_L$ increase with $L$, but
in a way that still depends on the set of allowed vertices.
Since it is, for the finite-size analysis, desirable to have as wide as
possible ranges of system sizes $L$ available for each parameter choice,
we have constructed separate coding algorithms for four applicable sets,
namely including $z$, $z$ and $x$, $z$ and $c$, and lastly $z$, $x$, and $c$.

\subsection{Some remarks on the actual coding methods}
The existing literature already contains much information about the
coding and decoding algorithms used in transfer-matrix calculations.
Here we present only a short characterization of, or references to, the
coding methods used in the present work. We first consider the case of even
$L$, such that connected dangling bonds occur only in even numbers.

\subsubsection{$x=c=0$, $z \ne 0$}
The coding of the $z$-type connectivities is part of a more complicated
problem that was already described  in some detail in Ref.~\onlinecite{BN},
namely the coding of so-called dilute well-nested O($n$) connectivities,
which contain, besides connected pairs, also ``vacant'' dangling bonds,
which are not occupied by a loop segment. This is the coding method used
for the  case $z \ne 0$, $c=x=0$.

\subsubsection{$c=0$, $z \ne 0$, $x \ne 0$}
For intersecting loop models in which also $x$ is nonzero, the
connectivities are represented by rows of integers occurring in pairs,
but no longer well-nested. The enumeration of these connectivities is
described in Ref.~\onlinecite{onc}.

\subsubsection{$x=0$, $z \ne 0$, $c \ne 0$}
The coding problem for the case that cubic vertices are present, in
the absence of $x$-type vertices, was already considered in
Ref.~\cite{GQBW}. Vacant bonds were included as well in that work.
The coding used there was basically the coding for random-cluster
connectivities \cite{BN82}, which is sufficient because the completely
packed cubic connectivities are a subset of the set of random-cluster
connectivities. However, for larger system sizes it is only a relatively
small subset, which leads to a reduction of the efficiency of the
algorithm and of the largest possible system size. For this reason 
we constructed a new coding algorithm for the completely packed
cubic connectivities without the use of the random-cluster algorithm.

The principle is summarized as follows.
Represent the configuration of dangling bonds by a row of $L$ integers,
such that equal integers describe connected bonds, and that unequal
integers describe unconnected bonds. We refer to this $L$-site cubic
connectivity as the level-zero connectivity. On the basis of these $L$
integers one may perform the following steps.
\begin{description}
\item[{\em (i)}]
Let the size of the cluster containing site 1 (the number of connected
dangling bonds that include dangling bond 1) be $n_1$. This cluster can be
characterized by a bitstring of $L$ bits containing $n_1$ ones. However, 
not all such bitstrings are allowed. The first bit is always 1 and may be
skipped, and since the number of zeroes interlaced between a pair of
consecutive ones can only be even, one may also skip one half of the
zeroes. The cluster containing site 1 is thus characterized by means of
a unique bitstring of length $(L+n_1)/2-1$ with an odd number of ones.
These bitstrings are simply coded and decoded in lexicographic order by
using the binomial distribution.
\item[{\em (ii)}]
Let the zeroes in the bitstring occur in $n_g$ groups separated by one or
more ones.  As a consequence of the well-nestedness property, the sites
in each of these groups cannot be connected to sites in other groups.
These groups are called ``level-1 connectivities'' are still represented
by rows of integers, which, for a given site, keeps the same value as
for the original level-0 connectivity. The degrees of freedom of the
level-0  connectivity that are not accounted for by the enumeration of
the level-0 bitstring are thus represented by $n_g$ connectivities on
less than $L$ points.
The coding of the original cubic connectivity is completely specified
by the level-0 bitstring code, supplemented with the $n_g$ level-1
connectivities.
\item[{\em (iii)}]
One can now analogously perform the operations specified in steps
{\em (i)} and {\em (iii)} for each of the $n_g$ level-1 connectivities. 
This will yield the enumeration of the level-1 bitstrings, and may also
lead to a number of level-2 connectivities, and so on. The process ends
at the level that yields 0 subgroups for the next level. 
\end{description}

This process generates a tree-like structure of which the relevant
data, i.e., the bitstring codes, the number of subgroups, their
length and the position of the first site of that subgroup, are stored
for all subsequent levels. After completion of the tree, this information
can be transformed into a unique number: the code the $L$-point connectivity.
For that purpose one has to define an ordering of the connectivities,
which can, on level 0, be done using the number of sites connected to
site 1, combined with the code of the bitstring describing the cluster
containing site 1. The same type of ordering is applied to subrows
at all levels. Those at the highest level are assigned the number 1,
and the ordering then determines the enumeration at the next-highest
level, and so on until level 0.

\subsubsection{$z \ne 0$, $x \ne 0$, $c \ne 0$}
In the general case that all three vertex types are simultaneously
present, the number of possible connectivities
for a given system size becomes even larger, but the coding algorithm
actually becomes much simpler. If site 1 belongs to a cluster of 
$n_1$ connected sites, the cluster is represented by a bitstring of
length $L-1$ with $n_1-1$ ones, which may sit on arbitrary positions.
This bitstring is enumerated according to lexicographic ordering, and
the coding assigns a number equal to the connectivity equal to the number
of connectivities with a smaller bitstring number, plus the number
associated with the coding of the remaining $L-n_1$-point connectivity.
The latter problem is entirely similar to the original problem, and can
thus, step by step, be further reduced, until all sites of the remaining
connectivity belong to one connected cluster. In that case we assign 
the number 1 to the remaining connectivity.

\subsubsection{Odd system sizes}
For $L$-point connectivities with odd $L$ we allow one odd group of
dangling bonds, containing one bond for $c=0$, and an arbitrary odd
number of  bonds for $c\ne 0$. Coding of these odd connectivities is
done by similar methods.

The largest transfer-matrix sizes used for the various types of coding
methods are listed in Table \ref{ncon}.

\begin{table}
\caption {The largest system sizes and the corresponding numbers of
connectivities (maximum linear size of the transfer matrix) used in
the present calculations, for several combinations of the allowed
vertex types listed in the first column.}
\label {ncon}
\renewcommand{\arraystretch}{1.5}
\begin{center}
\begin{tabular}{ |c | c || c | r | }
\hline
    vertex types& even/odd & $L_{\rm max}$ & $ ~C_{L_{\rm max}}$ \\ \hline
    \multirow{2}{*}{$z$}    & even & 30 & 9694845  \\ \cline{2-4}
                            & odd  & 27 & 20058300 \\ \hline
    \multirow{2}{*}{$z,c$}  & even & 22 & 8414640  \\ \cline{2-4}
                            & odd  & 19 & 6906900  \\ \hline
    \multirow{2}{*}{$z,x$ } & even & 16 & 2027025  \\ \cline{2-4}
                            & odd  & 15 & 2027025  \\ \hline
    \multirow{2}{*}{$z,c,x$}& even & 14 & 4373461  \\ \cline{2-4}
                            & odd  & 13 & 4373461  \\ \hline
\end{tabular}
\end{center}
\end{table}

\end{document}